\documentclass[12Pt]{article} 
\usepackage{amssymb,euscript,latexsym,amsmath} 
\usepackage{graphicx} 
\usepackage{epstopdf}
\usepackage{bm} 
\usepackage{enumerate} 
\usepackage{color} 
\usepackage{setspace} 
\usepackage{subfigure} 
\usepackage{hyperref}
\usepackage{array}
\newcolumntype{M}{>{\centering\arraybackslash}m{\dimexpr.215\linewidth-2\tabcolsep}}

\DeclareMathOperator{\sgn}{sgn}

\begin{document}
\title{Optimization of two- and three-link snake-like locomotion}
\author{Fangxu Jing and Silas Alben}
\maketitle

\begin{abstract}
We analyze two- and three-link planar snake-like locomotion and optimize the motion for efficiency. The locomoting system consists of two or three identical inextensible links connected via hinge joints, and the angles between the links are actuated as prescribed periodic functions of time. An essential feature of snake locomotion is frictional anisotropy: the forward, backward and transverse coefficients of friction are different. The dynamics are studied analytically and numerically for small and large amplitudes of the internal angles. Efficiency is defined as the ratio between distance traveled and the energy expended within one period, i.e. the inverse of the cost of locomotion. The optimal set of coefficients of friction to maximize efficiency consists of a large backward coefficient of friction and a small transverse coefficient of friction, compared to the forward coefficient of friction. For the two-link case with a symmetrical motion, efficiency is maximized when the internal angle amplitude is approximately $\pi/2$, for transverse coefficient sufficiently large. For the three-link case, the efficiency-maximizing paths are triangles in the parameter space of internal angles.
\end{abstract}

\section{Introduction} 
\label{sec:intro}

Snake locomotion has long been a topic that fascinates researchers, and has recently received a renewed wave of interest in the fields of robotics and control, as well as in organismal biology. Snakes are familiar organisms, but as limbless animals, their locomotion has special features~\cite{Ga1962}. Terrestrial snakes move using friction between the ground and their belly scales, which have anisotropic frictional properties~\cite{Gr1946}. It has been proposed that the cost of transport (energetic efficiency) for snake slithering is no greater than that of limbed animals~\cite{SeJaBe1992,WaJaBe1990}. 

Some works on modeling snake locomotion are oriented towards wheeled snake robots~\cite{BuRaCh1994, Ch2003, Ch2005, GuMa2008, Hi1993, Mi2002, OsBu1996}. These models are typically concerned with motion planning, and assume that the transverse coefficient of friction is high enough to prevent transverse motion, while the forward and backward coefficients of friction are the same. These models work well for wheeled robots and provide valuable insights into the locomotion of biological snakes. In experiments, Hu {\em et al.}~\cite{HuNiScSh2009} measured the frictional anisotropy of juvenile milk snakes, and found that the forward, backward and transverse coefficients are different but similar in magnitude. They also studied the effects of the snake's active modulation of its weight distribution on the ground. Hu \& Shelley~\cite{HuSh2012} analyzed the so-called ``lateral undulation" motion modeled as a family of sinusoidal traveling-wave shapes and calculated the dependence of speed and efficiency on amplitude and wave length of the kinematics as well as coefficients of friction. 

In this work, we adopt the model of~\cite{HuNiScSh2009}, but for bodies composed of two and three rigid links. Linked bodies are fundamental for robotic sliding systems. By specializing to bodies with two and three links, we consider the simplest such systems, which nonetheless have nontrivial behaviors. Two- and three-link locomoting bodies have been considered previously as swimmers at zero Reynolds number (Stokes flow). Purcell~\cite{Pu1977} described the physics of swimming in Stokes flow, and stated the {\em Scallop Theorem}: in Stokes flow, net locomotion is not possible if a swimmer deforms in a way that is invariant under the reversal of time~\cite{Ch1981}. Such is the case for periodic motions of a two-link body (``scallop''). He then proposed a three-link swimmer that moves only one link at a time, in a non-reciprocal motion that results in net locomotion. Subsequent studies have calculated efficiency-optimizing motions for Purcell's swimmer and similar systems. Becker \textit{et al.}~\cite{BeKoSt2003} calculated efficiency-optimizing stroke amplitudes for Purcell's swimmer, and considered different length ratios of the three links. Tam \& Hosoi~\cite{TaHo2007} extended the optimization to arbitrary kinematics (allowing both internal angles to change simultaneously) and arbitrary slenderness ratios. They found an optimal path in the parameter space of internal angles using a Fourier series representation, and showed that the high frequency modes are subdominant to the low frequency modes. Avron \& Raz~\cite{AvRa2008} developed a qualitative geometric approach by focusing on the curvature of the local connection matrix to study the same system. Hatton \& Choset~\cite{HaCh2011} further developed this technique, and suggested a systematic way of choosing the best body-fixed frame to approximate the inertial frame displacement while accounting for the overall rotation. They calculated optimal motions for other systems such as a three-link fish swimming in infinite Reynolds number (potential flow), which admits a similar formulation.  Kanso {\em et al.}~\cite{KaMaRoMe2005} and Melli {\em et al.}~\cite{MeRoRu2006} gave a geometric formulation for swimming in a potential flow and calculated optimal strokes. Jing \& Kanso~\cite{JiKa2011} used this formulation to study the effects of elasticity and body configuration on the stability of passive locomotion.

Here we study the locomotion of two- and three-link bodies not in a fluid but instead on a planar surface with sliding friction. Unlike the fluid studies, the dynamical equations are nonlinear with respect to body velocities, so the Scallop theorem and many other results no longer apply. Like the aforementioned studies, we focus on finding motions which optimize the efficiency of locomotion. The structure of the paper is as follows. In Section~\ref{sec:setup} we formulate the model for a two-link ``snake," nondimensionalize the system and derive the equations of motion. In Section~\ref{sec:2link} we use analysis and computation to optimize the two-link model with respect to kinematic parameters and coefficients of friction, for both small- and large-amplitude actuations. In Section~\ref{sec:3link} we analyze the three-link snake model and compute optimal kinematics of the relative angles at one realistic set of coefficients of friction using a Fourier series representation. In Section~\ref{sec:discussion}, we summarize our work and suggest directions for future study.

\section{Problem setup for two-link model} 
\label{sec:setup}

\begin{figure}
	[!htb] \centering 
	\includegraphics[width=0.4\textwidth]{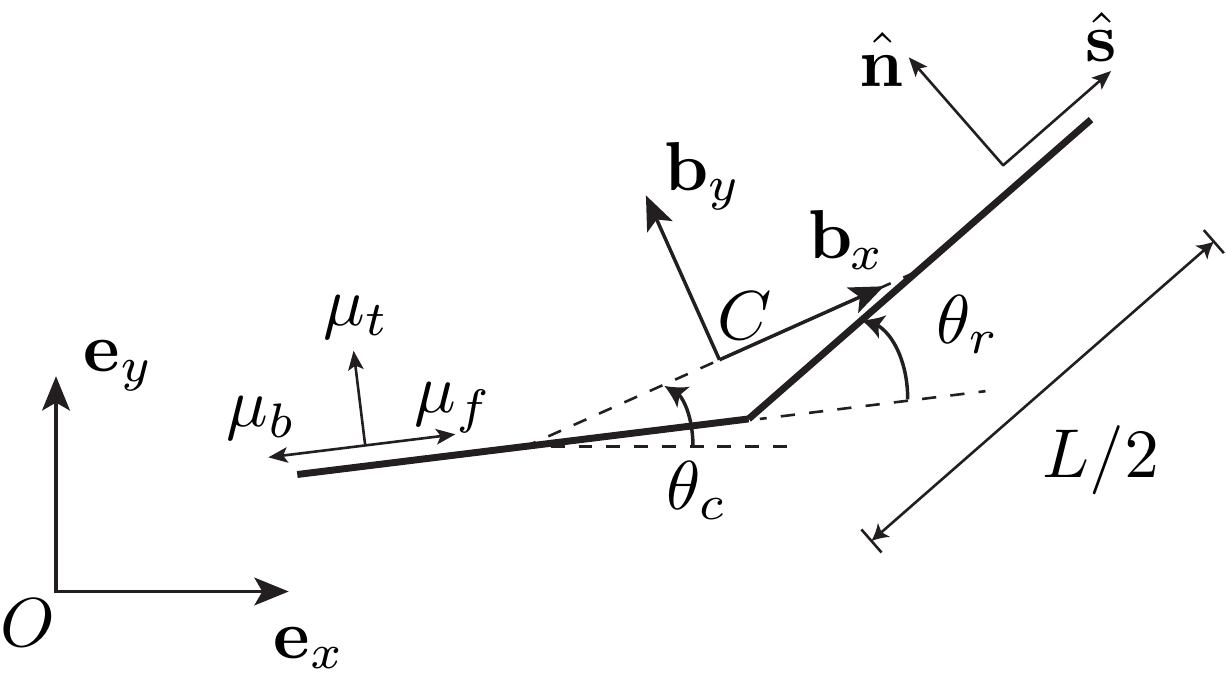} 
	\caption{\footnotesize Two-link snake model; see text for description.}\label{fig:2link_model} 
\end{figure}

The snake is modeled in 2D as two identical inextensible line segments ({\em links}) connected via a hinge joint as depicted in Figure~\ref{fig:2link_model}. The total length of the snake is $L$, and for each link is $L/2$. The snake shape is parametrized by the angle $\theta_{r}$ between the tail link (left) and the head link (right), with the positive direction of rotation being counterclockwise. Denoting the snake's mass per unit length as $\rho$, the total mass is $m =\rho L$. We denote $s$ as the arc length between any point of the snake and the tip of its tail, so $0 \leq s \leq L$. The tail tip, hinge joint and head of the snake correspond to $s = 0, L/2$ and $L$ respectively. 

Motion of the two-link snake can be observed both in an inertial frame $\{\mathbf{e}_x,\mathbf{e}_y\}$ with origin at a fixed point $O$, or in a body-fixed frame $\{\mathbf{b}_x,\mathbf{b}_y\}$ with origin at the center of mass $C$. The unit vector $\mathbf{b}_x$ is parallel to the line connecting the links' centers, and $\mathbf{b}_y$ is $\mathbf{b}_x$ rotated by 90 degrees. The position of $C$ in the inertial frame is $\mathbf{x}_c = (x_c , y_c)$, and the orientation $\theta_c$ is the angle from $\mathbf{e}_x$ to $\mathbf{b}_x$. In the inertial frame, the position of an arbitrary material point on the snake is denoted as $\mathbf{x} = (x ,y)$, and $\theta$ is the angle between the tangent to the snake at a given point and $\mathbf{e}_x$. The positive direction of rotation is counterclockwise. In the body-fixed frame, the position of the same material point is denoted $\mathbf{X} = (X, Y)$, and the tangent angle is $\Theta$. For a given material point, we define the configuration variable in the inertial frame $\mathbf{g} = [x\,,\,y\,,\,\theta]^T$, and in the body-fixed frame $\mathbf{G} = [X\,,\,Y\,,\,\Theta]^T$. Specifically, for $C$, we have $\mathbf{g}_c = [x_c\,,\,y_c\,,\,\theta_c]^T$ and $\mathbf{G}_c = [0\,,\,0\,,\,0]^T$. The relation between the configuration variable in both frames is
\begin{equation}
	\label{eq:relconfig}
	\mathbf{g} = \mathbf{g}_c + R_{\theta_c} \mathbf{G},\qquad
	R_{\theta_c} = \begin{pmatrix}
		\cos\theta_c & -\sin\theta_c & 0\\
		\sin\theta_c & \cos\theta_c & 0\\
		0 & 0 & 1
	\end{pmatrix},
\end{equation}
where $R_{\theta_c}$ is the transformation matrix. For the tail link, i.e. $0 \leq s \leq L/2$, the configuration in $\{\mathbf{b}_x,\mathbf{b}_y\}$ is given by
\begin{equation}
	\label{eq:2linktail}
	\mathbf{G}_t = \begin{pmatrix}
	X_t\\Y_t\\\Theta_t
	\end{pmatrix} = \begin{pmatrix}
		(s-1/2) \cos(\theta_r/2)\\
		-(s-1/4) \sin(\theta_r/2)\\
		-\theta_r/2
		\end{pmatrix},
\end{equation}
where the subscript $t$ indicates tail. For the head link, $L/2 \leq s \leq L$, the configuration is
\begin{equation}
	\label{eq:2linkhead}
	\mathbf{G}_h = \begin{pmatrix}
	X_h\\Y_h\\\Theta_h
	\end{pmatrix} = \begin{pmatrix}
		(s-1/2) \cos(\theta_r/2)\\
		(s-3/4) \sin(\theta_r/2)\\
		\theta_r/2
		\end{pmatrix},
\end{equation}
where $h$ represents head.

\begin{figure}
	[!tb]  
	\centering 
	\includegraphics[width=0.5\textwidth]{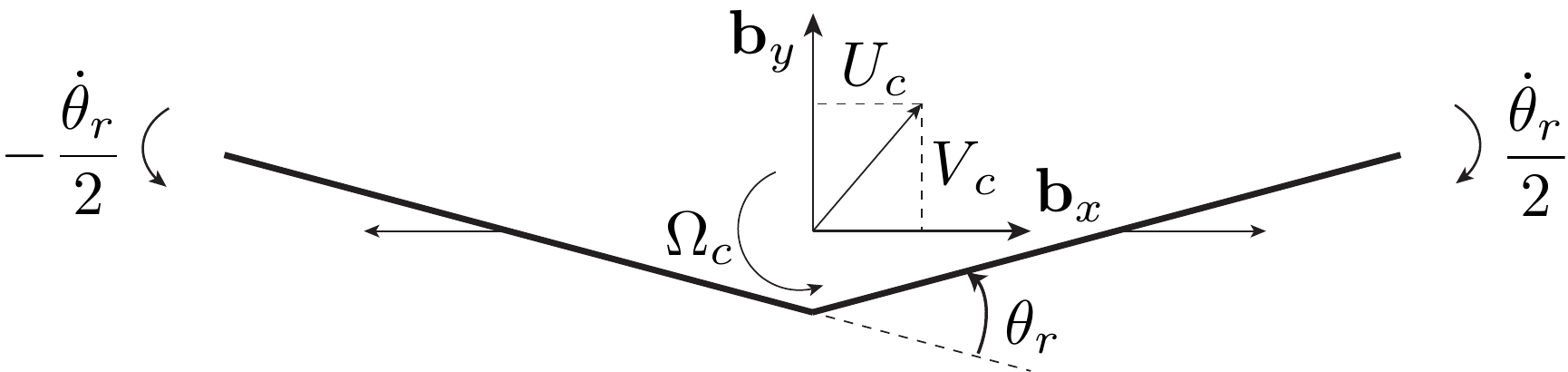}
	\caption{\footnotesize Schematic of the body-frame velocities for the two-link model. Center-of-mass velocity is $\boldsymbol{\xi}_c = [U_c\,,\,V_c\,,\,\Omega_c]^T$, the velocity due to rotation about $C$ is $\boldsymbol{\xi}_{rot} = [-\Omega_c Y\,,\,\Omega_c X\,,\,0]^T$, where $(X\,,\,Y)$ is the position of a material point in body frame, and the velocity due to shape change $\boldsymbol{\xi}_{shape}$ consists of a horizontal motion of the link centers and rotations about the link centers with angular velocities $\pm \dot{\theta}_r/2$ for the two links (in this example, $\theta_r > 0$ and $\dot{\theta}_r < 0$).} \label{fig:2link_sketch}
\end{figure}

In the inertial frame, the linear and angular velocities of any point on the snake are given by the time derivative of its configuration, that is $\dot{\mathbf{g}}$. In particular, the velocity of $C$ in the inertial frame is given by $\dot{\mathbf{g}}_c$. The velocity of $C$ {\em with respect to} the inertial frame can also be {\em expressed} in the body frame as $\boldsymbol{\xi}_c = [U_{c}\,,\,V_{c}\,,\,\Omega_c]^T$, where $\Omega_c = \dot{\theta}_c$. Similarly, the velocity of any material point with respect to the inertial frame can be expressed in the body frame as 
\begin{equation}
	\label{eq:bodyvelocity}
	\boldsymbol{\xi} \equiv \begin{pmatrix}
		U\\V\\\Omega
	\end{pmatrix} = \boldsymbol{\xi}_c + \boldsymbol{\xi}_{shape} +
	\boldsymbol{\xi}_{rot},\quad \boldsymbol{\xi}_{shape} = \frac{\partial
	\mathbf{G}(s,t)}{\partial t},\quad \boldsymbol{\xi}_{rot} = \begin{pmatrix}
		-\Omega_c Y\\
		\Omega_c X\\
		0
	\end{pmatrix},
\end{equation}
where $\boldsymbol{\xi}_{shape}$ is due to shape changes and $\boldsymbol{\xi}_{rot}$ is due to rotation of the snake about $C$ (see Figure~\ref{fig:2link_sketch}). The relation between the velocity in both frames is given by
\begin{equation}
	\label{eq:velocityrelation}
	\dot{\mathbf{g}} = R_{\theta_c} \boldsymbol{\xi},\quad \text{in
particular} \quad \dot{\mathbf{g}}_c = R_{\theta_c} \boldsymbol{\xi}_c.
\end{equation}

Perpendicular to the plane of motion, the forces on the snake are gravity and the supporting force from the ground, which balance each other. Within the plane of motion, the forces on the snake are external friction from the ground and internal forces. Since the coefficients of friction are anisotropic, the frictional force is decomposed into components in different directions. In the body frame, we denote the linear velocity of an arbitrary material point as $\boldsymbol{\xi}_{lin} = [U \, ,\,  V]^T$, the unit vector tangent to the snake as $\hat{\mathbf{s}} = (\cos\Theta , \sin\Theta)$, and the unit normal vector as $\hat{\mathbf{n}} = (-\sin\Theta , \cos\Theta)$. Our model for the snake mechanics is essentially the same as that in \cite{HuNiScSh2009}. The Coulomb frictional force density at a given point on the snake is
\begin{equation}
	\mathbf{f}(s,t) = \rho g \left\{ - \mu_t (\hat{\boldsymbol{\xi}}_{lin}
	\cdot \hat{\mathbf{n}}) \hat{\mathbf{n}} - \left[ \mu_f
	H(\hat{\boldsymbol{\xi}}_{lin} \cdot \hat{\mathbf{s}}) + \mu_b \left(1 -
	H(\hat{\boldsymbol{\xi}}_{lin} \cdot
	\hat{\mathbf{s}})\right)\right](\hat{\boldsymbol{\xi}}_{lin} \cdot
	\hat{\mathbf{s}}) \hat{\mathbf{s}}  \right\}, 
\end{equation}
where $\mu_f, \mu_b$ and $\mu_t$ are the forward, backward and transverse coefficients of friction respectively, $\hat{\boldsymbol{\xi}}_{lin} = \boldsymbol{\xi}_{lin} / \|\boldsymbol{\xi}_{lin}\|$, and $H(\cdot)$ is the Heaviside function, used to distinguish forward and backward friction \cite{HuNiScSh2009}. Note that the frictional force density depends on the {\em direction} of velocity but not the {\em magnitude}. In addition to the external force, there are also forces internal to the snake, and the internal force density is denoted as $\mathbf{f}_{in} (s,t)$. The torque density with respect to $C$ due to friction is given by $\tau(s,t) = \mathbf{X}^{\perp} \cdot \mathbf{f}$, while that due to internal force is $\tau_{in}(s,t) = \mathbf{X}^{\perp} \cdot \mathbf{f}_{in}$. The internal force and torque densities are due to a system of equal and opposite tension and shearing forces acting on adjacent sections of the snake across their interfaces~\cite{Se1987}, which makes the snake inextensible and enforces its shape. The integrals of internal forces and torques are zero
\begin{equation}
	\label{eq:internal}
	\int_0^L \mathbf{f}_{in}\, \text{d}s = \mathbf{0},\qquad
	\int_0^L \tau_{in}\,\text{d}s = 0.
\end{equation}

Now we nondimensionalize the variables. We consider actuations $\theta_r$ (and resulting snake motions) that are periodic in time with period $T$. Variables are nondimensionalized by scaling by the total length $L$, period $T$ and mass $m = \rho L$. 
Three important dimensionless numbers for the snake dynamics are
\begin{equation}
	\label{eq:nondnumbers}
	Fr \equiv \frac{L}{\mu_f g T^2}, \quad \tilde{\mu}_b \equiv
	\frac{\mu_b}{\mu_f}, \qquad \tilde{\mu}_t \equiv \frac{\mu_t}{\mu_f}.
\end{equation}
Here $Fr$ is the Froude number, which can be written as a ratio of snake inertia to the forward frictional force acting on it. The other two parameters are friction coefficient ratios. We assume the coefficients of friction are uniform along the snake, and define the forward direction as that with the smaller of the tangential friction coefficients (if $\mu_f \neq \mu_b$), so  $\tilde{\mu}_b \geq 1$ by definition. For real snakes, $Fr \ll 1$, which means that the snake's inertia is negligible compared to frictional forces~\cite{HuNiScSh2009, HuSh2012}.

For simplicity, we now drop the tildes with the understanding that all variables are dimensionless in the remainder of this work. The dimensionless point-wise frictional force density is given in the body frame by
\begin{equation}
	\label{eq:f}
	\mathbf{f} = - \mu_t (\hat{\boldsymbol{\xi}}_{lin} \cdot \hat{\mathbf{n}})
	\hat{\mathbf{n}} - \left\{ H(\hat{\boldsymbol{\xi}}_{lin} \cdot
	\hat{\mathbf{s}}) + \mu_b \left[ 1 - H(\hat{\boldsymbol{\xi}}_{lin} \cdot
	\hat{\mathbf{s}})\right] \right\}(\hat{\boldsymbol{\xi}}_{lin} \cdot
	\hat{\mathbf{s}}) \hat{\mathbf{s}}.
\end{equation}
It can be transformed into the inertial frame via the transformation matrix $R_{\theta_c}$. The governing equations are given by the linear and angular balance laws,
\begin{equation}
	\label{eq:balancelaw}
	R_{\theta_c}\int_0^1 (\mathbf{f} + \mathbf{f}_{in})\,\text{d}s = Fr \,\ddot{\mathbf{x}}_c\ ,\qquad 
	\int_0^1 \mathbf{X}^{\perp}\cdot(\mathbf{f} + \mathbf{f}_{in})\,\text{d}s = \frac{\text{d}}{\text{d}t}\int_0^1 (Fr\, 	\boldsymbol{\xi}_{lin}\times\dot{\boldsymbol{\xi}}_{lin})\,\text{d}s\ .
\end{equation}
Since the inertia term is negligible, the Froude number is assumed to be zero: $Fr = 0$. Therefore, substituting~\eqref{eq:internal} into~\eqref{eq:balancelaw}, the governing equations are reduced to the integrals of frictional force and torque densities equaling zero,
\begin{equation}
	\label{eq:eom} 
	\int_0^1 \mathbf{f}\, \text{d}s = \mathbf{0},\qquad
		\int_0^1 \mathbf{X}^\perp \cdot \mathbf{f}\,\text{d}s = 0. 
\end{equation}
The above equations give three independent scalar equations for the three unknowns $\boldsymbol{\xi}_c = [U_c\,,\,V_c\,,\,\Omega_c]^T$, and the solution depends on the parameters $\mu_b$ and $\mu_t$, as well as the prescribed relative angle $\theta_r(t)$. Notice that~\eqref{eq:eom} are intrinsically nonlinear, and the nonlinearity primarily lies in the form of friction given in~\eqref{eq:f}. An explicit derivation of the equations of motion is given in Appendix~\ref{appen:a}.

Since $\mathbf{f}$ only depends on the direction of the velocity, and no inertia term is present in the equations, the only time scale in the problem is $T$. For a given $\theta_r(t)$, if $T$ is doubled, the speed of the motion is reduced by half, but the snake will trace exactly the same trajectory, as does Purcell's three-link swimmer in Stokes flow. This is referred to as the {\em rate independence} or {\em time invariance} of inertialess systems: if a body undergoes a deformation, the trajectory traveled by the body between two different shape configurations does not depend on the rate of deformation, but only on the sequence of deformation~\cite{BeKoSt2003, Ch1981, Pu1977}. On the other hand, the Scallop theorem indicates that inverting the shape change sequence corresponds to inverting time for the original shape change sequence (note this is {\em not} equivalent to time invariance). A corollary~\cite{AvRa2008, HaCh2011} is: in the body-fixed frame, the velocity of the center of mass is proportional to the velocity of the shape change. This is known as the {\em kinematic reconstruction equation} in the geometric mechanics literature
\begin{equation}
	\label{eq:normalrecon}
	\boldsymbol{\xi}_c = A(\boldsymbol{\theta}_r)
	\dot{\boldsymbol{\theta}}_r,
\end{equation}
where $\boldsymbol{\theta}_r$ is a shape change vector for systems with multiple degrees of freedom, $A(\boldsymbol{\theta}_r)$ is the {\em local connection matrix}, which is an $n\times m$ matrix that relates an $m\times 1$ shape change velocity $\dot{\boldsymbol{\theta}}_r$ to an $n\times 1$ velocity of $C$ in the body frame $\boldsymbol{\xi}_c$. For the two-link model, $\theta_r$ is a scalar and $\boldsymbol{\xi}_c$ is a $3\times 1$ vector, hence $\mathbf{A}(\theta_r)$ is a $3\times 1$ vector. Note that $A(\boldsymbol{\theta}_r)$ does not depend on $\dot{\boldsymbol{\theta}}_r$. 

However, equation~\eqref{eq:normalrecon} does {\em not} apply to our snake model. This is due to the anisotropy of coefficients of friction, specifically, the Heaviside function in the force equation~\eqref{eq:f}, that causes the irreversibility of shape change. Instead, for our system, the local connection matrix also depends on the direction (or {\em sign}) of $\dot{\boldsymbol{\theta}}_r$, denoted by $\mathbf{S}_r = \sgn(\dot{\boldsymbol{\theta}}_r)$, but not the magnitude $\|\dot{\boldsymbol{\theta}}_r\|$. If there were no Heaviside function in~\eqref{eq:f}, and force were only decomposed into tangential and transverse directions, then the system would become very similar to the multi-link fish problem in a potential flow, for which it is known that with the added inertia decomposed into tangential and transverse directions~\eqref{eq:normalrecon} holds~\cite{KaMaRoMe2005}. Consequently, the techniques of analyzing the local connection matrix developed in~\cite{AvRa2008, HaCh2011} cannot be directly implemented here since it is based on~\eqref{eq:normalrecon}. In general, the {\em modified} kinematic reconstruction equation for the snake model can be written as
\begin{equation}
	\label{eq:recon}
	\boldsymbol{\xi}_c = A(\boldsymbol{\theta}_r , \mathbf{S}_r) \dot{\boldsymbol{\theta}}_r.
\end{equation}
For the two-link model where $\theta_r$ is a scalar, the above equation is reduced to $\boldsymbol{\xi}_c = \mathbf{A}(\theta_r , S_r) \dot{\theta}_r$. Note when relative angle $\theta_r(t)$ is prescribed,~\eqref{eq:recon} is a nonlinear algebraic equation rather than a differential equation, and the nonlinearity lies in the form of $\mathbf{A}(\theta_r , S_r)$. The solution of~\eqref{eq:eom} or equivalently~\eqref{eq:recon} is the velocity of $C$ expressed in the body frame, $\boldsymbol{\xi}_c$. Without loss of generality, assume the snake starts from the origin with zero initial orientation angle, i.e. $\mathbf{g}_c(0) = \mathbf{0}$. The configuration in the inertial frame is then 
\begin{equation}
	\label{eq:displacement}
	\mathbf{g}_c(t) = \int_0^t \dot{\mathbf{g}}_c\, \text{d}t = \int_0^t
	R_{\theta_c} \boldsymbol{\xi}_c \,\text{d}t,
\end{equation}
which is an {\em iterative} integral in time since $\theta_c$ is only known from the integral $\theta_c = \int_0^t \Omega_c(\tilde{t}) \, \text{d}\tilde{t}$. The distance traveled by $C$ during one period observed in the inertial frame is given by
\begin{equation}
	\label{eq:distance} d = \|\mathbf{x}_c(T) - \mathbf{x}_c(0)\|.
\end{equation}
The work generated by the snake during one period is equal to the energy dissipation due to friction since the system is inertialess, i.e. 
\begin{equation}
	\label{eq:energy} 
	W = \int_0^T \!\!\!\int_0^L -\mathbf{f} \cdot	\boldsymbol{\xi}_{lin} \,\text{d}s \,\text{d}t,
\end{equation}
here it is more convenient to express $\mathbf{f}$ in the body frame. The efficiency of locomotion is defined as the ratio between distance and work,
\begin{equation}
	\label{eq:efficiency}
	e = \frac{d}{W}. 
\end{equation}
This efficiency $e$, after nondimensionalization, is equivalent to the inverse of the {\em cost of locomotion} commonly seen in the animal locomotion literature~\cite{Ch1981}. Intuitively speaking, $e$ is analogous to the concept of ``miles-per-gallon".

\begin{figure}
	[!htb] \centering 
	\subfigure[$\theta_r$ vs $t$]{\includegraphics[width=0.26\textwidth]{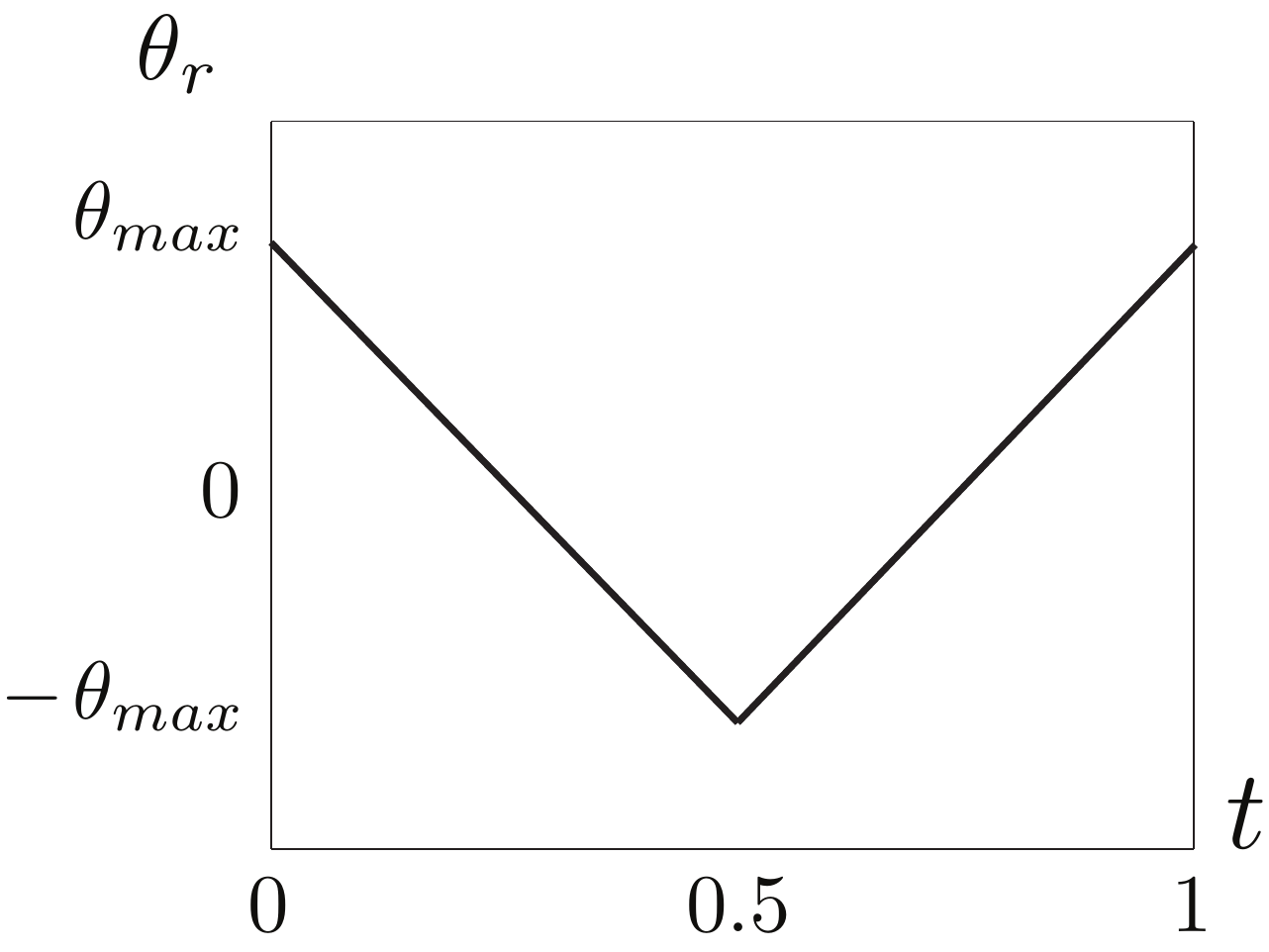}}\qquad \quad
	\subfigure[$\dot{\theta}_r$ vs $t$]{\includegraphics[width=0.265\textwidth]{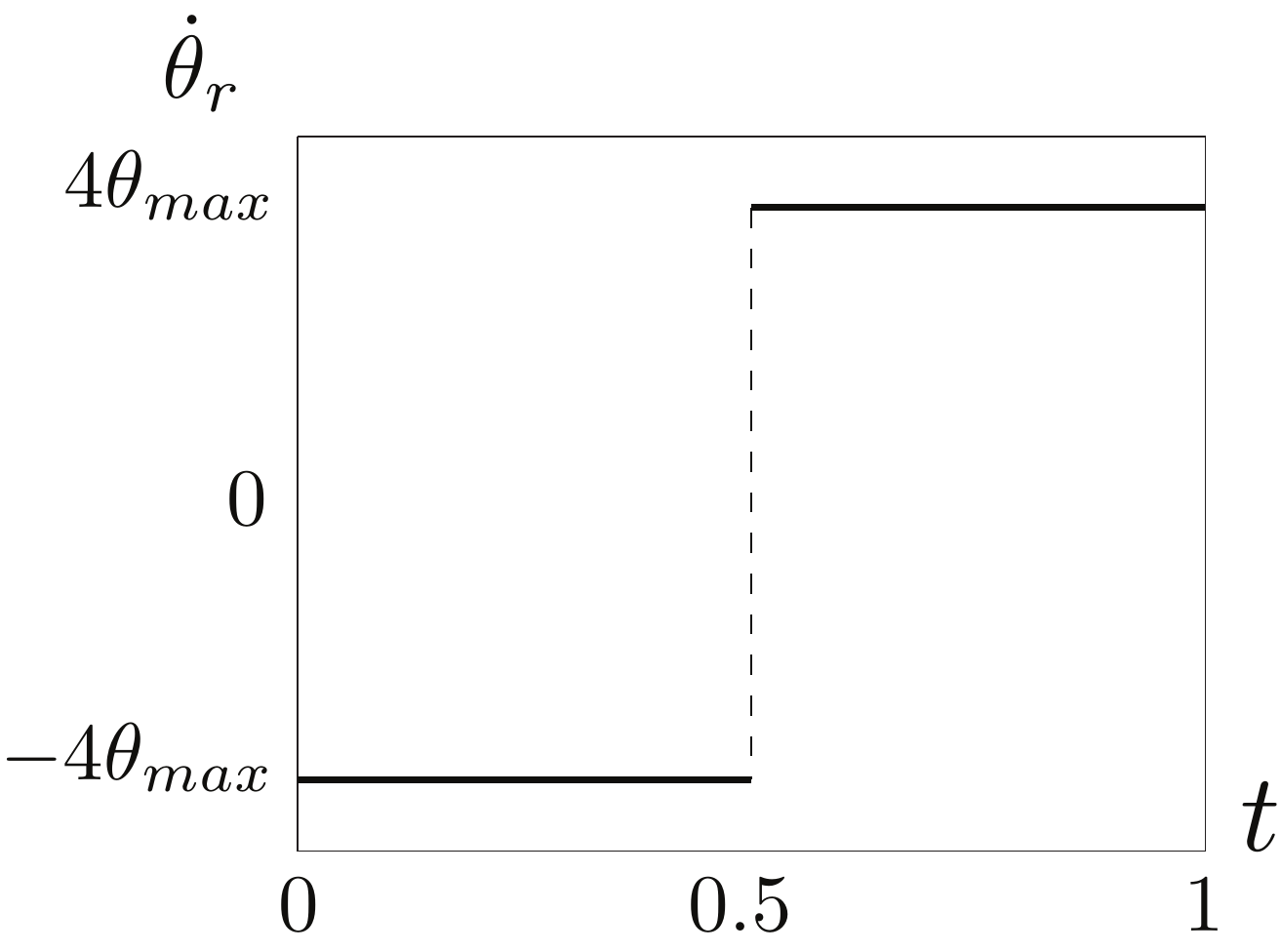}} 
	\caption{\footnotesize (a) Relative angle $\theta_r$ prescribed as a triangular wave with period $T = 1$ and amplitude $\theta_{max}$. (b) Angular velocity for relative angle $\dot{\theta}_r$.}\label{fig:2linktri_th} 
\end{figure}
\section{Two-link model analysis} 
\label{sec:2link}

\begin{figure}
	[!htb] \centering 
	\subfigure[trajectory of $C$ and snapshots]{\includegraphics[width=0.7\textwidth]{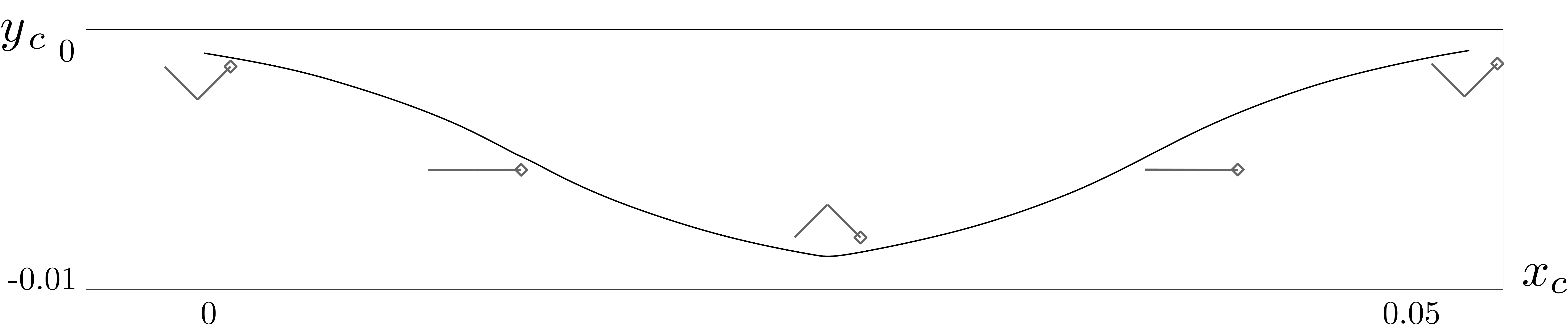}}\qquad
	\subfigure[orientation $\theta_c$ vs. $t$]{\includegraphics[width=0.191\textwidth]{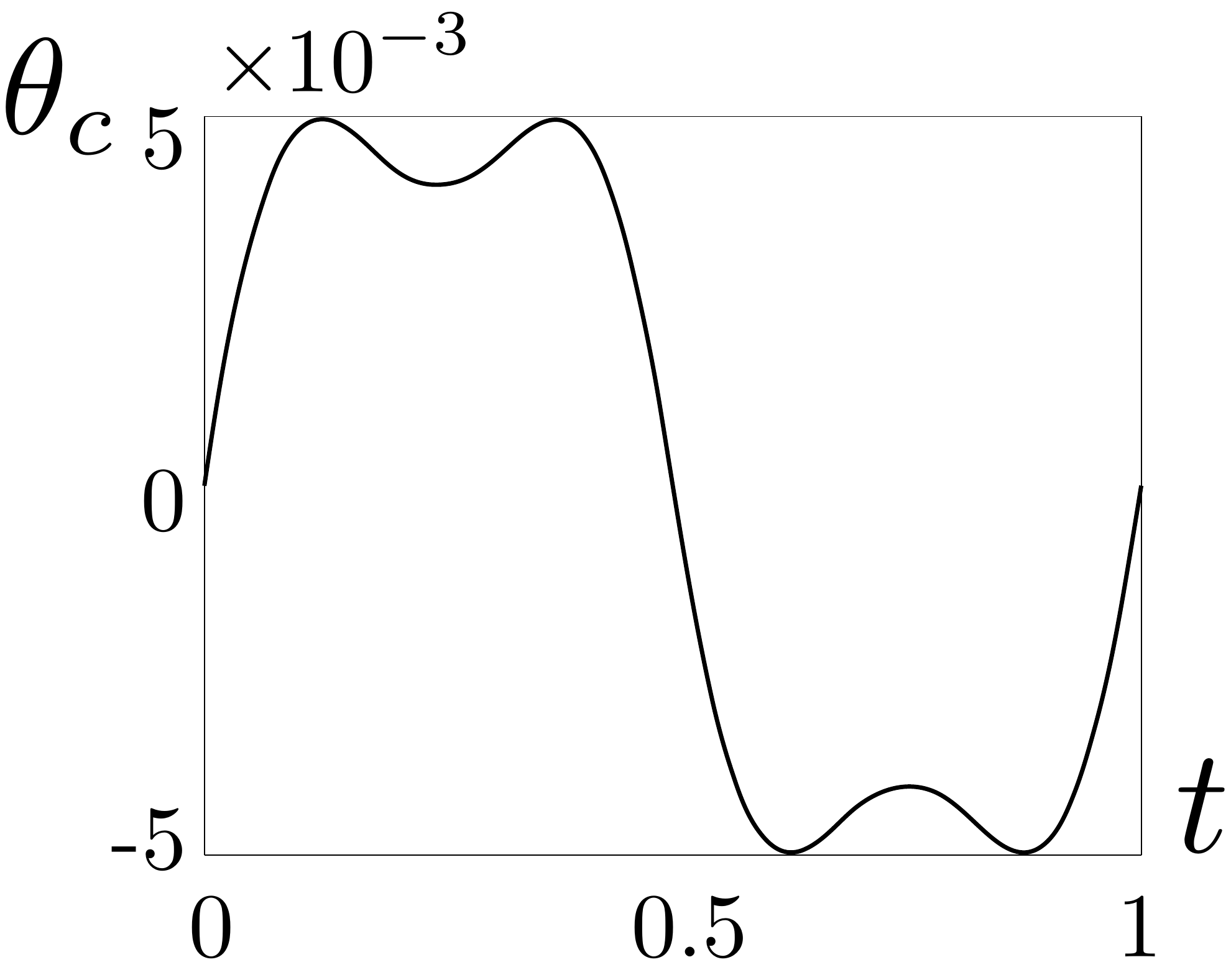}}
	\caption{\footnotesize Motion of two-link snake when $\theta_r$ is prescribed as a triangular wave given in~\eqref{eq:thetar} and $\theta_{max} = \pi/2$: (a) Trajectory of center of mass $C$ in the inertial frame with snapshots of the snake at $t = 0, 0.25, 0.5, 0.75$ and 1; (b) Orientation $\theta_c$ as a function of time. The coefficients of friction are $\mu_b = 1.3$ and $\mu_t = 1.7$.}\label{fig:2link_ss} 
\end{figure}

\begin{figure}
	[!htb] \centering \subfigure[$\dot{x}_c$ vs. $t$]{ 
	\includegraphics[width=0.32
	\textwidth]{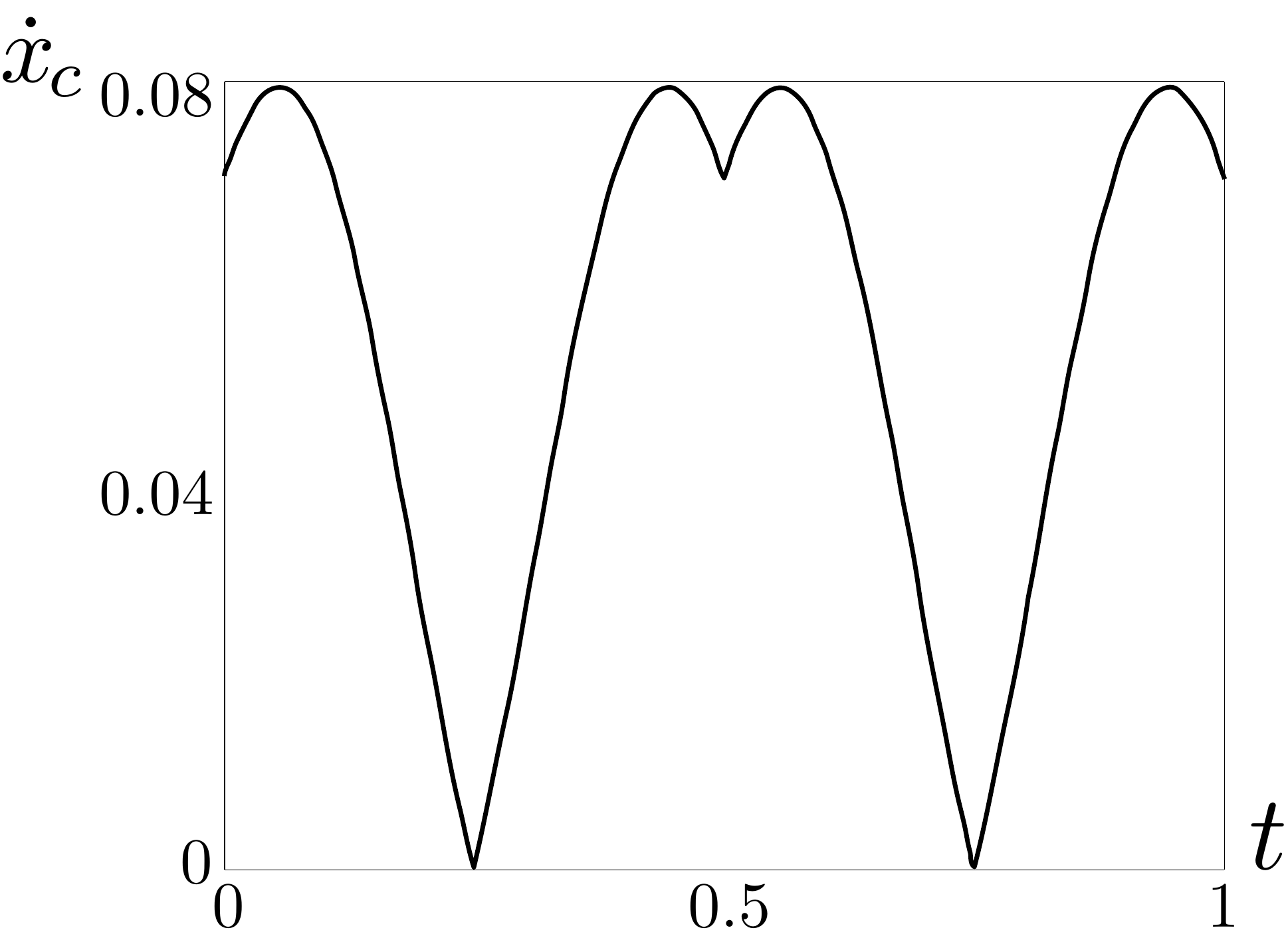}}\subfigure[$\dot{y}_c$ vs. $t$]{ 
	\includegraphics[width=0.32
	\textwidth]{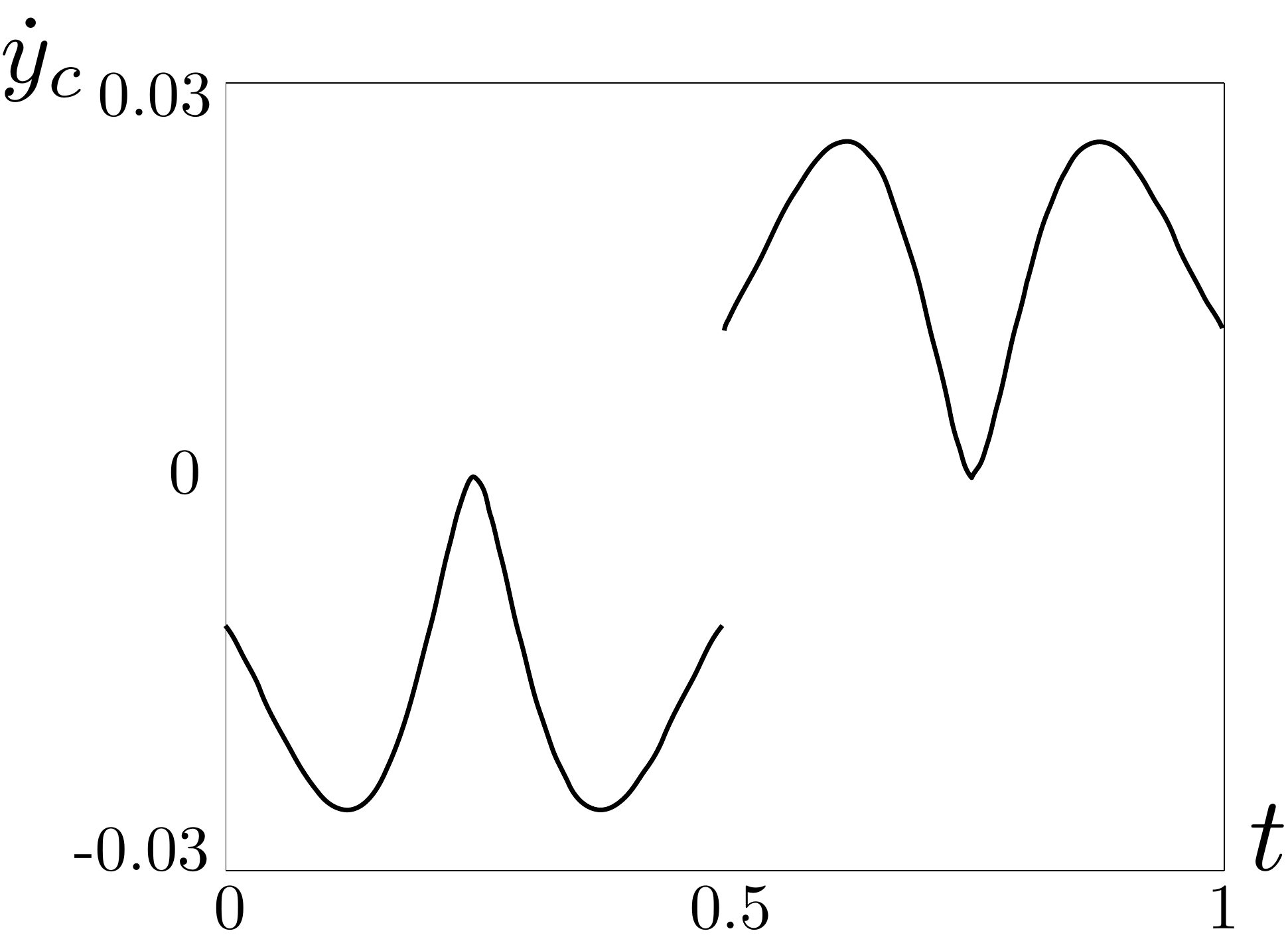}} \subfigure[$\dot{\theta}_c$ vs. $t$]{ 
	\includegraphics[width=0.31
	\textwidth]{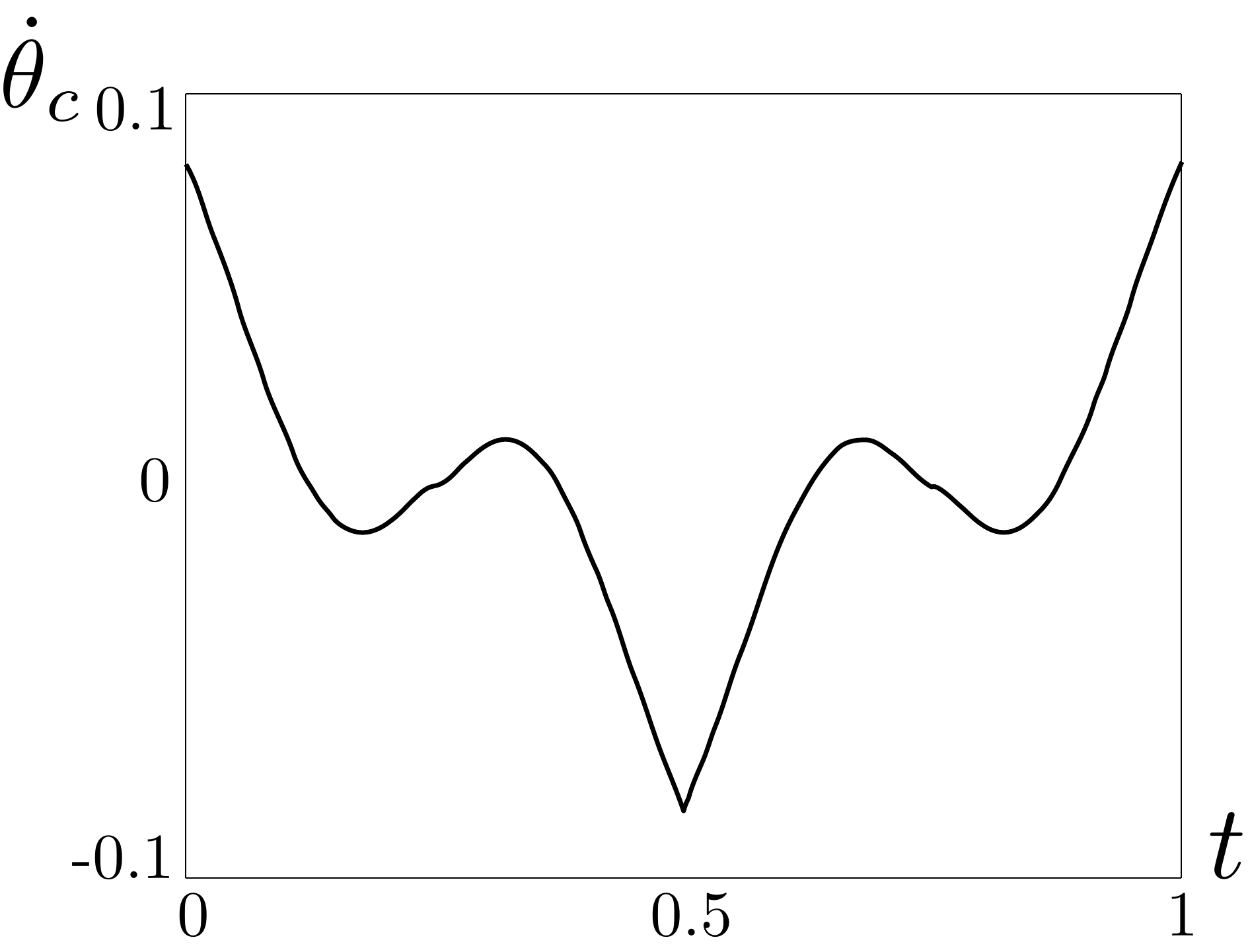}} \caption{\footnotesize Inertial frame velocities of the center of mass $C$ for the two-link model with $\theta_r$ given by~\eqref{eq:thetar} and $\theta_{max} = \pi/2$. The coefficients of friction are $\mu_b = 1.3$ and $\mu_t = 1.7$.}
\label{fig:2linkvelpiover2} 
\end{figure}

We first look at an example of two-link snake locomotion. We prescribe the relative angle $\theta_r$ as a triangular wave with period $T = 1$ and amplitude $\theta_{max}$, and with $\theta_r (0) = \theta_{max}$, i.e. 
\begin{equation}
	\label{eq:thetar} \theta_r (t) = \begin{cases}
	\theta_{max} (1 - 4 t), & 0 \leq t < 0.5,\\
	\theta_{max}(-3 + 4t), & 0.5 \leq t \leq 1,
	\end{cases}
	\qquad \qquad
	\dot{\theta}_r (t) = \begin{cases}
		-4\theta_{max}, & 0 \leq t < 0.5,\\
		4\theta_{max}, & 0.5 \leq t \leq 1. 
	\end{cases}
\end{equation}
Figure~\ref{fig:2linktri_th} depicts $\theta_r$ and $\dot{\theta}_r$ within one period. For concreteness, let $\mu_b = 1.3$ and $\mu_t = 1.7$, which are taken from the experimental measurement in~\cite{HuSh2012}. The equations of motion~\eqref{eq:eom} are solved numerically using the subroutine {\tt fsolve} in MATLAB, which implements a Trust-Region-Dogleg algorithm. When $\theta_{max} = \pi/2$, the trajectory of the center of mass $C$ is shown in Figure~\ref{fig:2link_ss}(a) with five snapshots of the snake at $t = 0, 0.25, 0.5, 0.75$ and 1 overlaid (the head of the snake is represented by $\Diamond$), and the orientation $\theta_c$ as a function of time is depicted in Figure~\ref{fig:2link_ss}(b). One can see from the trajectory plot that the displacement in the $x$ direction is larger than that in the $y$ direction. And even for the large amplitude of actuation $\theta_{max} = \pi/2$, the rotation is very small: $\|\theta_c\| < 5\times10^{-3} = 0.3$ degrees for all time. The distance traveled during the period is $d = 0.0535$, the work is $W = 0.7868$, and the efficiency is $e = 0.0680$. Figure~\ref{fig:2linkvelpiover2} shows the velocity of the center of mass $\dot{\mathbf{g}}_c$ in the inertial frame. The horizontal velocity $\dot{x}_c$ is always non-negative. The vertical velocity $\dot{y}_c$ is non-negative during the first half-period, and non-positive during the second half. The two half periods nearly cancel out and result in nearly zero net displacement in the $y$ direction as shown in Figure~\ref{fig:2link_ss}(a). The angular velocity $\dot{\theta}_c$ alternates between positive and negative during the period, and is symmetric about $t = 0.5$. All components of velocity become 0 when $t = 1/4$ and $t= 3/4$, which corresponds to $\theta_r = 0$. From Figures~\ref{fig:2link_ss} and~\ref{fig:2linkvelpiover2}, one can see the resulting velocities depend nonlinearly on $\dot{\theta}_r$. They can only be solved numerically. The nonlinearity arises from both the form of the friction in~\eqref{eq:f}, and the large amplitude of actuation, in this case $\theta_{max} = \pi/2$. In order to focus on understanding the former nonlinearity, we now analyze an actuation with small amplitude.

\paragraph{Small amplitude analysis} 
\label{par:two_link_model_small_amplitude_analysis}
For a general kinematics $\theta_r(t)$, assume $\sup_t \|\theta_r\| = \epsilon \ll 1$, and consequently $\theta_r,\, \dot{\theta_r} \sim O(\epsilon)$. One can show that $U_c,\, V_c,\, \Omega_c \ll O(\epsilon)$ as follows. Specifically, when the motion of the snake is as depicted in Figure~\ref{fig:2link_sketch}, that is $\theta_r >0$ and $\dot{\theta}_r < 0$, we show
in Appendix~\ref{appen:a} that the integral form of the equations of motion~\eqref{eq:eom} results in
\begin{equation}
	\label{eq:2linklin}
	\begin{cases}
		\mu_b \left(U_c +
\dfrac{1}{16}\dot{\theta}_r\theta_r\right)\ln\left|\dfrac{-\dot{\theta}_r}{4U_c+\dot{
\theta}_r\theta_r/4}\right|+\left(U_c -
\dfrac{1}{16}\dot{\theta}_r\theta_r\right)\ln\left|\dfrac{-\dot{\theta}_r}{4U_c-\dot{
\theta}_r\theta_r/4}\right| \approx 0,\\[3ex]
		-\mu_t \left(2V_c + \dfrac{1}{16}\dot{\theta}_r\theta^2_r\right)
- \mu_b \left(U_c\theta_r +
\dfrac{1}{16}\dot{\theta}_r\theta^2_r\right)\ln\left|\dfrac{-\dot{\theta}_r}{4U_c+\dot{
\theta}_r\theta_r/4}\right| \approx 0,\\[3ex]
		-\dfrac{3}{4} \mu_t\left(2U_c\theta_r - \Omega_c\right) +
\left[4\mu_t(1+\mu_b) \dfrac{U_c}{\dot{\theta}_r} +
\dfrac{\mu_t}{4}(1-\mu_b)\theta_r\right]\left(U_c +
\dfrac{1}{16}\dot{\theta}_r\theta_r\right)\ln\left|\dfrac{-\dot{\theta}_r}{4U_c+\dot{
\theta}_r\theta_r/4}\right| \approx 0.
	\end{cases}
\end{equation}
The solutions for equations~\eqref{eq:2linklin} are given by
\begin{equation}
	\label{eq:2linkvellin}
	\begin{cases}
		U_c \approx
-\dfrac{1}{16}(1+\beta)\dot{\theta}_r\theta_r,\\[2.5ex]
		V_c \approx \dfrac{1}{32}\left(-1 +
\dfrac{\mu_b}{\mu_t}\beta\ln\left|\dfrac{4}{\beta\theta_r}\right|\right)\dot{\theta}
_r\theta^2_r,\\[2.5ex]
		\Omega_c \approx \left\{-\dfrac{1}{8}(1+\beta) -
\dfrac{1}{48}\beta\left[\beta (1+\mu_b) +
2\mu_b\right]\ln\left|\dfrac{4}{\beta\theta_r}\right|\right\}\dot{\theta}_r\theta^2_r,
	\end{cases}
\end{equation}
where $\beta$ is the solution of the following transcendental equation
\begin{equation}
		 \left[(\mu_b + 1)\beta(t) + 2\right]  \left[2\ln2 - \ln\left|\theta_r(t)\right|
	\right] - \mu_b \beta(t) \ln |\beta(t)| - \left[\beta(t) + 2\right]
	\ln\left|\beta(t)+2\right| \approx 0.
\end{equation}
Notice $\beta(t)$ depends on $\mu_b$ and $\theta_r$ but not on $\mu_t$. For $\theta_r \sim O(\epsilon) \ll 1$, $\beta(t)$ can be approximated by 
\begin{equation}
	\label{eq:beta}
	\beta \approx -\frac{2}{\mu_b + 1},
\end{equation}
which is between 0 and -1 for $\mu_b \geq 1$. For details of the derivation of~\eqref{eq:2linklin}  and~\eqref{eq:2linkvellin}, as well as some numerical results regarding~\eqref{eq:beta}, see Appendix~\ref{appen:a}. From~\eqref{eq:2linkvellin}, one has $U_c \sim O(\epsilon^2)$ and $V_c, \,\Omega_c \sim O(\epsilon^3)$ since $\theta_r, \,\dot{\theta_r} \sim O(\epsilon)$, which is consistent with the observation mentioned before: $U_c,\,V_c,\, \Omega_c \ll O(\epsilon)$. Since $\Omega_c = \dot{\theta}_c \sim O(\epsilon^3) \ll 1$, one has $\theta_c \ll O(1)$ for a period of time $t\sim O(1)$. Hence, the transformation matrix $R_{\theta_c}$ given in~\eqref{eq:relconfig} is approximately an identity matrix, which means the inertial frame velocity can be approximated by the body frame velocity, i.e. $\dot{\mathbf{g}}_c \approx \boldsymbol{\xi}_c$. One can easily obtain the velocities for other combinations of the signs of $\theta_r$ and $\dot{\theta}_r$ by symmetry.

\begin{figure}[!tb]  
	\centering 
	\subfigure[$\dot{x}_c$ vs. $t$]{
	\includegraphics[width=0.31\textwidth]{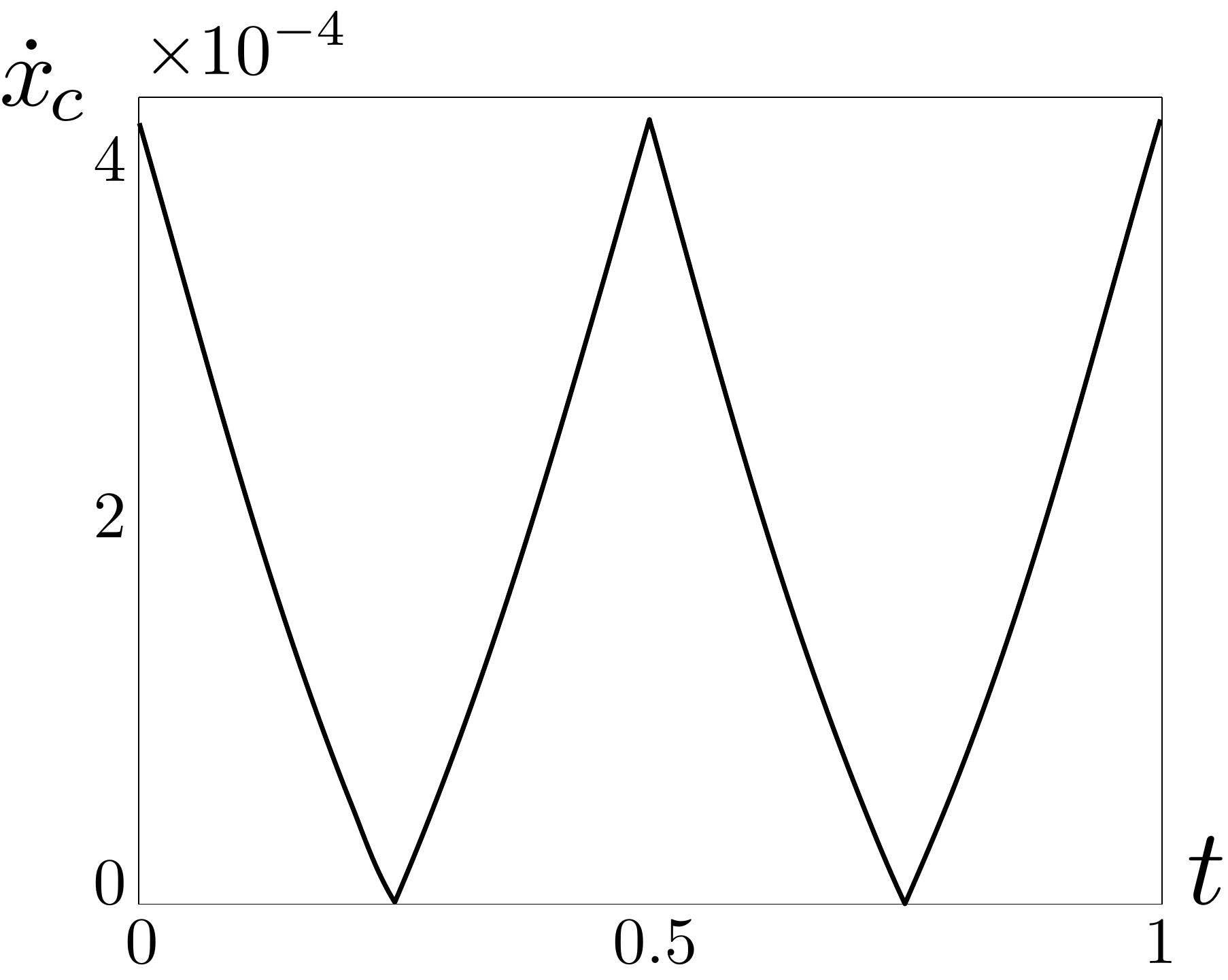}}
	\subfigure[$\dot{y}_c$ vs. $t$]{
	\includegraphics[width=0.31\textwidth]{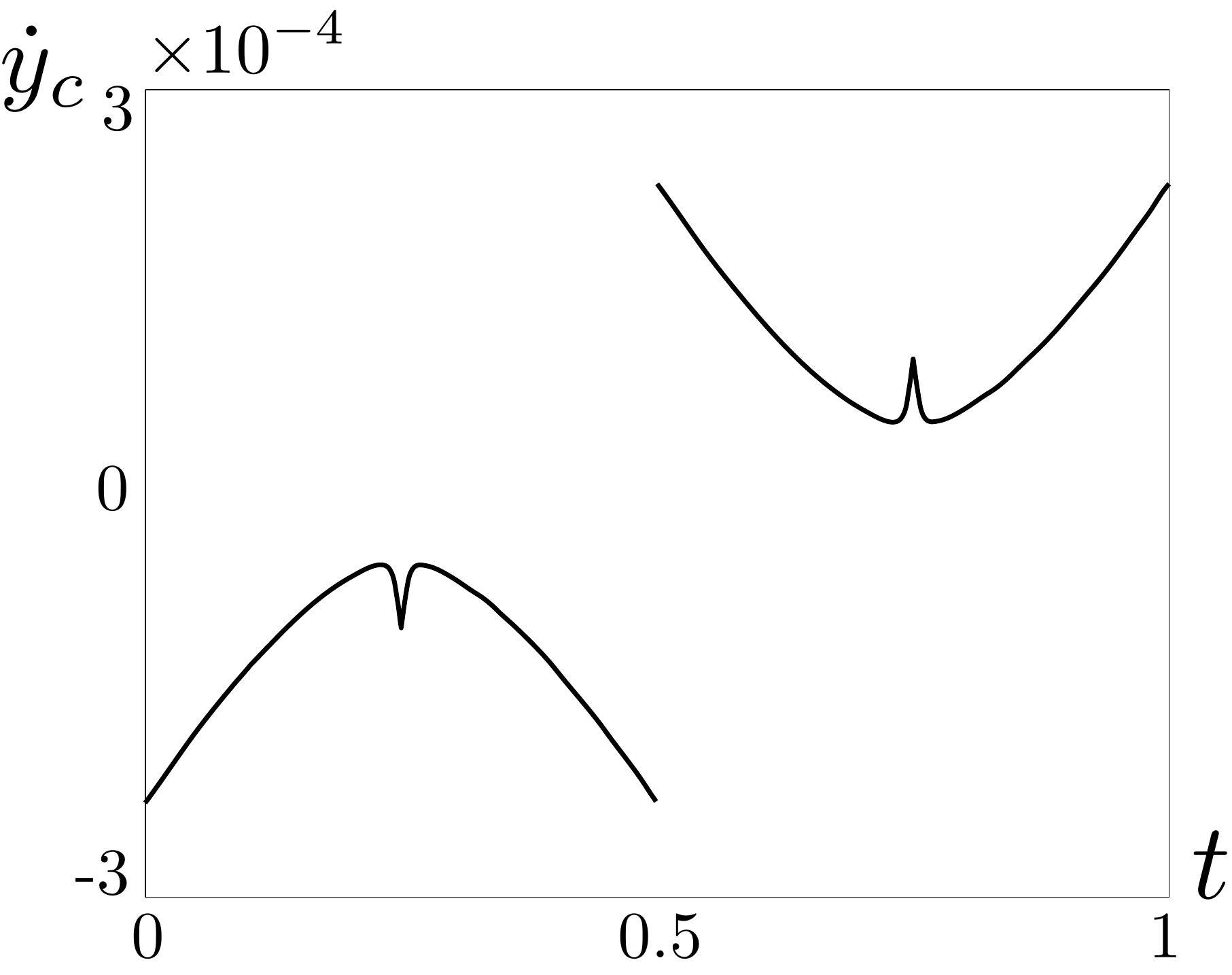}}
	\subfigure[$\dot{\theta}_c$ vs. $t$]{
	\includegraphics[width=0.31\textwidth]{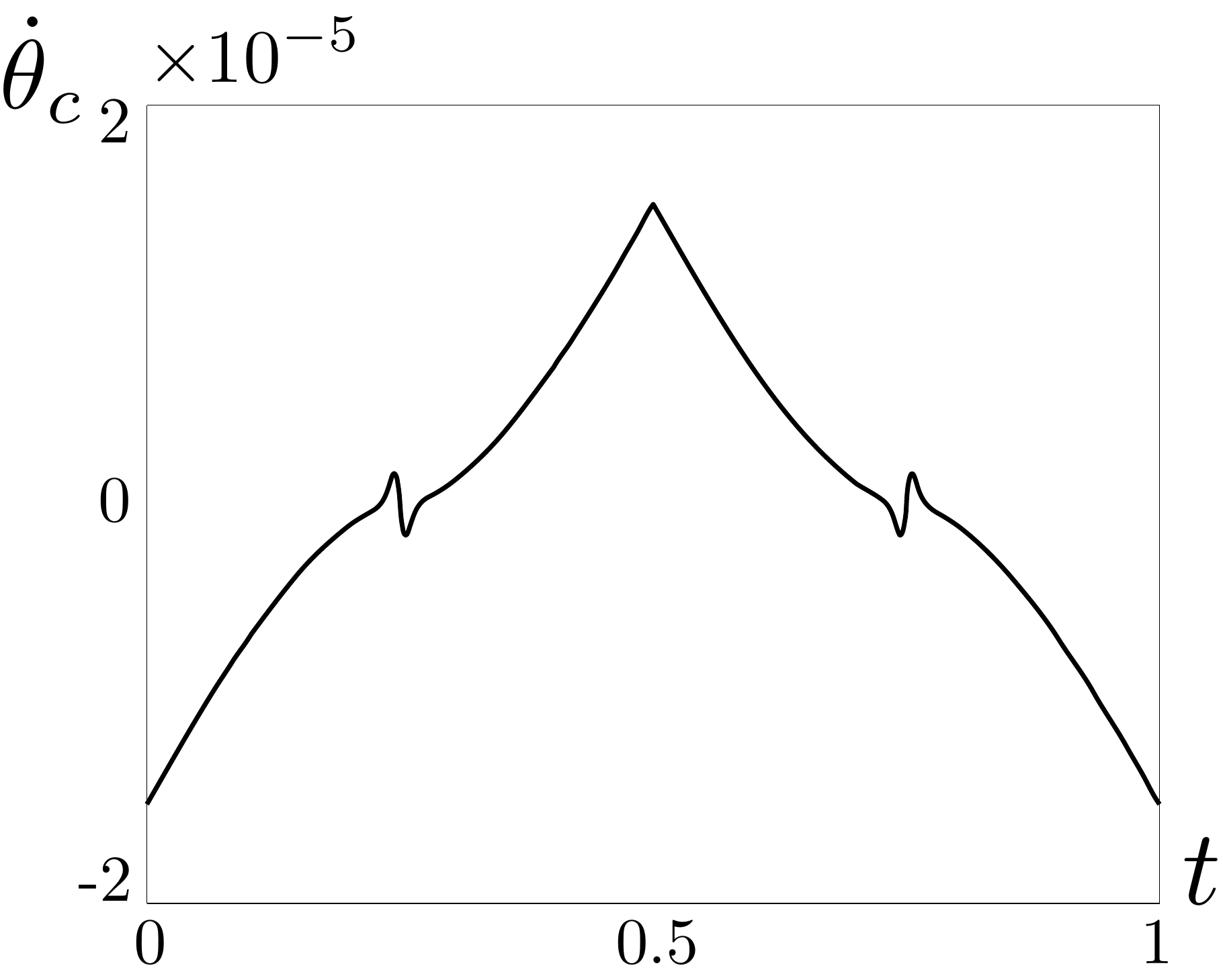}}
	\caption{\footnotesize Inertial frame velocities of the center of mass $C$ for the two-link model with $\theta_r$ given by~\eqref{eq:thetar} and small amplitude of actuation
$\theta_{max} = 0.1$. The coefficients of friction are $\mu_b = 1.3$ and $\mu_t = 1.7$.} \label{fig:2linkvel0.1}
\end{figure}

As an example, when $\theta_r$ is given by~\eqref{eq:thetar} and $\theta_{max} = \epsilon = 0.1$, the velocity of $C$ is numerically solved from the full nonlinear equations~\eqref{eq:eom}, and the result is plotted in Figure~\ref{fig:2linkvel0.1}. The large-amplitude solutions plots (Figure~\ref{fig:2linkvelpiover2}) approximately contain those in Figure~\ref{fig:2linkvel0.1}. Specifically, $\theta_r$ for $\theta_{max} = 0.1$ nearly coincides with the two subintervals where $\theta_r$ is between $[-0.1 \ \ 0.1]$ for the case with $\theta_{max} = \pi/2$, when time is dilated by a constant factor. Hence, the results in Figure~\ref{fig:2linkvel0.1} also nearly coincide with the portions around zero in Figure~\ref{fig:2linkvelpiover2}. One can observe that the order of magnitude of $\dot{\mathbf{g}}_c$ is much smaller than $\epsilon = 0.1$. For the first quarter period, $\dot{x}_c \propto (1-4t)$, $\dot{y}_c \propto (1-4t)^2$ and $\dot{\theta}_c \propto (1-4t)^2$, which is consistent with the analytical solutions in~\eqref{eq:2linkvellin}. When $t$ is close to $1/4$ and $3/4$, the velocities can no longer be approximated by linear or quadratic functions. This is because $\beta(t)$ can no longer be approximated by a constant since $\dot{\theta}_r$ is discontinuous, but the velocities are still bounded. Due to the symmetry in $\theta_r$ in the four quarter periods, the distance traveled by $C$ and the total work during one period are given by
\begin{equation}
	\label{eq:2linksmall_dW}
	d \approx 4 \int_0^{\frac{1}{4}} \dot{x}_c \,\text{d}t \approx \frac{\mu_b -
1}{8(\mu_b + 1)} \epsilon^2, \qquad
	W = 4\int_0^{\frac{1}{4}}\!\!\!  \int_0^1 - \mathbf{f}\cdot
\boldsymbol{\xi}_{lin} \,\text{d}s\,\text{d}t \approx \frac{1}{4}\mu_t\epsilon,
\end{equation}
since $\int_0^1 \dot{y}_c \,\text{d}t\approx 0$. Therefore, the efficiency for small-amplitude actuation is approximately
\begin{equation}
	\label{eq:2linksmall_e}
	e = \frac{d}{W} \approx \frac{\mu_b - 1}{2 \mu_t (\mu_b + 1)} \epsilon.
\end{equation}
This shows that the efficiency is maximized when $\mu_t \rightarrow 0$ and $\mu_b \rightarrow \infty$ (i.e. $(\mu_b - 1)/(\mu_b + 1) \rightarrow 1)$. Note that the linearization is only valid when $\mu_t \gg O(\epsilon^2)$. When $\mu_t$ is comparable to $\epsilon^2$, which means the body is almost frictionless in the transverse direction, the terms in $W$ which are of higher order in $\epsilon$ cannot be ignored, and these higher-order terms depend on $\mu_b$ as well. This means that transverse friction provides the leading-order contribution to the work, and tangential (forward and backward) friction is comparable at higher order (see Appendix~\ref{appen:a} for details). However,~\eqref{eq:2linksmall_e} is valid for $\epsilon \rightarrow 0$ with fixed $\mu_t$, which indicates $e \rightarrow 0$, or small amplitude actuation is energetically inefficient, and the efficiency increases linearly with $\epsilon$. Thus, maximum efficiency occurs at large amplitude, where geometric nonlinearities play a role.

\paragraph{Large amplitude optimization}
\label{par:large_amplitude_optimization}

We now investigate the case when the amplitude of actuation is not small in general. Consider a periodic actuation of the relative angle that varies in the interval $\theta_r \in [\theta_{min} , \theta_{max}]$ during $t \in [0 , 1]$. Without loss of generality, assume $\theta_r(0)  = \theta_r(1) = \theta_{max}$, and $\theta_r(t_{min}) = \theta_{min}$. To simplify the analysis, we consider the case that $\theta_{max}$ and $\theta_{min}$ are the only extrema of $\theta_r$ during the period. In other words, $\theta_r$ varies monotonically between the maximum and minimum, i.e. $\dot{\theta}_r \leq 0$ when $0 < t < t_{min}$, and $\dot{\theta}_r \geq 0$ when $t_{min} < t < 1$. In general, if more local extrema exist, one can always divide the period into multiple parts at the extrema, and the following analysis can be modified accordingly. Recall that our system has the modified kinematic reconstruction equation~\eqref{eq:recon}. That is, for the two-link problem,
\begin{equation}
	\boldsymbol{\xi}_c \equiv \begin{pmatrix}
	U_c\\
	V_c\\
	\Omega_c		
	\end{pmatrix} = \begin{pmatrix}
		U^*(\theta_r,S_r)\\
		V^*(\theta_r,S_r)\\
		\Omega^*(\theta_r,S_r)
	\end{pmatrix} \dot{\theta}_r \equiv \mathbf{A}(\theta_r, S_r) \dot{\theta}_r,
\end{equation}
where the exact forms of the components $U^*, V^*$ and $\Omega^*$ can be derived from the equations of motion~\eqref{eq:eom}. We will show mathematically that the trajectory of $C$ only depends on the path of $\theta_r$ but not on the speed $\dot{\theta}_r$ along the path. For a prescribed $\theta_r$, one has
\begin{equation}
	\label{eq:thc}
	\begin{split}
		\theta_c(t) & = \int_0^t \dot{\theta}_c(\tilde{t}) \,\text{d}\tilde{t} = \int_0^t \Omega_c(\tilde{t}) \,\text{d}\tilde{t} = \int_0^t \Omega^*(\theta_r,S_r) \dot{\theta}_r \,\text{d}\tilde{t} = \int_{\theta_{max}}^{\theta_r(t)} \Omega^*(\tilde{\theta}_r,S_r) \,\text{d}\tilde{\theta}_r\\
		& =
		\begin{cases}	\int_{\theta_{max}}^{\theta_r(t)} \Omega^*(\tilde{\theta}_r,-1)\,\text{d}\tilde{\theta}_r, & \text{for}\  0 < t \leq t_{min},\\[1ex]
			\int_{\theta_{max}}^{\theta_{min}} \Omega^*(\tilde{\theta}_r,-1)\,\text{d}\tilde{\theta}_r + \int_{\theta_{min}}^{\theta_r(t)} \Omega^*(\tilde{\theta}_r,1)\,\text{d}\tilde{\theta}_r, & \text{for}\ t_{min} \leq t < 1,
			\end{cases}
	\end{split}
\end{equation}
where $\tilde{t}$ and $\tilde{\theta}_r$ are integration variables. Once $\theta_c(t)$ is obtained, the position of $C$ can be given by ~\eqref{eq:displacement},
\begin{equation}
	\label{eq:xcyc}
	\begin{split}
		\begin{pmatrix}
			x_c(t)\\
			y_c(t)
		\end{pmatrix} & = \int_0^t R_{\theta_c} \boldsymbol{\xi}_c \, \text{d}\tilde{t} = \int_0^t \begin{pmatrix}
			\cos\theta_c & -\sin\theta_c\\
			\sin\theta_c & \cos\theta_c
		\end{pmatrix}
		\begin{pmatrix}
			U_c\\
			V_c
		\end{pmatrix}\,\text{d}\tilde{t}
		 = \int_0^t \begin{pmatrix}
			\cos\theta_c & -\sin\theta_c\\
			\sin\theta_c & \cos\theta_c
		\end{pmatrix}
		\begin{pmatrix}
			U^*(\theta_r,S_r)\\
			V^*(\theta_r,S_r)
		\end{pmatrix}\dot{\theta}_r\,\text{d}\tilde{t}\\
		&  = \int_{\theta_{max}}^{\theta_r(t)} \begin{pmatrix}
			\cos\theta_c & -\sin\theta_c\\
			\sin\theta_c & \cos\theta_c
		\end{pmatrix}
		\begin{pmatrix}
			U^*(\tilde{\theta}_r,S_r)\\
			V^*(\tilde{\theta}_r,S_r)
		\end{pmatrix}\,\text{d}\tilde{\theta}_r,
	\end{split}
\end{equation}
where the integration can be evaluated similarly to~\eqref{eq:thc}. Therefore, $\mathbf{g}_c(t)$ only depends on the path of $\theta_r$ via the integration limits, and the speed $\dot{\theta}_r$ does not explicitly appear in the expression (although its sign does appear). For distance $d$, work $W$ and consequently efficiency $e$, the integration is over the whole period from $t = 0$ to 1, and hence they only depend on the extrema of $\theta_r$ during the period: $\theta_{max}$ and $\theta_{min}$. This is because the parameter space of shapes is only one dimensional, so the path of $\theta_r$ is defined by the endpoints $\theta_{max}$ and $\theta_{min}$.

Due to the nonlinearity of the system, the calculation of $\mathbf{g}_c(t)$ via \eqref{eq:thc} and \eqref{eq:xcyc} is not trivial. However, from the numerical results of $\theta_{max} = 0.1$ and $\pi/2$, one can observe that the orientation of the snake $\theta_c$ is usually very small during the locomotion. This is due to the symmetry in shape about the $\mathbf{b}_y$ axis, and the fact that the
$\mu_b$--$\mu_f$ asymmetry has little effect on rotation. Indeed, even for a large amplitude $\theta_{max} = \pi - 0.01$, one still has $\sup_t \|\theta_c\| < 1.4$ degrees for all time. Hence, one can closely approximate the problem by assuming $\theta_c \approx 0$ during the period. Therefore, from~\eqref{eq:bodyvelocity}, the velocities of $C$ in both frames are approximately the same, i.e. $\dot{\mathbf{g}}_c \approx \boldsymbol{\xi}_c$, in which the non-zero components are the linear velocity,
\begin{equation}
	\label{eq:2link_reconcomponent}
 	\begin{pmatrix}
 		u_c\\
 		v_c
 	\end{pmatrix}
	\approx \begin{pmatrix}
		U_c\\
		V_c
	\end{pmatrix}
	 = \begin{bmatrix}
		U^*(\theta_r,S_r)\\
		V^*(\theta_r,S_r)
	\end{bmatrix} \dot{\theta}_r.
\end{equation}
Since the Heaviside function appears in the tangential but not the transverse direction, comparing the force equation in~\eqref{eq:eom} for the same $\theta_r$ and opposite $\dot{\theta}_r$ gives
\begin{equation}
	\label{eq:2link_symmetry}
		U^*(\theta_r, 1) = -U^*(\theta_r, -1), \quad V^*(\theta_r, 1) =
V^*(\theta_r, -1).
\end{equation}
This is also evident from the results shown in Figure~\ref{fig:2linkvelpiover2}(a) and (b): for two instants symmetric about 0.5, which have equal $\theta_r$ but opposite $\dot{\theta}_r$, the corresponding $u_c$ are nearly equal but $v_c$ are nearly opposite. To simplify the notation, denote $u(\theta_r) \equiv U^*(\theta_r , S_r = 1)$, and $v(\theta_r) \equiv V^*(\theta_r , S_r = 1)$. They are nonlinear functions of $\theta_r$ only. From~\eqref{eq:xcyc} and~\eqref{eq:2link_reconcomponent}, integrating the velocity in the inertial frame for the whole period yields
\begin{equation}
	\begin{split}
	x_c(1) = \int_0^1 u_c \, \text{d}t & \approx \left(\int_0^{t_{min}} +
	\int_{t_{min}}^1\right) U^*(\theta_r, S_r)\,\dot{\theta}_r \,
	\text{d}t= \int_0^{t_{min}} - u(\theta_r) \,\dot{\theta}_r \, \text{d}t
	+ \int_{t_{min}}^1  u(\theta_r)\,\dot{\theta}_r \, \text{d}t \\
	& = \left(- \int^{\theta_{min}}_{\theta_{max}} +
	\int_{\theta_{min}}^{\theta_{max}}\right) u(\tilde{\theta}_r) \, \text{d}\tilde{\theta}_r 
	 = 2 \mathbb{X}(\theta_{max}) - 2 \mathbb{X}(\theta_{min}),\\[2ex]
		y_c(1) = \int_0^1 v_c \, \text{d}t & \approx \left(\int_0^{t_{min}} +
	\int_{t_{min}}^1\right) V^*(\theta_r, S_r)\,\dot{\theta}_r \, \text{d}t
	= \int_0^{t_{min}} v(\theta_r) \,\dot{\theta}_r \, \text{d}t +
	\int_{t_{min}}^1  v(\theta_r)\,\dot{\theta}_r \, \text{d}t \\
	& = \left( \int^{\theta_{min}}_{\theta_{max}} +
	\int_{\theta_{min}}^{\theta_{max}}\right) v(\tilde{\theta}_r) \, \text{d}\tilde{\theta}_r
	 = 0,
	\end{split}
\end{equation}
where \(\mathbb{X}(\theta) = \int_0^{\theta} u(\tilde{\theta})\, \text{d}\tilde{\theta}\) depends on the form of $u$ and the integration limit $\theta$. The distance $d$ is given by
\begin{equation}
	\label{eq:2linkd}
	d = \sqrt{x_c(1)^2 + y_c(1)^2} \approx 2\mathbb{X}(\theta_{max}) -
2\mathbb{X}(\theta_{min}),
\end{equation}
since $x_c(0) = y_c(0) = 0$. The power, or rate of work, done by the snake at a given time can be written
\begin{equation}
		P(\theta_r, \dot{\theta}_r) \equiv \int_0^1
	-\mathbf{f}\cdot\boldsymbol{\xi}_{lin} \,\text{d}s = 
	p^*(\theta_r,S_r)\dot{\theta}_r.
\end{equation}
Power is identical for the instants with the same value of $\theta_r$ and same magnitude but opposite sign of $\dot{\theta}_r$. That is, $P(\theta_r, \dot{\theta}_r) = P(\theta_r, -\dot{\theta}_r)$. Therefore, one has $p^*(\theta_r,1) = -p^*(\theta_r,-1)$. Similarly to our definition of $u(\theta_r)$, we define $p(\theta_r) \equiv p^*(\theta_r,1)$. The total work is given by integrating power over the period,
\begin{equation}
	\label{eq:2linkW}
		W = \int_0^1 P(\theta_r, \dot{\theta}_r)\, \text{d}t =  \left(-
\int^{\theta_{min}}_{\theta_{max}} + \int_{\theta_{min}}^{\theta_{max}}\right)
p(\theta_r) \, \text{d}\theta_r = 2 \mathbb{W}(\theta_{max}) - 2
\mathbb{W}(\theta_{min}),
\end{equation}
where $\mathbb{W}(\theta) = \int_0^{\theta} p(\tilde{\theta})\, \text{d}\tilde{\theta}$. The integrals $\mathbb{X}$ and $\mathbb{W}$ only explicitly depend on their upper integration limits (and implicitly on the parameters $\mu_b$ and $\mu_t$). By definition, they are odd functions, for instance, $\mathbb{X}(\theta) = -\mathbb{X}(-\theta)$. For symmetric cases, $\theta_{max} = -\theta_{min}$ (e.g. the one shown in Figure~\ref{fig:2linktri_th}), and the efficiency is simplified to
\begin{equation}
	\label{eq:2linke}
	 e = \frac{2\mathbb{X}(\theta_{max}) -
2\mathbb{X}(\theta_{min})}{2\mathbb{W}(\theta_{max}) - 
2\mathbb{W}(\theta_{min})} = \frac{\mathbb{X}(\theta_{max}) +
\mathbb{X}(\theta_{max})}{\mathbb{W}(\theta_{max}) +  \mathbb{W}(\theta_{max})}
= \frac{\mathbb{X}(\theta_{max})}{\mathbb{W}(\theta_{max})}.
\end{equation}
\begin{figure}
	[!tb] \centering
	\subfigure[$\mathbb{X}$ vs. $\theta_{max}$]{
	\includegraphics[width=0.32\textwidth]{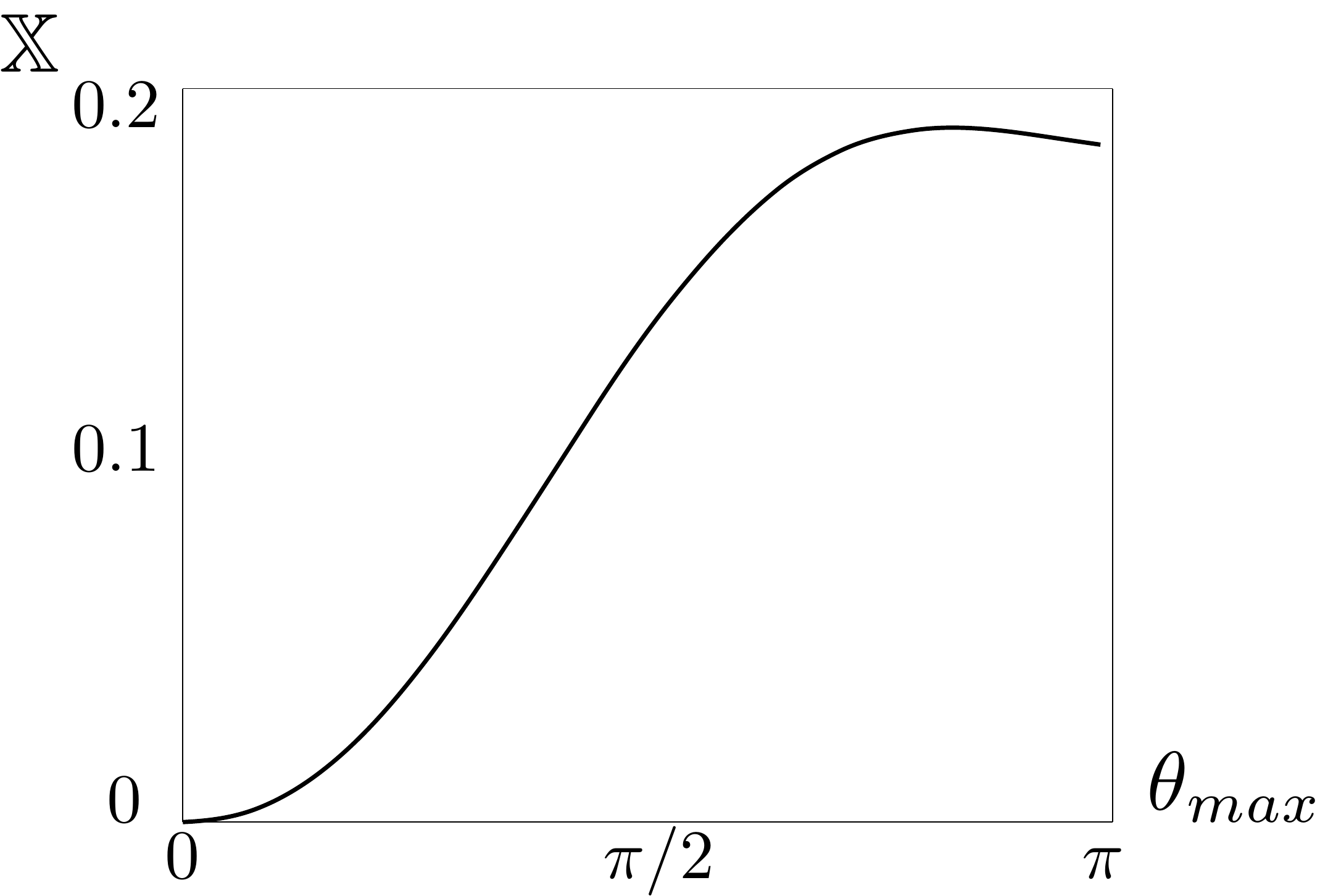}}
	\subfigure[$\mathbb{W}$ vs. $\theta_{max}$]{
	\includegraphics[width=0.31\textwidth]{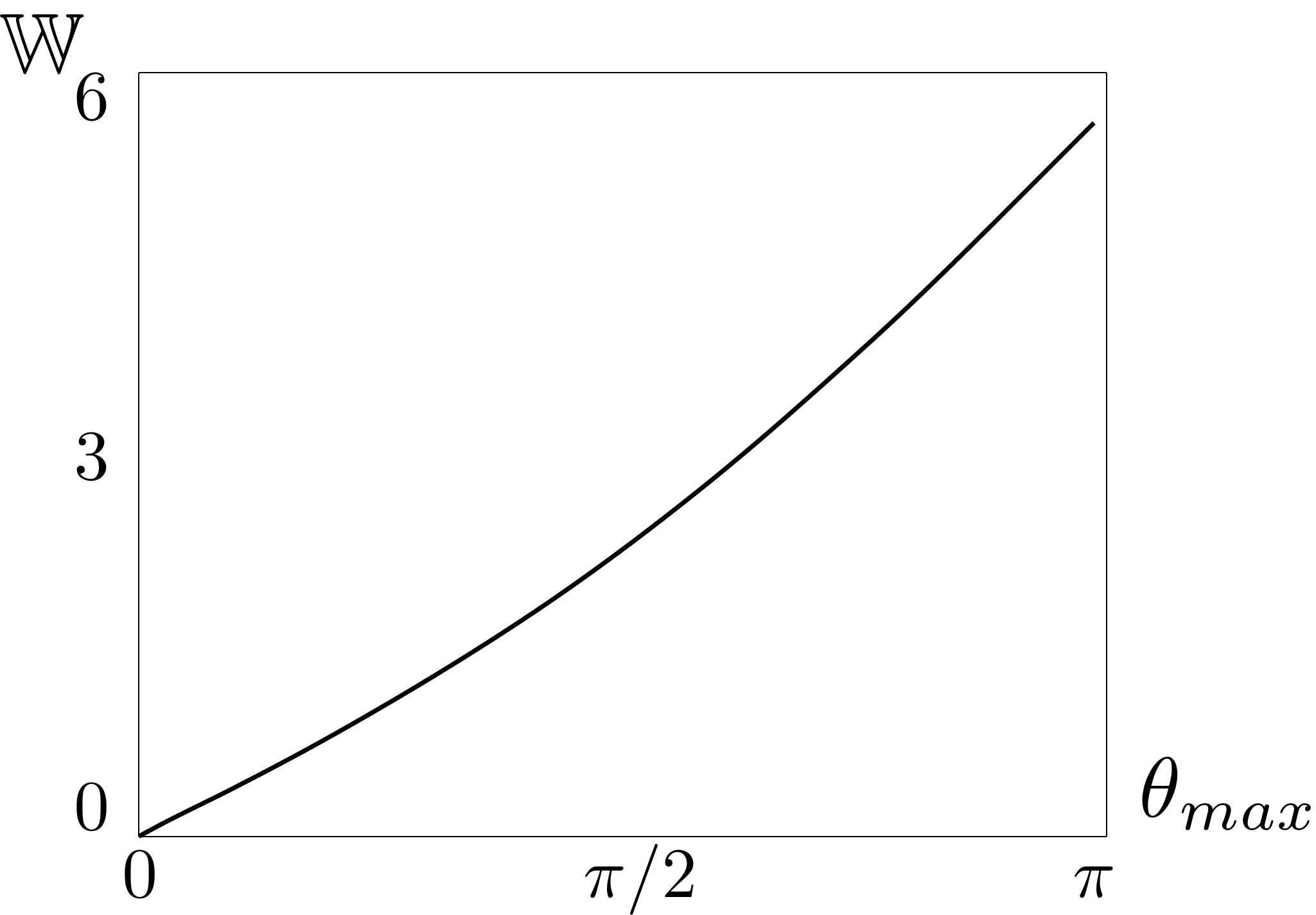}}
	\subfigure[$e$ vs. $\theta_{max}$]{
	\includegraphics[width=0.32\textwidth]{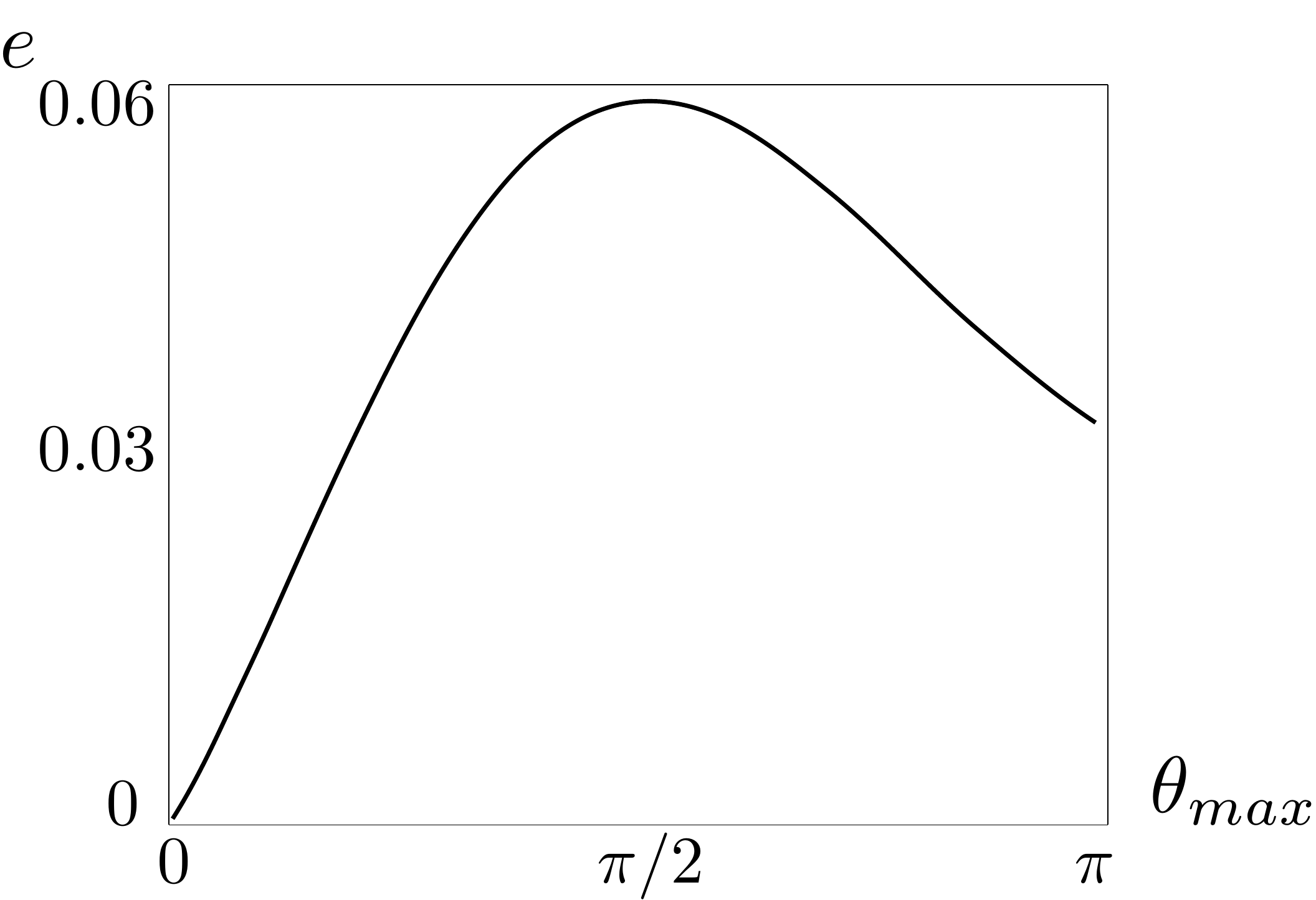}}
\caption{\footnotesize Two-link model (a) $\mathbb{X}$, (b) $\mathbb{W}$ and (c) $e$ as
functions of maximum amplitude of actuation $\theta_{max}$ for $\mu_b = 1.3$ and $\mu_t =
1.7$.}\label{fig:twolink_e} 
\end{figure}
Figure~\ref{fig:twolink_e} shows $\mathbb{X}, \mathbb{W}$ and $e$ as functions of the maximum amplitude $\theta_{max}$ for parameters $\mu_b = 1.3$ and $\mu_t = 1.7$. One can see that $\mathbb{X}$ is maximized around $\theta_{max} \approx 3\pi/4$, but since $\mathbb{W}$ increases faster at larger $\theta_{max}$, the efficiency $e$ is maximized around $\theta_{max} \approx \pi/2$. We emphasize that, since the parameter space of shapes is only one dimensional, as long as $\theta_{max}$ is the same, the efficiency and trajectory traveled by $C$ are the same. For example, any function $\theta_r$ that has the same maximum and minimum (and varies monotonically in between) as the one shown in Figure~\ref{fig:2linktri_th}, for example a cosine function with period 1 and amplitude $\pi/2$, will also result in the trajectory shown in Figure~\ref{fig:2linkvelpiover2}(a) and have the same $e$.

\begin{figure}
	[!b] \centering
	\subfigure[$e$ for $\mu_b = 1.3$ and various $\mu_t$]{
	\includegraphics[width=0.4\textwidth]{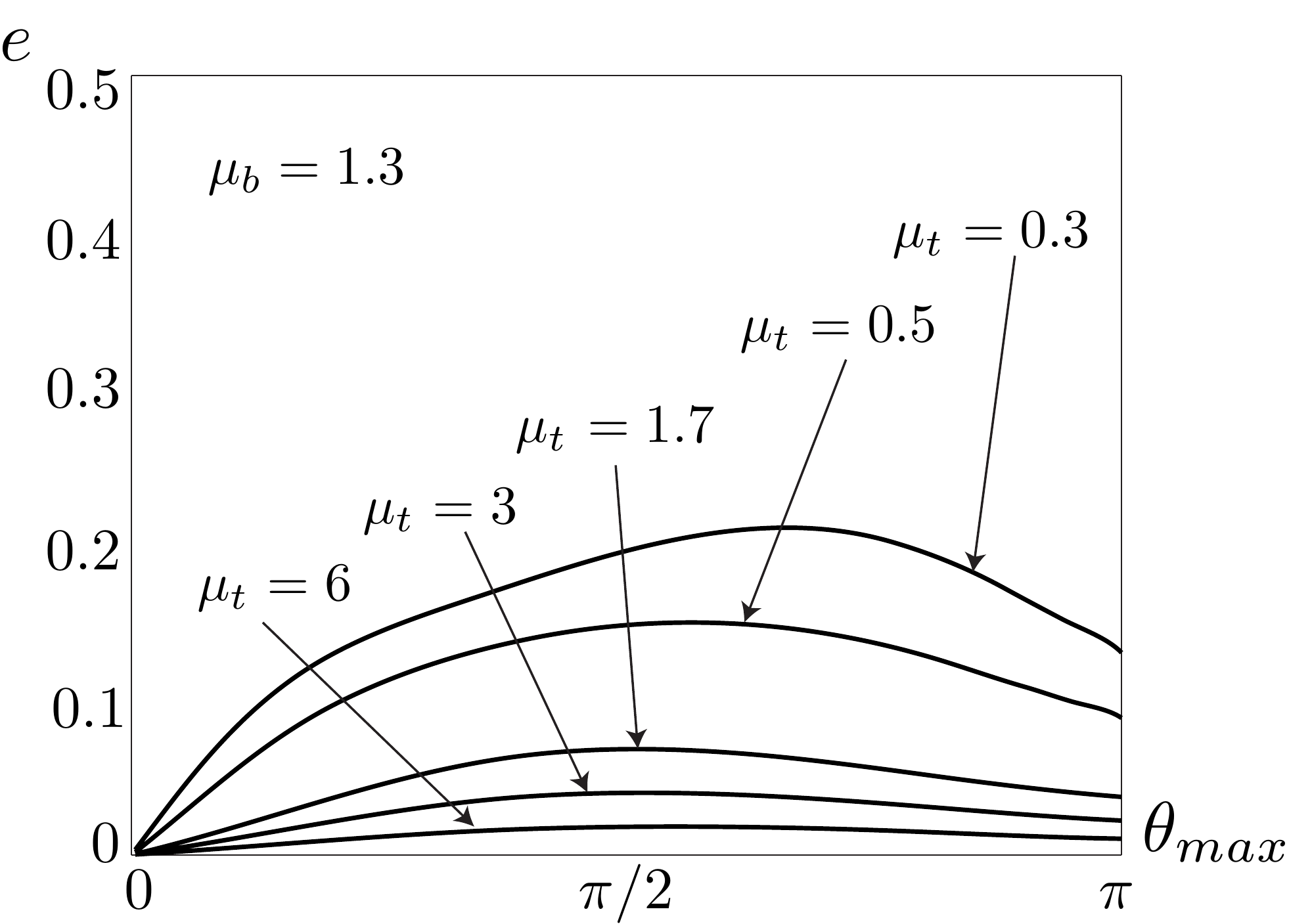}}\qquad
	\subfigure[$e$ for $\mu_t = 1.7$ and various $\mu_b$]{
	\includegraphics[width=0.4\textwidth]{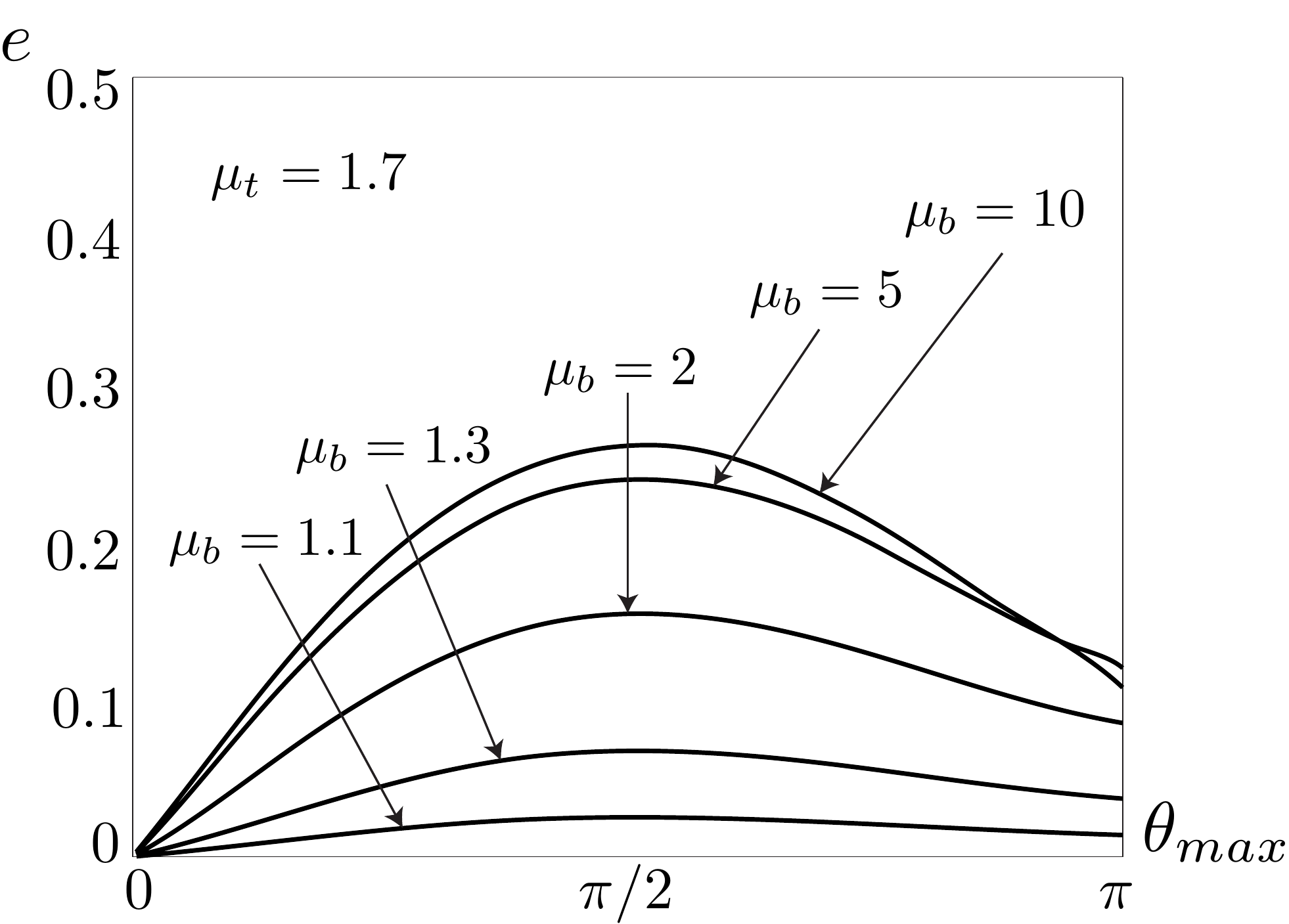}}
	\caption{\footnotesize Two-link efficiency $e$ as a function of actuation amplitude $\theta_{max}$ for various $\mu_b$ and $\mu_t$: (a) fixed $\mu_b = 1.3$ and various $\mu_t$ and (b) fixed $\mu_t = 1.7$ and various $\mu_b$.}\label{fig:twolink_eff} 
\end{figure}

We repeat the process above to calculate $e$ as a function of $\theta_{max}$ for various $\mu_b$ and $\mu_t$, and the results are shown in Figure~\ref{fig:twolink_eff} for (a) $\mu_b = 1.3$ and various $\mu_t$ and (b) $\mu_t = 1.7$ and various $\mu_b$. In (a), the efficiency-maximizing $\theta_{max} \approx \pi/2$ for $\mu_t > 1$, and increases as $\mu_t$ drops below 1, in which case there is a smaller increase in work for the additional transverse motion associated with a larger amplitude. In (b), one can see that for $\mu_t = 1.7 > 1$, the efficiency-maximizing $\theta_{max} \approx \pi/2$ regardless of the value of $\mu_b$. Note that when $\theta_{max}$ is small, $e$ varies almost linearly with $\theta_{max}$, which agrees with the small-amplitude result in~\eqref{eq:2linksmall_e}.

\begin{figure}
	[!htb] \centering
	\subfigure[$d$ contour]{
	\includegraphics[width=0.325\textwidth]{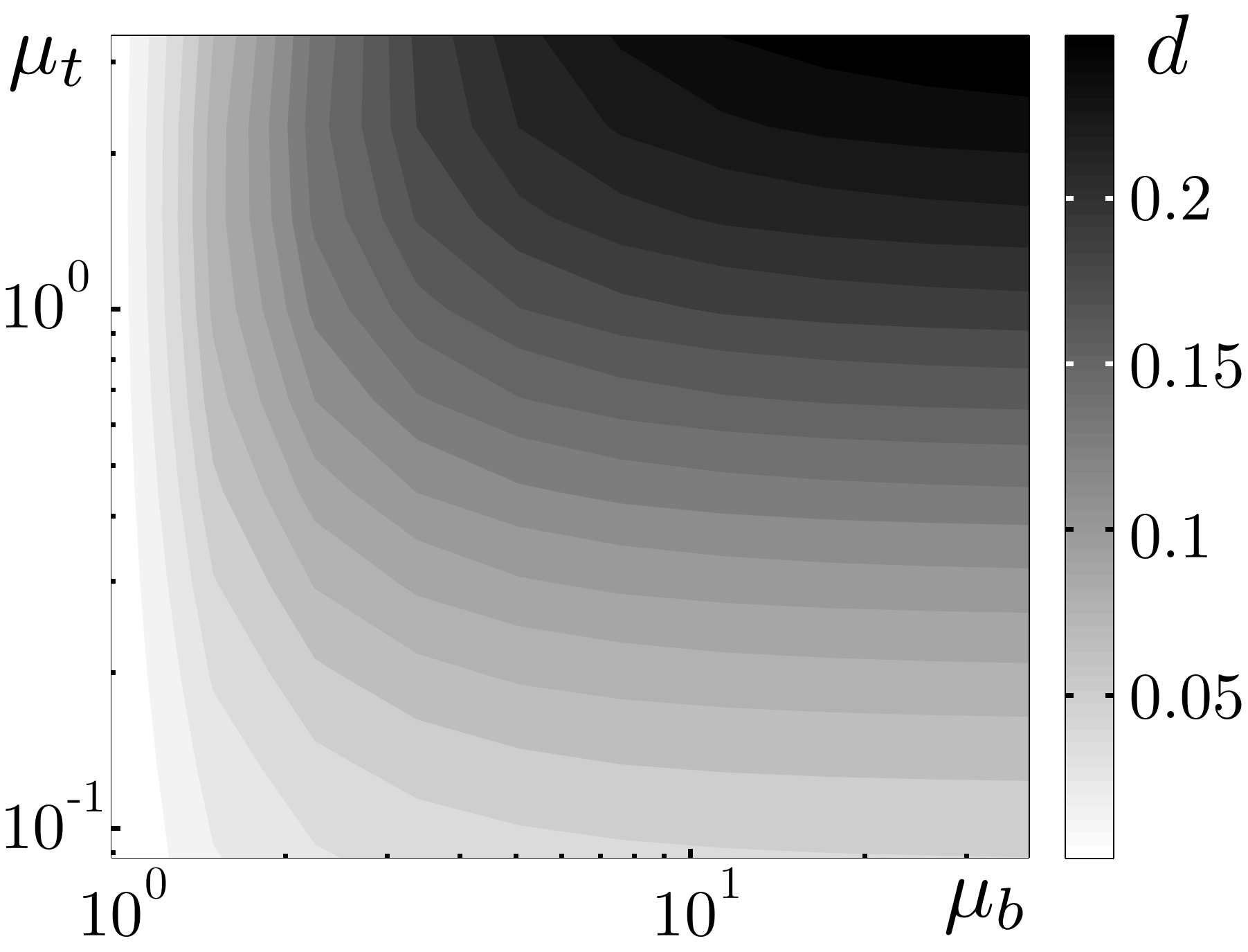}}
	\subfigure[$W$ contour]{
	\includegraphics[width=0.32\textwidth]{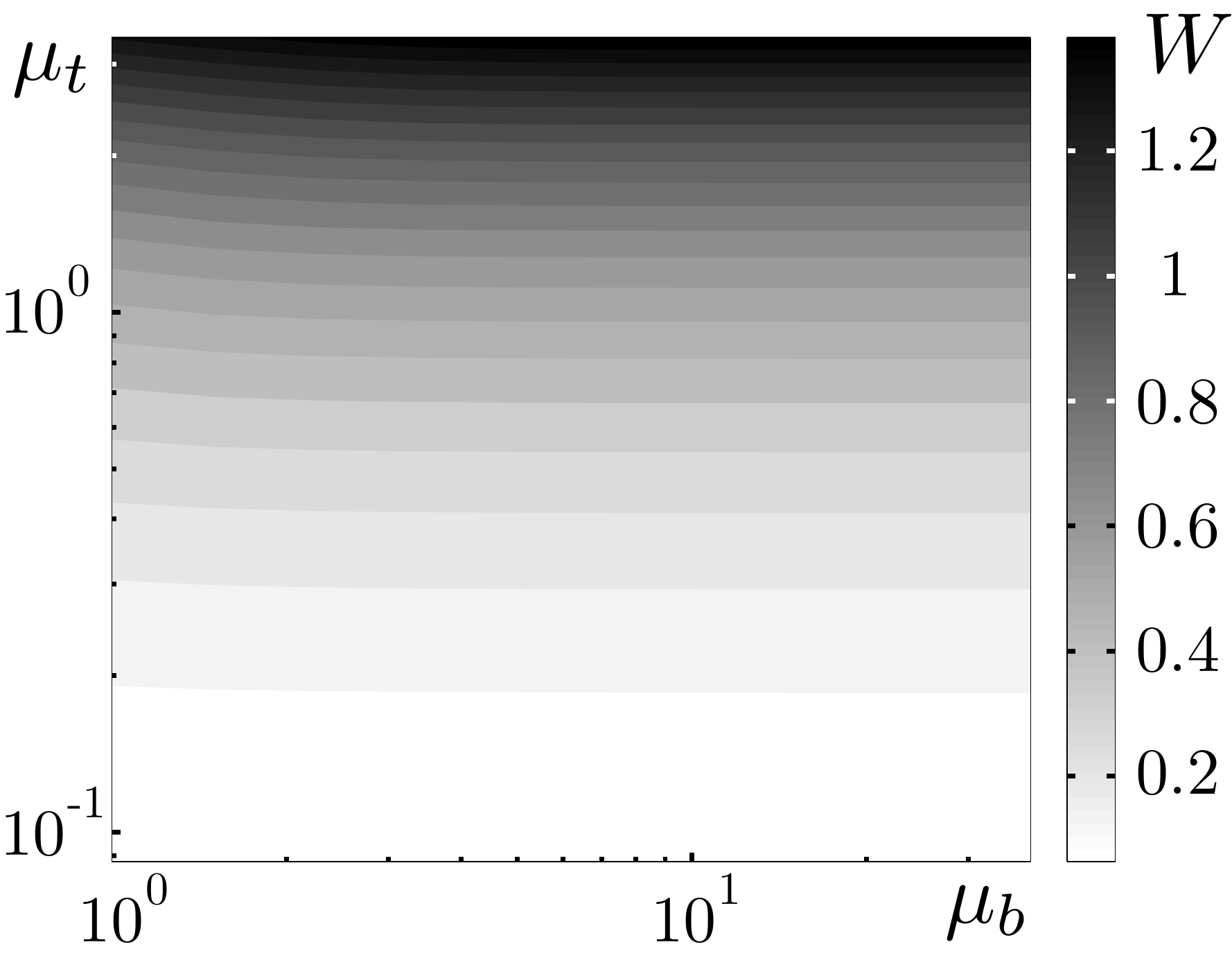}}
	\subfigure[$e$ contour]{
	\includegraphics[width=0.315\textwidth]{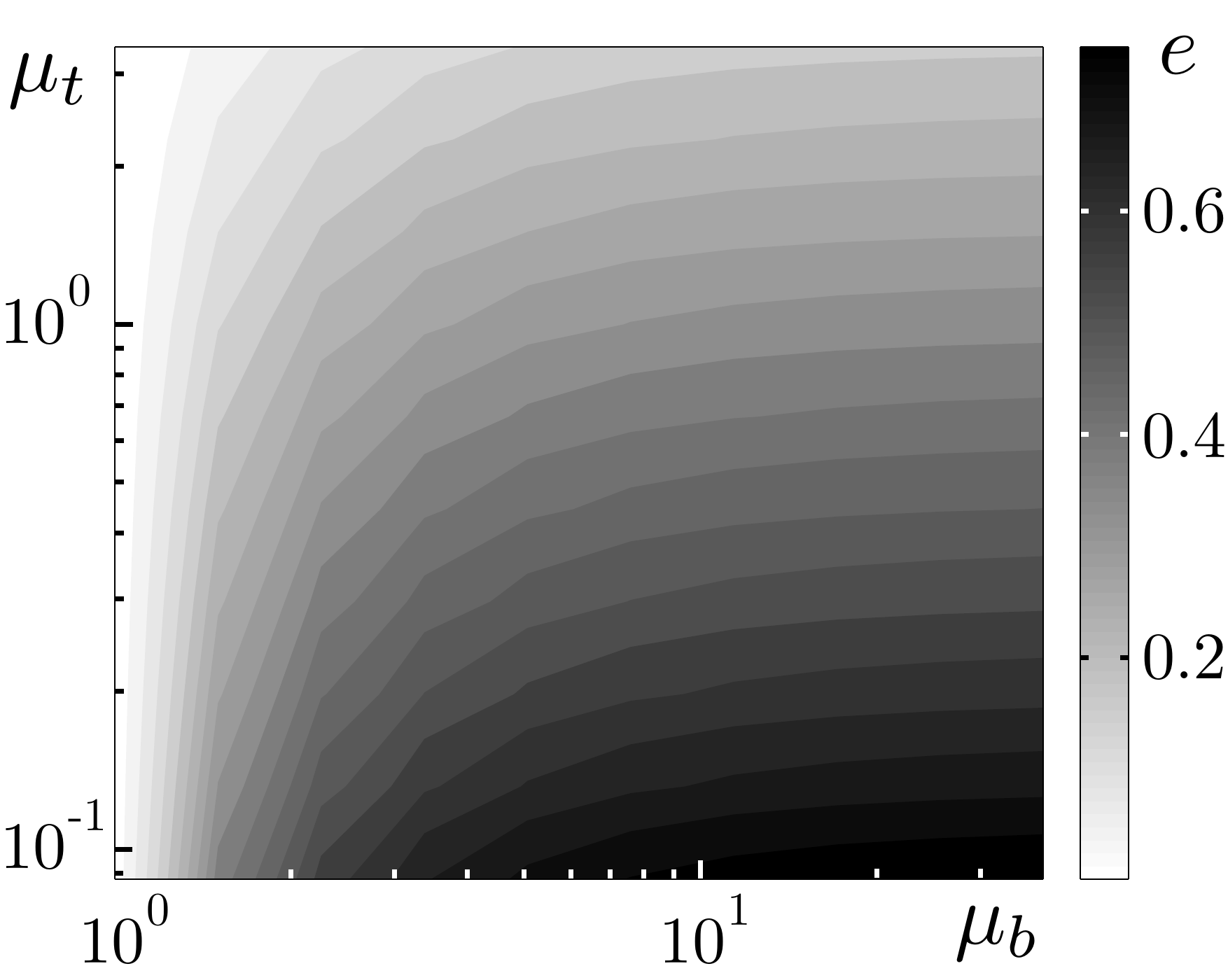}}
\caption{\footnotesize Contour maps of (a) distance $d$, (b) work $W$ and (c) efficiency $e$ as functions of $\mu_b$ and $\mu_t$ (on a log-log scale), for $\theta_{max} = \pi/2$.}\label{fig:twolink_contour} 
\end{figure}

We now determine the values of $\mu_b$ and $\mu_t$ that maximize $e$. Figure~\ref{fig:twolink_contour}(c) shows a contour plot of $e$ as a function of $\mu_b$ and $\mu_t$ when $\theta_{max} = \pi/2$. Here $e$ is evaluated at points on a $10\times 10$ logarithmic grid with nodes spaced by factors of 1.5, and $\mu_b$ ranges from $1.5^0 = 1$ to $1.5^9 \approx 38.44$ and $\mu_t$ from $1.5^{-6} \approx 0.088$ to $1.5^3 \approx 3.375$. The highest efficiency occurs at the largest $\mu_b$ and the smallest $\mu_t$ in this range. In Figure~\ref{fig:twolink_contour}(a), the distance $d$ increases when $\mu_b$ and $\mu_t$ increase. By contrast, work $W$ is mostly independent of $\mu_b$ and increases when $\mu_t$ increases, as shown in (b). Note this is consistent with the small amplitude result in~\eqref{eq:2linksmall_dW}. The trends shown in Figure~\ref{fig:twolink_contour} hold for a wider range of parameters, which suggests that the efficiency-maximizing parameters are $\mu_b$ large and $\mu_t$ small. 



\section{Three-link model} 
\label{sec:3link}

\begin{figure}
	[!htb] \centering 
	\includegraphics[width=0.45\textwidth]{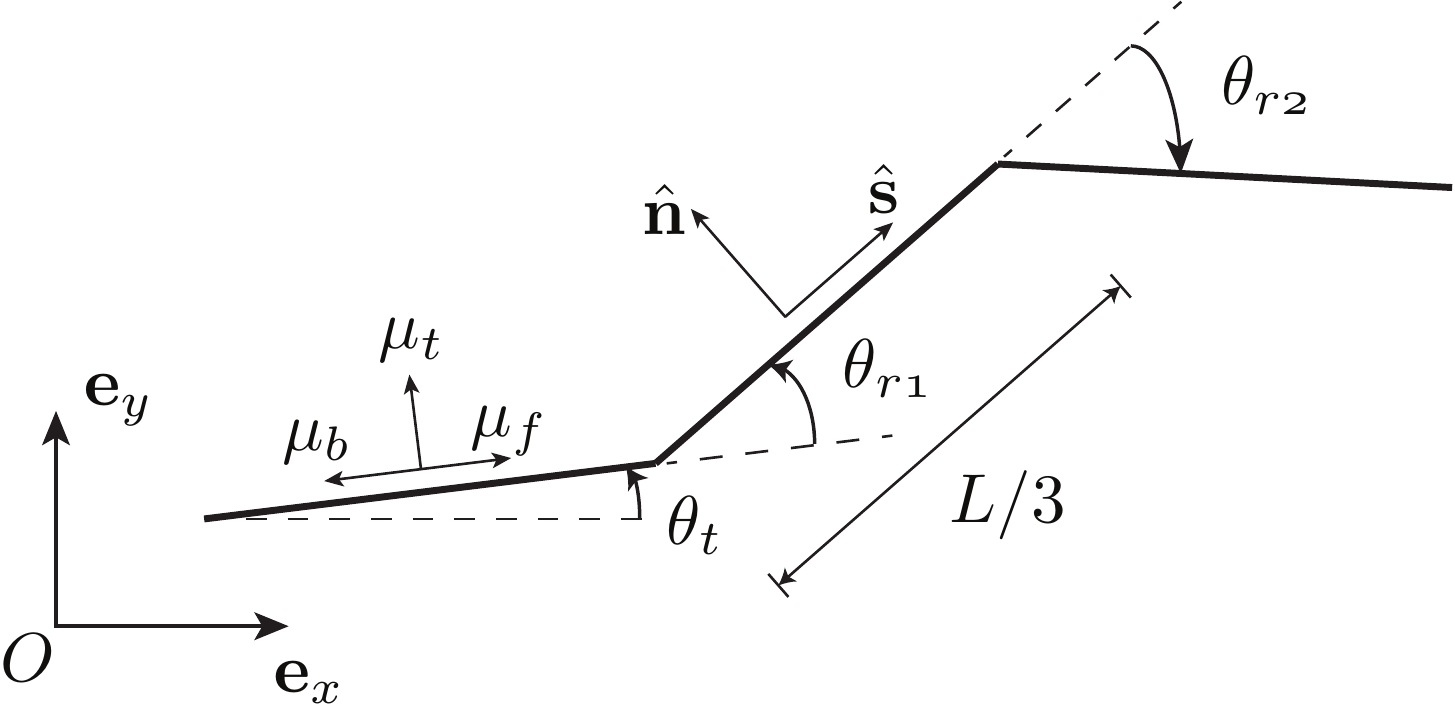} \caption{\footnotesize Three-link snake model; see text for description.}\label{fig:model3link} 
\end{figure}
We now consider a snake model with three links (each with length $1/3$) connected by hinge joints as depicted in Figure~\ref{fig:model3link}. The shape is now given by two relative angles, $\theta_{r1}$ and $\theta_{r2}$. Since the links cannot penetrate each other, the angles are constrained to lie in the set
\begin{equation}
	\label{eq:constraintset}
	S_{\theta_{r1},\theta_{r2}} = \left\{ (\theta_{r1},\theta_{r2})\in
(-\pi,\pi)\times(-\pi,\pi) \bigcap \left(
	\begin{array}{lr}
				\theta_{r2} > -\pi - \theta_{r1}/2, & -\pi <
\theta_{r1} \leq -2\pi/3\\
				\theta_{r2} > -2\pi - 2\theta_{r1}, & -2\pi/3
\leq \theta_{r1} \leq -\pi/2\\
				\theta_{r2} < 2\pi + 2\theta_{r1}, & \pi/2 \leq
\theta_{r1} \leq -2\pi/3\\
				\theta_{r2} < \pi - \theta_{r1}/2, & 2\pi/3 <
\theta_{r1} < \pi
			\end{array}\right)\right\}.
\end{equation}
\begin{figure}
	[!htb] \centering 
	\includegraphics[width=0.4\textwidth]{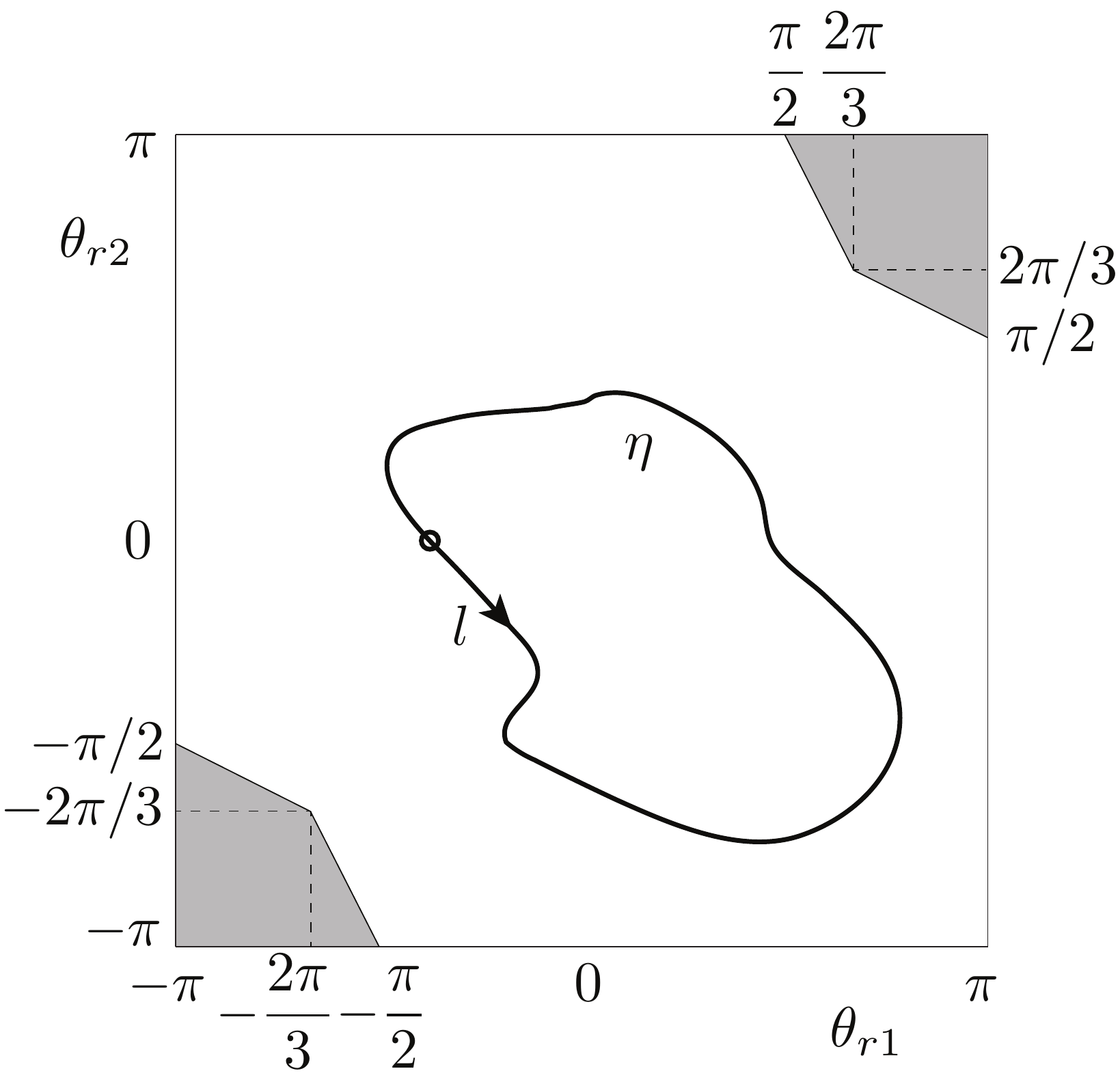}
	\caption{\footnotesize Shape change parameter plane $(\theta_{r1},\theta_{r2})$. The shaded areas are infeasible due to the mutual avoidance of the links. A periodic kinematics is a directional closed path $\eta$, with the initial state marked by $\circ$.}\label{fig:3link_th1th2constraint} 
\end{figure}
In Figure~\ref{fig:3link_th1th2constraint} the infeasible regions are shaded at the upper right and lower left corners in the $(\theta_{r1}\,,\,\theta_{r2})$ plane. One can describe the motion in the inertial frame $\{\mathbf{e}_x, \mathbf{e}_y\}$ or in the body-fixed frame $\{\mathbf{b}_x, \mathbf{b}_y\}$, with the angle between $\mathbf{b}_x$ and $\mathbf{e}_x$ now given by
\begin{equation}
	\label{eq:3linkthc}
	\theta_c \equiv \frac{1}{3}\left(\theta_t + \theta_m + \theta_h\right) =
\theta_m - \frac{1}{3}\theta_{r1} + \frac{1}{3}\theta_{r2}.
\end{equation}
Here $\theta_t$, $\theta_m$ and $\theta_h$ are the orientations of the tail, middle and head links in the inertial frame, respectively. In the body frame, the configuration variable on each of the three links is
\begin{equation*}
	\label{eq:3linktail}
	\mathbf{G}_t = \begin{pmatrix}
	(s-5/18) \cos\Theta_t - 1/6 \cos\Theta_m - 1/18\cos\Theta_h\\
	(s - 5/18) \sin \Theta_t - 1/6 \sin \Theta_m - 1/18 \sin \Theta_h\\
	\Theta_t \equiv -2\theta_{r1}/3 - \theta_{r2}/3
\end{pmatrix},
\end{equation*}
\begin{equation}
	\label{eq:3linkmiddle}
	\mathbf{G}_m = \begin{pmatrix}
	1/18 \cos \Theta_t + (s - 1/2) \cos \Theta_m - 1/18 \cos\Theta_h\\
	1/18 \sin \Theta_t + (s - 1/2) \sin \Theta_m - 1/18 \sin \Theta_h\\
	\Theta_m \equiv \theta_{r1}/3 - \theta_{r2}/3
\end{pmatrix},
\end{equation}
\begin{equation*}
	\label{eq:3linkhead}
	\mathbf{G}_h = \begin{pmatrix}
	1/18 \cos \Theta_t + 1/6 \cos \Theta_m + (s - 13/18) \cos\Theta_h\\
	1/18 \sin \Theta_t + 1/6 \sin \Theta_m + (s - 13/18) \sin \Theta_h\\
	\Theta_h \equiv \theta_{r1}/3 + 2\theta_{r2}/3
\end{pmatrix}.
\end{equation*}
The linear and angular velocities in the body and inertial frames and the transformations between
them are again given by~\eqref{eq:bodyvelocity} and \eqref{eq:velocityrelation}. The expressions for forces, the equations of motion and the definitions of distance, work and efficiency are all in the same as in the two-link case.

In this section, to keep the parameter space tractable, we set $\mu_b = 1.3$ and $\mu_t = 1.7$ (based on the experimental measurement in~\cite{HuSh2012}) and search for the efficiency-maximizing shape change in terms of $\theta_{r1}(t)$ and $\theta_{r2}(t)$. Now the path of the shape change over one period in $(\theta_{r1}\,,\,\theta_{r2})$ space is a directional closed curve ({\em path}) denoted by $\eta$ (an example of which is shown in Figure~\ref{fig:3link_th1th2constraint}). 

The reconstruction equation~\eqref{eq:recon} is now
\begin{equation}
	\label{eq:3linkrecon}
	\boldsymbol{\xi}_c \equiv \begin{pmatrix}
	U_c\\
	V_c\\
	\Omega_c		
	\end{pmatrix} = \begin{pmatrix}
		U_1^*(\theta_{r1},S_{r1},\theta_{r2},S_{r2}) & U_2^*(\theta_{r1},S_{r1},\theta_{r2},S_{r2})\\
		V_1^*(\theta_{r1},S_{r1},\theta_{r2},S_{r2}) & V_2^*(\theta_{r1},S_{r1},\theta_{r2},S_{r2})\\
		\Omega_1^*(\theta_{r1},S_{r1},\theta_{r2},S_{r2}) & \Omega_2^*(\theta_{r1},S_{r1},\theta_{r2},S_{r2})
	\end{pmatrix} 
	\begin{pmatrix}
		\dot{\theta}_{r1}\\
		\dot{\theta}_{r2}
	\end{pmatrix} \equiv A(\boldsymbol{\theta}_r, \mathbf{S}_r) \dot{\boldsymbol{\theta}}_r.
\end{equation}
As in the two-link case, most quantities of interest are independent of how time is
parametrized. For example,
\begin{equation}
	\label{eq:3linkthc}
	\begin{split}
			\theta_c(t) & = \int_0^t \Omega_c \,\text{d}\tilde{t} = \int_0^t \left[\Omega_1^*(\theta_{r1},S_{r1},\theta_{r2},S_{r2}) \dot{\theta}_{r1} + \Omega_2^*(\theta_{r1},S_{r1},\theta_{r2},S_{r2}) \dot{\theta}_{r2} \right]\text{d}\tilde{t} \\
			& = \int_{\eta_0^t} \left[\Omega_1^*({\theta}_{r1},S_{r1},\theta_{r2},S_{r2})\,\text{d}{\theta}_{r1} + \Omega_2^*(\theta_{r1},S_{r1},{\theta}_{r2},S_{r2})\,\text{d}{\theta}_{r2} \right],
		\end{split}
\end{equation}
where $\eta_0^t$ is the portion of the shape-change path connecting
$(\theta_{r1}(0),\theta_{r2}(0))$ to $(\theta_{r1}(t),\theta_{r2}(t))$.
Hence, $\theta_c(t)$ does not depend on the speeds of the shape change, $\dot{\theta}_{r1}$ and $\dot{\theta}_{r2}$. Similar results hold for $x_c(t)$ and $y_c(t)$, and consequently also for the distance $d$, work $W$ and efficiency $e$. Hence, $e$ is only a function of the path $\eta$. In contrast to the two-link case, the shape of the three-link model is not symmetric in general. Therefore, the orientation $\theta_c$ given by~\eqref{eq:3linkthc} is no longer always small, and the initial and final orientations in one period are not necessarily the same.

The extrema of $e$ occur when the variation of $e$ with respect to $\eta$ is zero, i.e.
\begin{equation}
	\delta_{\eta} e = 0.
\end{equation}
Our goal is to find the $\eta$ which are local and/or global maximizers of $e$. 

\paragraph{Triangular waves for $\theta_{r1}$ and $\theta_{r2}$} 
\label{par:triangular_waves_for_r1_and_r2}
\begin{figure}
	[!b] \centering 
	\subfigure[$e$ contour]{\includegraphics[width=0.42\textwidth]{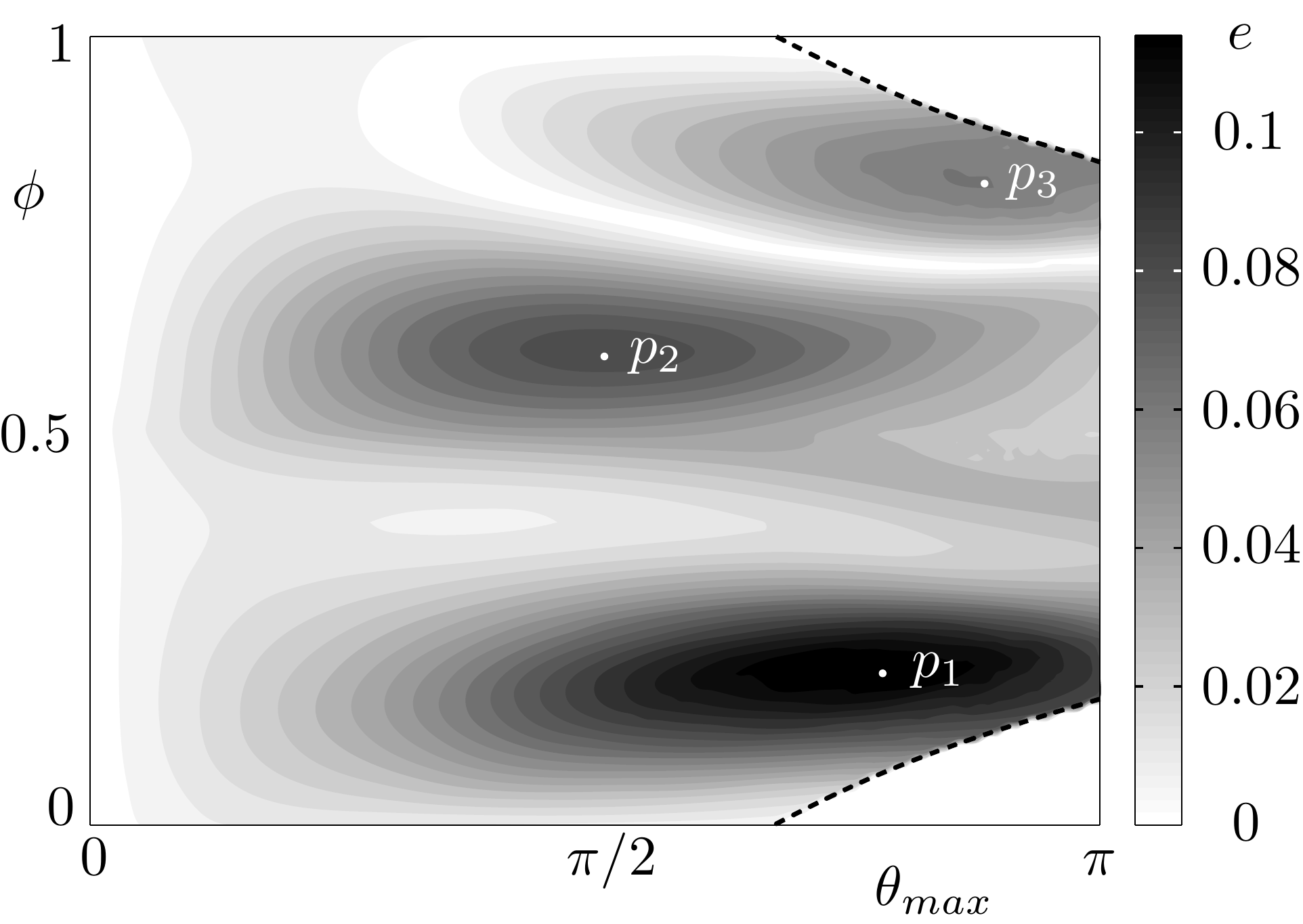}}\qquad
	\subfigure[paths in $(\theta_{r1} , \theta_{r2})$ space]{\includegraphics[width=0.285\textwidth]{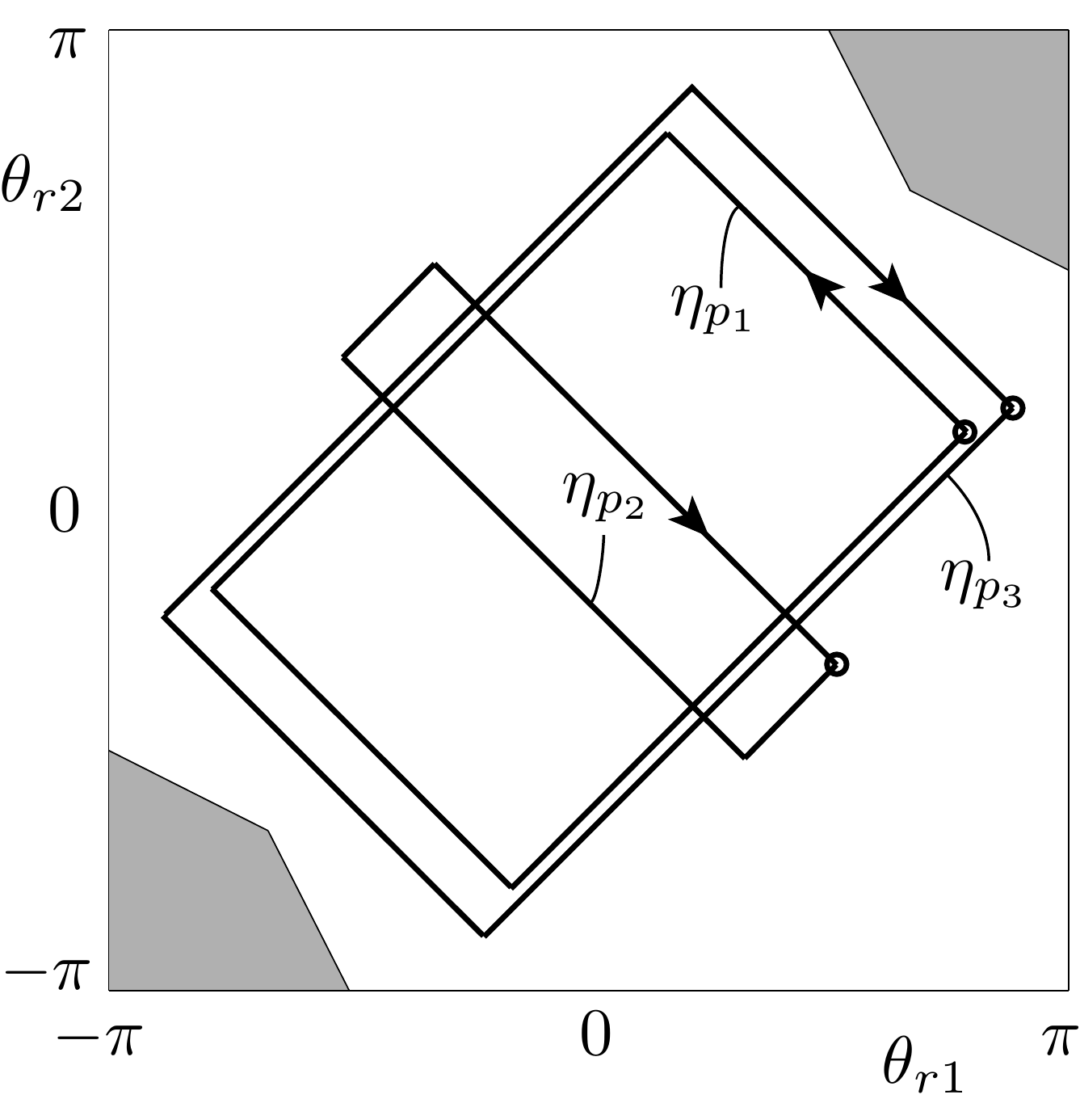}}
	\caption{\footnotesize (a) Contour plot of $e$ as function of $\theta_{max}$ and $\phi$ for the shape change prescribed in~\eqref{eq:3linkrect}. Areas bounded by dashed lines and boundaries at the upper and lower right corners are infeasible due to the mutual avoidance of the links. The three local maxima of $e$ are marked $p_1, p_2$ and $p_3$: $e_{p_1} = 0.1197$ (global maximum), $e_{p_2} = 0.0834$ and $e_{p_3} = 0.0634$. (b) Corresponding paths $\eta$ in the $(\theta_{r1} , \theta_{r2})$ parameter space.}\label{fig:3linkrecteff}
\end{figure}
We start by maximizing within low-dimensional subspaces of the space of
all feasible paths. First, we prescribe $\theta_{r1}$ and $\theta_{r2}$ to be triangular waves with period 1 (similar to the two-link case), and equal amplitudes: $\theta_{r1_{max}}  = \theta_{r2_{max}} = -\theta_{r1_{min}} = -\theta_{r2_{min}} \equiv \theta_{max}$. There is a phase delay $\phi$ between the two angles: $\theta_{r1}$ is assumed to be given by Figure~\ref{fig:2linktri_th}(a) and $\theta_{r2}$ is shifted $\phi$ behind $\theta_{r1}$ in time. In general, $0<\theta_{max}<\pi$ and $0 <\phi<1$, but the constraint given in~\eqref{eq:constraintset} also applies. The angular velocities are given by
\begin{equation}\label{eq:3linkrect}
	\dot{\theta}_{r1} = \begin{cases}
	- 4 \theta_{max}, & 0 \leq t < 0.5,\\
	4 \theta_{max}, & 0.5 \leq t \leq 1,
	\end{cases}
	\quad
	\dot{\theta}_{r2} = \begin{cases}
	- 4 \theta_{max}, & 0 \leq t - \phi < 0.5\ \ \text{or} \ \ -1 < t - \phi
< -0.5,\\
	4 \theta_{max}, & 0.5 \leq t  - \phi < 1\ \ \text{or} \ \ -0.5 \leq t -
\phi < 0.
	\end{cases}
\end{equation}
Such a path $\eta$ is a rectangle in $(\theta_{r1}, \theta_{r2})$, centered at the origin with the edges at $\pm45$ degrees to the $\theta_{r1}$ axis, examples of which are shown
in Figure~\ref{fig:3linkrecteff}(b). The shapes and orientations of the rectangular paths lie in a two-dimensional space parametrized by $(\theta_{max},\phi)$. We perform an exhaustive search for the globally optimal shape change in this case: discretize $(\theta_{max},\phi)$ on a $100\times 100$ mesh on the range $[0.01\,,\,\pi-0.01]\times[0.01\,,\,0.99]$, and calculate the efficiency at each node by solving the equations of motion given in~\eqref{eq:eom}. The results are depicted as a contour plot of $e$ as a function of $\theta_{max}$ and $\phi$ in Figure~\ref{fig:3linkrecteff}(a). The areas bounded by the dashed lines and the boundaries at the upper and lower right corners are infeasible due to the constraint given in~\eqref{eq:constraintset}. One finds 3 local maxima of $e$, located at $p_1, p_2$ and $p_3$. At $p_1$, $\theta_{max} = 2.4694$, $\phi = 0.1981$, and $e = 0.1197$, which is the global maximum for this family of kinematics. The second and third local minima are $e_{p_2} = 0.0834$ at $\theta_{max} = 1.6181, \phi = 0.5940$ and $e_{p_3} = 0.0634$ at $\theta_{max} = 2.7847, \phi = 0.8118$.
\begin{figure}
	[!tb] \centering 
	\includegraphics[width=\textwidth]{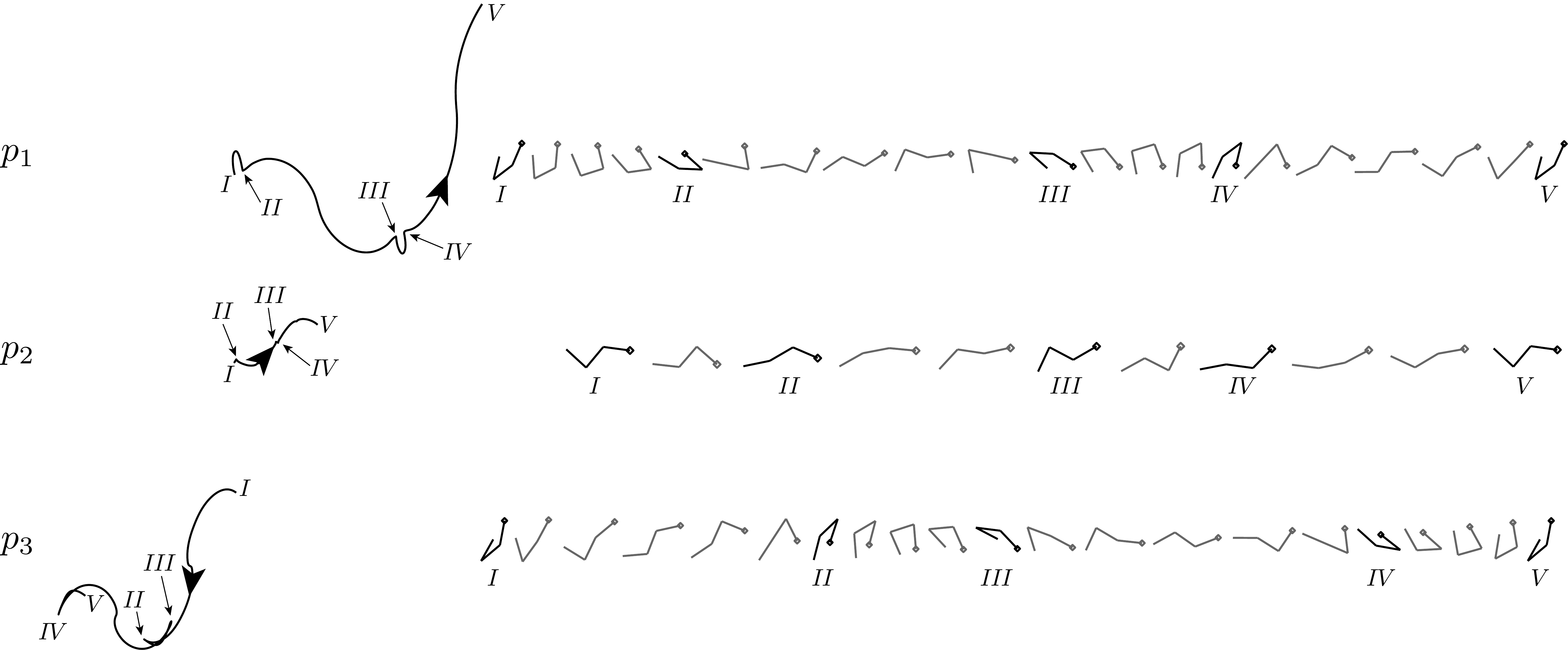}
	\caption{\footnotesize Trajectories of $C$ (left) and snapshots (right) in the inertial frame for $p_1 , p_2$ and $p_3$. Starting at $(0 , 0)$ at $t = 0$, the positions of $C$ at $t = 1$ are: for $p_1$, $(0.1649, 0.0504)$; for $p_2$, $(0.0536 , 0.0105)$; for $p_3$, $(-0.0973 , -0.0300)$. The snapshots are given at time increments of 0.05 for $p_1$ and $p_3$ and $0.1$ for $p_2$. The head of the snake is represented by $\Diamond$. The orientations of the snapshots in the inertial frame are preserved, while the centers are shifted to a straight line for better illustration.}\label{fig:3linkrect}
\end{figure}
The corresponding trajectories $\eta$ in the $(\theta_{r1}\,,\,\theta_{r2})$ plane are shown in Figure~\ref{fig:3linkrecteff}(b). Note that $\eta_{p_1}$ is counterclockwise, and the other two are clockwise, because $\phi < 0.5$ for $p_1$ and $\phi > 0.5$ for the others. Figure~\ref{fig:3linkrect} shows trajectories of $C$ and snapshots of the snake at these local
optima. The arrows indicate the directions of travel along the trajectories. We note that on average the snake moves forward for $p_1$ and $p_2$, but backward for $p_3$ (keep in mind that distance is the norm of net displacement). It seems suboptimal to move backwards (since $\mu_b > 1$), and, indeed, $e_{p_3}$ is not a global maximum. One can see two small ``loops" in the trajectory of $C$ for $p_1$ and they correspond to the first and third edges of the rectangle, during which the head and tail links are almost parallel to each other during the motion. The other two edges correspond to the head and tail links rotating in opposite directions, which results in much greater displacements compared to the small loops. During the period, the orientation of the middle link does not vary as much as the other two links. The differences among the edges in the work done are not very large but the differences in distance traveled are. The first and third edges are somewhat analogous to recovery strokes, with
the second and fourth edges analogous to power strokes. 

\paragraph{More general shape changes} 
\label{par:general_shape_change}
Now, we optimize in a more general class of shape changes. Since $\theta_{r1}$ and $\theta_{r2}$ are periodic functions with period 1, they can be represented as Fourier series,
\begin{equation}\label{eq:fourier}
	\theta_{ri}(t) = \frac{a_0^i}{2} + \sum_{j=1}^{n} \left[a_j^i\cos\left(2
j \pi t\right) + b_j^i\sin\left(2 j \pi t\right)\right],
\end{equation}
where $a_j^i$ and $b_j^i$ are Fourier coefficients, their upper-scripts $i = 1$ or $i = 2$ correspond to $\theta_{r1}$ or $\theta_{r2}$, respectively, and the lower-scripts $j$ indicate the mode in Fourier series. One needs $n \rightarrow \infty$ to represent a general function. Following~\cite{TaHo2007}, we start with $n = 1$ and increase $n$ systematically, observing the resulting change in the optima.

\begin{figure}
	[!tb] \centering 
	\subfigure[path $\eta$ in $(\theta_{r1} , \theta_{r2})$ space]{\includegraphics[width=0.3\textwidth]{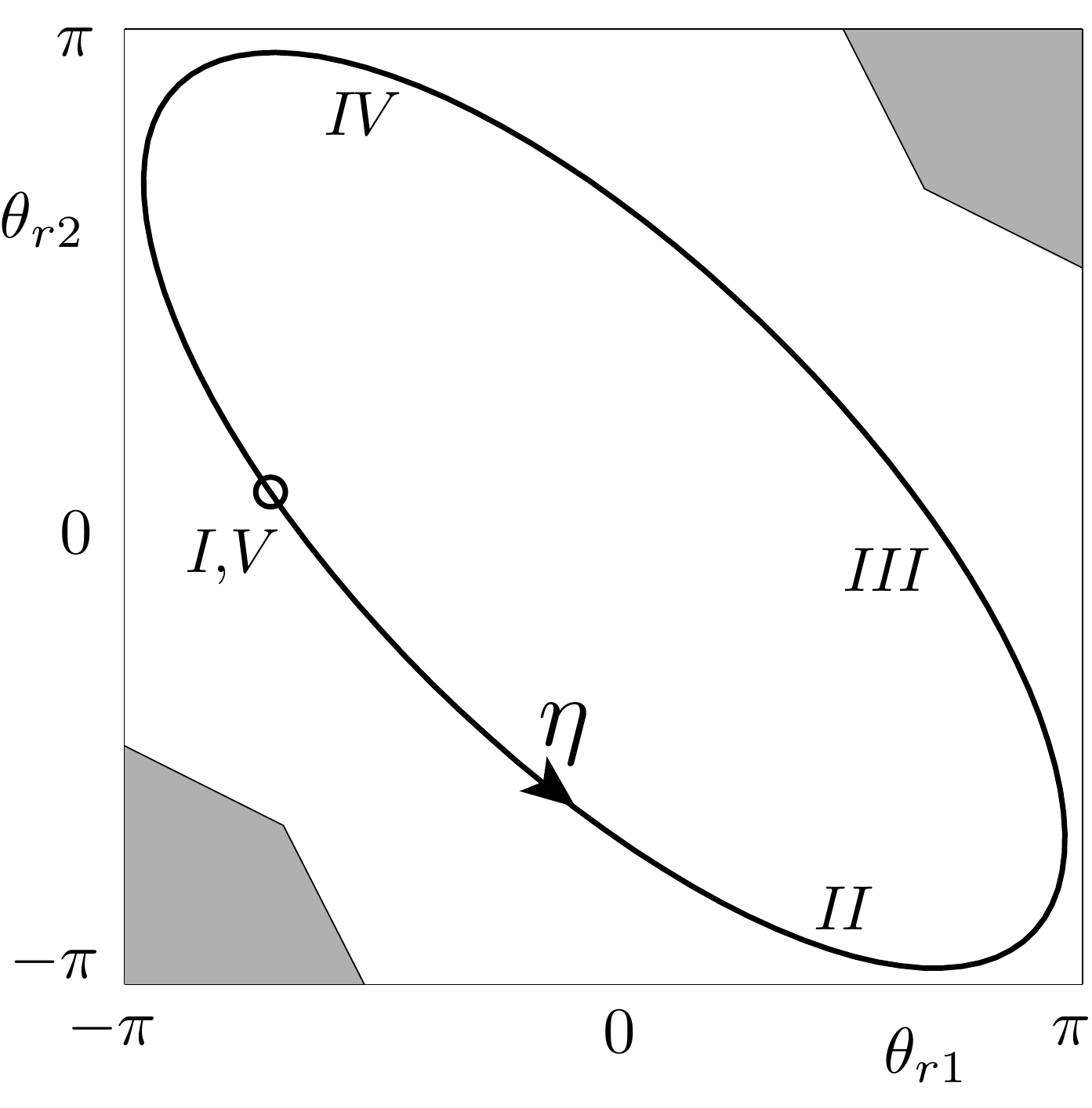}}\qquad\qquad
	\subfigure[trajectory of $C$]{\includegraphics[width=0.25\textwidth]{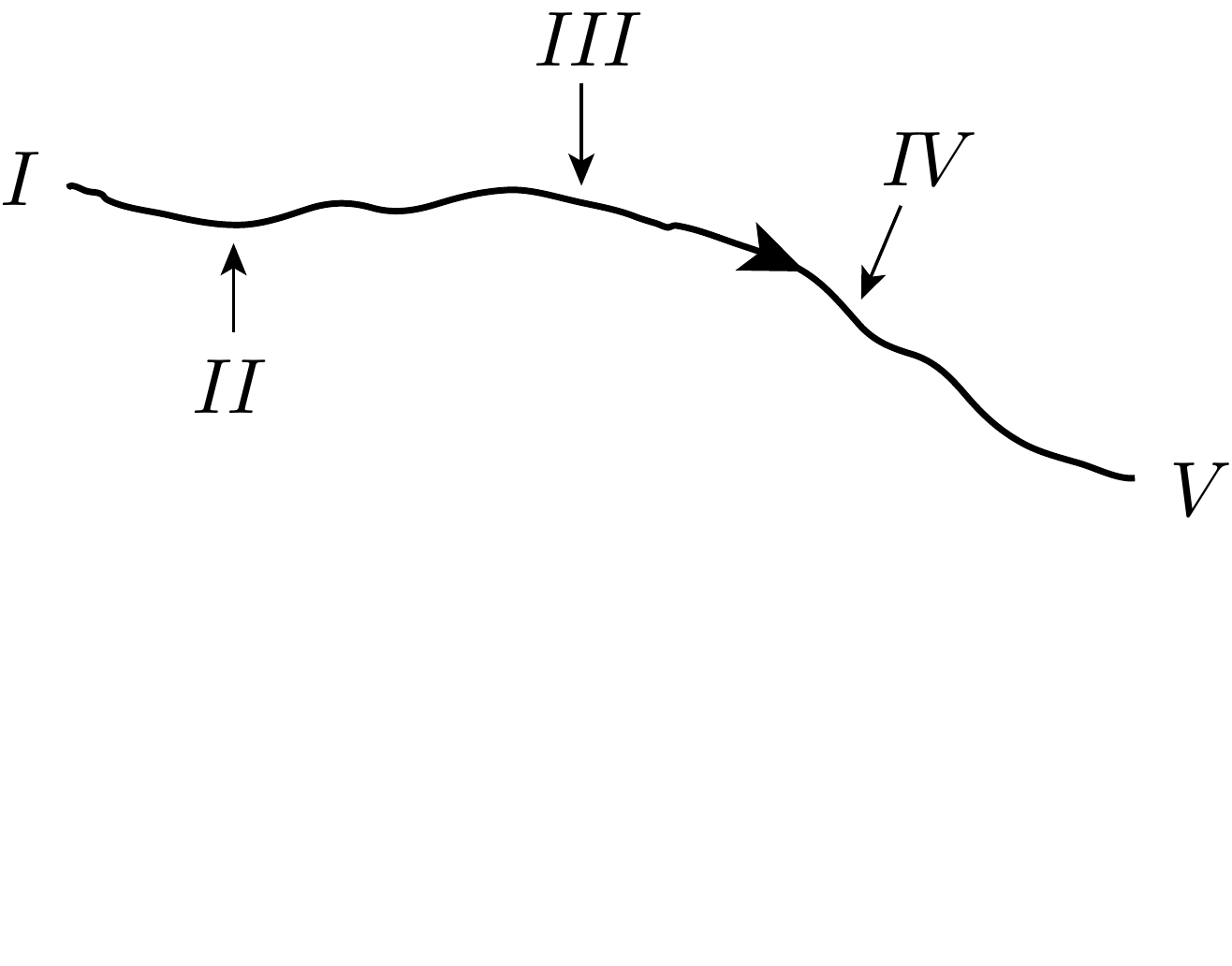}}
	\subfigure[snapshots]{\includegraphics[width=0.75\textwidth]{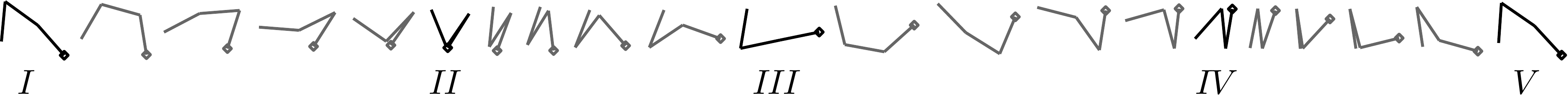}}
	\caption{\footnotesize (a) Global optimal path $\eta$ in $(\theta_{r1} , \theta_{r2})$ space for 1 mode in the Fourier series. (b) Trajectory of the center of mass $C$ in the inertial frame, which moves from $(0 , 0)$ at $t = 0$ to $(0.3381 , -0.0780)$ at $t=1$, with an arrow showing the direction of locomotion. (c) Snapshots of three-link snake are given at time increments of 0.05, with the head of snake represented by $\Diamond$.}\label{fig:3link1mode}
\end{figure}

For $n = 1$, the relative angles are given by $\theta_{r1}(t) = a_0^1/2 + a_1^1\cos\left(2\pi t\right) + b_1^1\sin\left(2\pi t\right)$ and $\theta_{r2}(t) = a_0^2/2 + a_1^2\cos\left(2\pi t\right) + b_1^2\sin\left(2\pi t\right)$. The corresponding $\eta$ is a relatively smooth curve,
an example of which is shown in Figure~\ref{fig:3link1mode}(a). For compactness, we 
denote $\mathbf{q} = [a_0^1\,,\,a_0^2\,,\,a_1^1\,,\,a_1^2\,,\,b_1^1\,,\,b_1^2]^T$ as the coefficient vector. Mathematically, the optimization problem can be stated as the following,
\begin{equation}
	\label{eq:optform}
	\begin{array}{lc}
		\text{Objective} & \underset{\mathbf{q}}{\max} \ e(\eta),\\[3ex]
		\text{Variables} & \mathbf{q} = [a_0^1\,,\,a_0^2\,,\,a_1^1\,,\,a_1^2\,,\,b_1^1\,,\,b_1^2]^T,\\[2.5ex]
		& \eta : \theta_{ri}(t) = a_0^i/2+ a_1^i\cos\left(2\pi t\right)
+ b_1^i\sin\left(2\pi t\right), \ i = 1, 2,\\[2.5ex]
		 \text{Equations} &	\int_0^1 \mathbf{f}\, \text{d}s =
\mathbf{0},\quad	\int_0^1 \mathbf{X}^{\perp}\cdot \mathbf{f}\,\text{d}s = 0,
\quad \forall t \in [0 \,,\, 1 ],\\[2ex]
		\text{Constraints} & (\theta_{r1},\theta_{r2}) \in
S_{\theta_{r1},\theta_{r2}}\ ,
		\end{array}
\end{equation}
where $S_{\theta_{r1},\theta_{r2}}$ is given by~\eqref{eq:constraintset}. Since $\mathbf{q}$ has 6 degrees of freedom, performing an exhaustive search is more expensive now. Instead, we utilize the subroutines provided by the {\tt Global Optimization Toolbox} in MATLAB. The procedure is briefly described here. For a random initial guess $\mathbf{q}_0$, a local constrained optimization function {\tt fmincon}, which implements a Sequential Quadratic Programming (SQP) algorithm, is called to solve for a {\em local} optimal $\mathbf{q}_{local}$ that results in a local maximum of efficiency $e_{local}$. To find the globally optimal variable $\mathbf{q}_{global}$, two approaches are taken: (i) repeat the local search 2000 times with random initial guesses pre-filtered such that the new searches do not fall back to the immediate vicinity of exploited results; then pick the largest $e_{local}$ as a candidate for the global maximum (using {\tt MultiStart}); (ii) numerically calculate the basins of attraction of the local maxima and then pick the largest $e_{local}$ as a candidate when the basins cover the variable space (using {\tt GlobalSearch}). Each approach is repeated several times, and the largest candidate is regarded as the global maximum. We verify the global maximum is indeed a local maximum by numerically computing the gradient using {\tt PatternSearch}. 

The globally optimal coefficients are $\mathbf{q}_{global} = [-0.1635, -0.2274, -2.0246, -0.1079, 2.2364, -2.9886]^T$, and the corresponding $\eta$ is depicted in Figure~\ref{fig:3link1mode}(a), with the starting point marked by $\circ$ and the direction shown by the arrow. Figure~\ref{fig:3link1mode}(b) shows the trajectory of $C$ in the inertial frame, which moves from the origin when $t= 0$ to $(0.3381 , -0.0780)$ when $t = 1$. Figure~\ref{fig:3link1mode}(c) shows the snapshots of the snake at time increment $\Delta t = 0.05$ from $t = 0$ to 1. The five instants $t = 0, 0.25, 0.5, 0.75, 1$ correspond to the five states $I$ to $V$, respectively. The distance is $d = 0.3470$, the work is $W = 1.5123$, and the efficiency is $e = 0.2295$. In comparison to the optimal solution in the triangular wave case ($e = 0.1197$), the efficiency is much higher. The loops in the trajectory in Figure~\ref{fig:3linkrect} no longer exist, the distance is much larger, and the trajectory is smoother than in the triangular wave case.

\begin{figure}
	[!tb] \centering 
	\subfigure[paths in parameter space]{\includegraphics[width=0.3\textwidth]{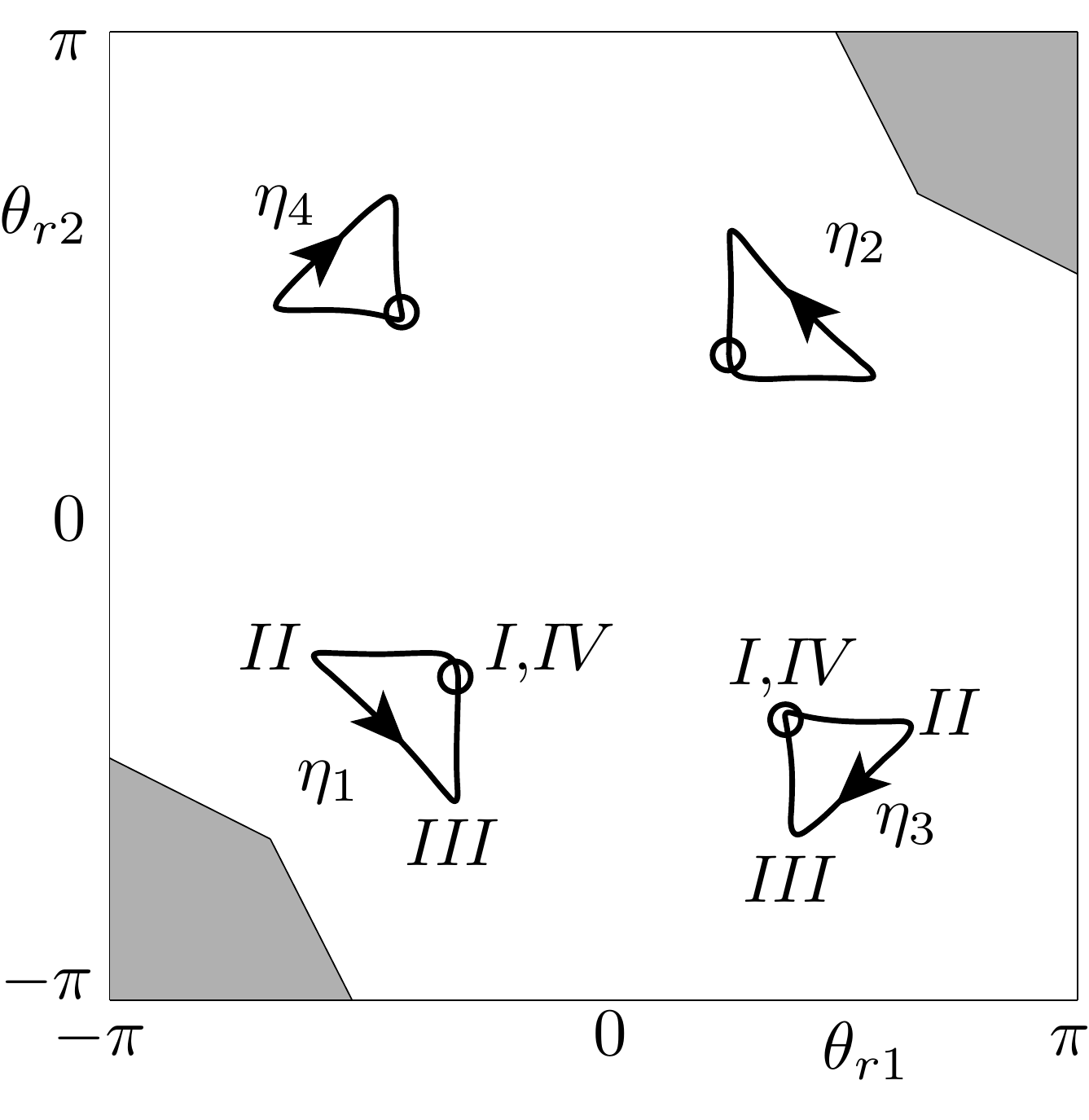}}\qquad\quad
	\subfigure[trajectories and snapshots of $\eta_1$ and $\eta_3$]{\includegraphics[width=0.6\textwidth]{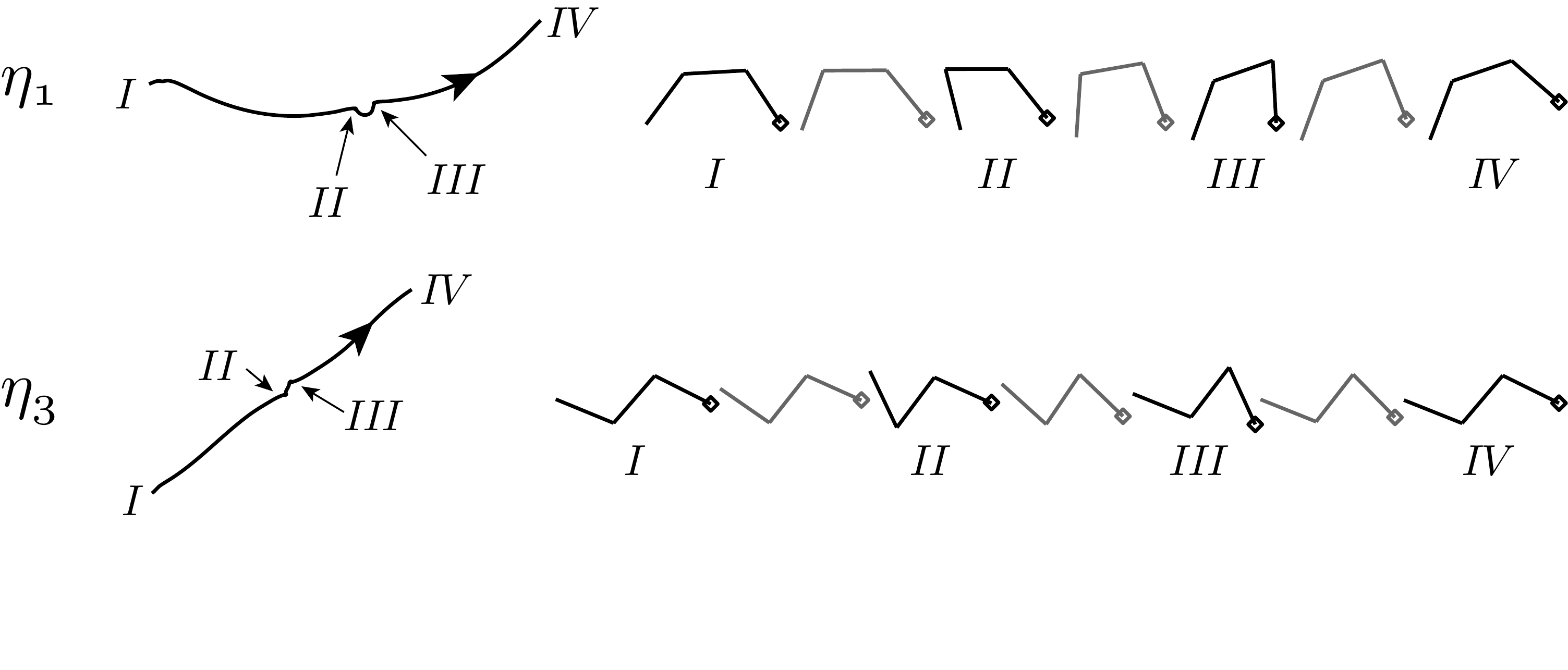}}
	\caption{\footnotesize (a) Optimal paths $\eta_i, i = 1, 2, 3, 4$ in $(\theta_{r1} , \theta_{r2})$ space for $n=2$ mode in the Fourier series; $e_1 = 0.3253$ (global optimum), $e_2 = 0.3252, e_3 = 0.2893$ and $e_4 = 0.2882$. (b) Trajectories of the centers of mass $C$ in the inertial frames for $\eta_1$ (top), for which $C$ moves from $(0 , 0)$ at $t = 0$ to $(0.0884 , 0.0144)$ at $t = 1$. The snapshots are at $t = 0, 0.185, 0.37, 0.555, 0.74, 0.87$ and 1. For $\eta_3$ (bottom), $C$ moves from $(0 , 0)$ at $t = 0$ to $(0.0586 , 0.0454)$ at $t = 1$. The snapshots are at $t = 0, 0.19, 0.38, 0.555, 0.73, 0.875$ and 1.}\label{fig:3link2mode}
\end{figure}

For $n = 2$, the coefficient variable is $\mathbf{q} = [a_0^1\,,\,a_0^2\,,\,a_1^1\,,\,a_1^2\,,\,b_1^1\,,\,b_1^2\,,\,a_2^1\,,\,a_2^2\,,\,b_2^1\,,\,b_2^2]^T$, and the 4 additional coefficients correspond to the second mode in the Fourier series. Following the aforementioned procedure, we obtain the global optimal coefficient $\mathbf{q} = [-2.7831, -2.0203, 0.3941, 0.0653, -0.3580,  0.4929, -0.0156, 0.1473,\linebreak 0.1589, -0.0305]^T$. Note that the second-mode coefficients are smaller on average than the first-mode coefficients. Besides this global maximum, we also find 3 other local maxima of interest. Their paths in the parameter space are plotted in Figure~\ref{fig:3link2mode}(a). We denote the global optimal path at the lower left corner as $\eta_1$, and the local optimal results at the upper right, lower right and upper left corners as $\eta_2, \,\eta_3$ and $\eta_4$, respectively. The corresponding efficiencies are $e_1 = 0.3253, e_2 = 0.3252, e_3 = 0.2893$ and $e_4 = 0.2882$. Note that they are all larger than the global maximum with one mode. One can see that $\eta_1$ and $\eta_2$ are very similar in shape and are counterclockwise. Their efficiencies are almost equal. Similar results hold for $\eta_3$ and $\eta_4$, only they are clockwise. In all four cases, the path shapes are close to 45-45-90 right triangles. The locations of the centers of $\eta_1$ and $\eta_2$ are close to the $\theta_{r1} = \theta_{r2}$ line and those of $\eta_3$ and $\eta_4$ are close to the $\theta_{r1} = -\theta_{r2}$ line. Figure~\ref{fig:3link2mode}(b) shows the trajectories of $C$ in the inertial frame and snapshots of $\eta_1$ and $\eta_3$. For $\eta_2$ and $\eta_4$, they are nearly mirror images of those for $\eta_1$ and $\eta_3$, respectively. In all four cases the motion is the following: moving just the head or just the tail link first, then moving both, then just the other link. The orientation of the middle link does not change much in all cases. It is interesting to note that for the 2-mode case, more efficient kinematics correspond to a ``contraction-expansion'' motion, which is reminiscent of the ``concertina'' mode of snake locomotion~\cite{MaHu2012}.

For $n > 2$, so far we have not found any motion that is more efficient that the ones found for $n = 2$. The numerical error in our calculation of efficiency is less than $10^{-5}$, based on the time
discretization used. The higher dimensionality of the parameter space at larger $n$ implies more computational time is needed to locate local optima, which may also increase in number. Nevertheless, based on our computations for $n > 2$, the global optimum obtained in the 2-mode study seems likely to be close to the global efficiency-maximizing shape change for the three-link model.

\section{Discussion} 
\label{sec:discussion}

To summarize, we have adapted a model for the slithering locomotion of snakes from~\cite{HuNiScSh2009} to the case of
two-link and three-link bodies sliding in 2D. Two dimensionless numbers, the ratios of the coefficients of friction, are the key physical parameters. Because of the frictional anisotropy, the local connection matrix in the reconstruction equation depends on both the relative angles and the directions of their velocities, which breaks the symmetry of time reversal and shape reversal equivalence, and the Scallop theorem does not apply. We maximized the efficiency $e$, the ratio of the distance traveled to the work done by the snake during one period of actuation, under various kinematic assumptions. For a two-link snake in the limit of small-amplitude actuations, $e$ is maximized when $\mu_b \gg 1$ and $\mu_t \ll 1$. Simulations of large-amplitude actuations show that the optimal shape change occurs when the relative angle amplitude $\theta_{max} \approx \pi/2$ except for small $\mu_t$. In real snakes, $\mu_t$ is usually not very small. We note that when $\mu_t \ll 1$, the distance traveled is also very small, as shown in Figure~\ref{fig:twolink_contour}. Although the body can increase its actuation frequency by decreasing $T$ to achieve higher $d$ in a given time, the actuation frequency may be constrained in biological or robotic snakes. If so, a very small $\mu_t$ may constrain the speed of locomotion to be small, which may be undesirable for biological or robotic snakes. We also note that $\mu_b$ can be increased by real snakes by altering the angles at which their scales contact the ground~\cite{MaHu2012}. For the three-link model, we searched for optimal motions in terms of the two relative angles $\theta_{r1}$ and $\theta_{r2}$, and assumed $\mu_b = 1.3$, $\mu_t = 1.7$, which is a set of coefficients measured for juvenile milk snakes~\cite{HuSh2012}. We first studied the family of kinematics with relative angles constrained to a rectangular path in parameter space. Three local optima were found, and the global optimum results in a trajectory with sharp changes in center-of-mass velocity, shown in Figure~\ref{fig:3linkrect}. We then considered paths parametrized by Fourier series for $\theta_{r1}$ and $\theta_{r2}$. If only the first mode (lowest frequency) is allowed, the optimal shape-change path in the parameter space resembles an ellipse elongated in the diagonal direction, shown in Figure~\ref{fig:3link1mode}. If more modes are allowed, then the optimal paths are instead close to right triangles that are small compared to the ellipse, shown in Figure~\ref{fig:3link2mode}. The globally optimal motion is found to be reminiscent of the {\em concertina} mode of snake locomotion~\cite{MaHu2012}. Table~\ref{tab:sum} highlights some of our findings.

\begin{table}[!htb]
\centering
\begin{tabular}{cMMMM}
& Model & Optimal\linebreak  shape change & Friction\linebreak
coefficients used & Optimal\linebreak friction coefficients\\  \hline
Two-link & 
	\includegraphics[width=0.12\textwidth]{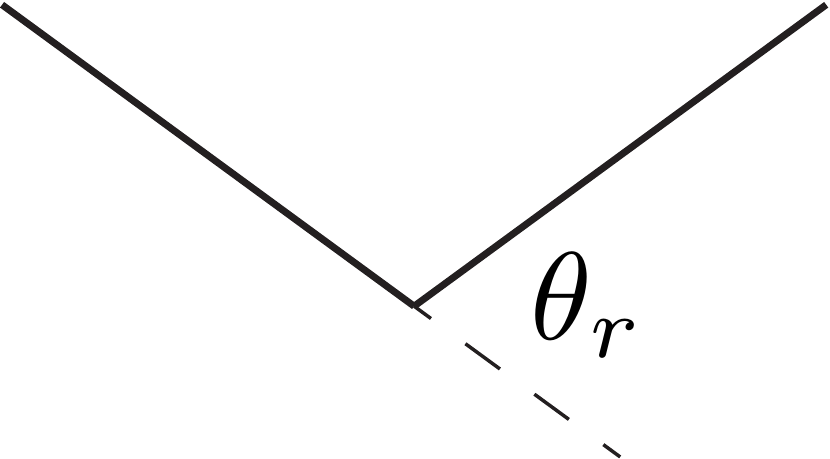} 
& $\theta_{max} \approx \pi/2$ $\mbox{unless}\; \mu_t \ll 1$
&  $1 < \mu_b/\mu_f < \infty$   $0 < \mu_t/\mu_f < \infty$
 & $\mu_b/\mu_f \gg 1$   $\mu_t/\mu_f \ll 1$
\\\hline
Three-link & 
\includegraphics[width=0.2\textwidth]{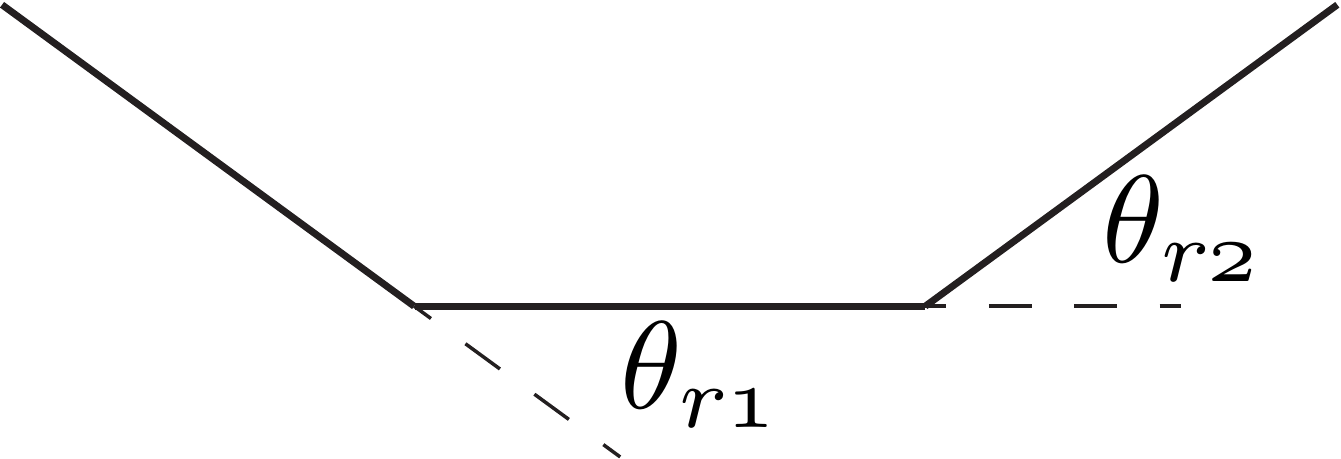} 
& 
\includegraphics[width=0.2\textwidth]{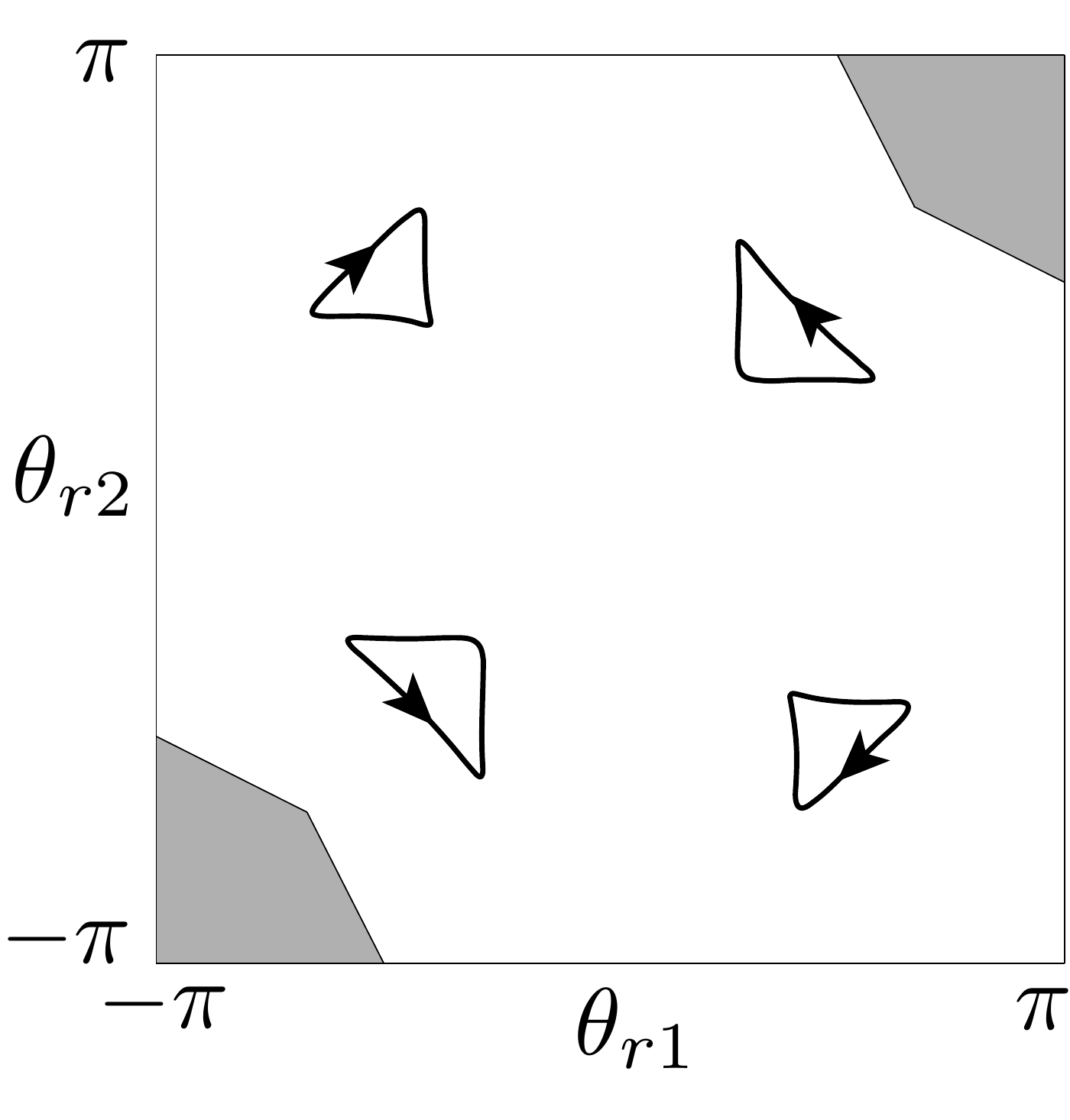} 
& $\mu_b/\mu_f = 1.3 \linebreak \mu_t/\mu_f = 1.7$ &
\end{tabular}
\caption{\footnotesize Highlights of results for two-link and three-link snake models. $\theta_r$ or $\theta_{r1}$
and $\theta_{r2}$ are the angles between the links, and $\mu_f, \mu_b$ and $\mu_t$ are the forward, backward and transverse
coefficients of friction, respectively. In the two-link model, the optimal coefficients are
obtained through analysis; in the three-link model, the coefficients of friction are fixed to
be those from an experiment~\cite{HuSh2012}.}
\label{tab:sum}
\end{table}

Although the results obtained in this work are based on a simplified model of snake locomotion, the system is still mathematically challenging due to its nonlinear nature. A key feature of snake locomotion -- frictional anisotropy -- is present in this work, but to keep the analysis tractable, other biological and dynamical aspects of snake slithering are omitted. One aspect is the dynamic load distribution, i.e. snakes lifting part of their body during locomotion~\cite{HuNiScSh2009}. To take this into consideration, the current model can be modified such that coefficients of friction can change in time and along the snake. This modification will result in a more complex optimization problem. The assumption of Coulomb friction could also be replaced by more complex and nonlinear friction models such as those discussed in~\cite{VeYe2008}, which would also present new challenges in solving the dynamical equations as well as the optimization problems.

In the three-link case, we focused on the optimal shape change with fixed coefficients of friction; a natural extension is to vary these. Another natural extension is to systematically increase the number of links $N$ and see how the optima change. For large $N$ it would interesting to compare optimal motions with those found in a two-parameter space of smooth shapes by \cite{HuSh2012}. 

\section{Acknowledgements} 
\label{sec:acknowledgements}
We would like to acknowledge the support of a Sloan Research Fellowship and 
NSF Division of Mathematical Sciences
Grant 1022619.

\appendix
\section{Derivation of the equations of motion and small-amplitude analysis} 
\label{appen:a}
The reader is first reminded that the velocity of an arbitrary point on any link can be decomposed into the velocity of the link center and a rotation about that center (see Figure~\ref{fig:2link_sketch}). For the tail link, in the body frame, the link center velocities in the tangential and transverse directions are 
\begin{equation}
	\label{eq:2linkut}
	\begin{pmatrix}
		U_{t}^s\\[2ex]
		U_{t}^n
	\end{pmatrix}
	\equiv 
	\begin{pmatrix}
		\cos\dfrac{\theta_r}{2} & -\sin\dfrac{\theta_r}{2}\\[2ex]
		\sin\dfrac{\theta_r}{2} & \cos\dfrac{\theta_r}{2}
	\end{pmatrix}
	\begin{pmatrix}
		U_c +\dfrac{1}{8}\dot{\theta}_r\sin\dfrac{\theta_r}{2}\\[2ex]
		V_c -\dfrac{1}{4}\Omega_c\cos\dfrac{\theta_r}{2}
	\end{pmatrix},
\end{equation}
and the rotation rate is $\dot{\Theta}_t = -\dot{\theta}_r/2$. The unit tangent vector of the tail link is $\hat{\mathbf{s}}_t = (\cos(\theta_r/2)\,,\,-\sin(\theta_r/2))$, and the unit transverse vector is $\hat{\mathbf{n}}_t = (\sin(\theta_r/2)\,,\,\cos(\theta_r/2))$. Therefore the linear velocity of any point on the tail link is given by
\begin{equation}
	\label{eq:2linkxit}
	\boldsymbol{\xi}_{t,lin} = U_{t}^s\,\hat{\mathbf{s}}_t + \left[U_{t}^n +
\left(s - \frac{1}{4}\right) \dot{\Theta}_t\right]\hat{\mathbf{n}}_t,\quad
\text{for} \quad 0\leq s \leq \frac{1}{2}.
\end{equation}
Similarly, for the head link, the link center velocity is
\begin{equation}
	\label{eq:2linkuh}
	\begin{pmatrix}
		U_{h}^s\\[2ex]
		U_{h}^n
	\end{pmatrix}
	\equiv 
	\begin{pmatrix}
		\cos\dfrac{\theta_r}{2} & \sin\dfrac{\theta_r}{2}\\[2ex]
		-\sin\dfrac{\theta_r}{2} & \cos\dfrac{\theta_r}{2}
	\end{pmatrix}
	\begin{pmatrix}
		U_c - \dfrac{1}{8}\dot{\theta}_r\sin\dfrac{\theta_r}{2}\\[2ex]
		V_c + \dfrac{1}{4}\Omega_c\cos\dfrac{\theta_r}{2}
	\end{pmatrix},
\end{equation}
and $\dot{\Theta}_h = \dot{\theta}_r/2$, the unit vectors $\hat{\mathbf{s}}_h = (\cos(\theta_r/2)\,,\,\sin(\theta_r/2))$ and $\hat{\mathbf{n}}_h = (-\sin(\theta_r/2)\,,\,\cos(\theta_r/2))$. Therefore the velocity of any point on the head link is given by
\begin{equation}
	\label{eq:2linkxih}
	\boldsymbol{\xi}_{h,lin} = U_h^s\, \hat{\mathbf{s}}_h + \left[U_h^n +
\left(s - \frac{3}{4}\right)\dot{\Theta}_h\right]\hat{\mathbf{n}}_h,\quad
\text{for} \quad \frac{1}{2}\leq s \leq 1.
\end{equation}
The unit velocity vector for the tail link is
\begin{equation}
	\label{eq:2linktailveldir}
	\hat{\boldsymbol{\xi}}_{t,lin} = \frac{1}{\sqrt{\left[U_t^n -(s - \frac{1}{4})\frac{\dot{\Theta}_t}{4} \right]^2 + (U_t^s)^2}}\left[ U_t^s \,,\,U_{t}^n +
	\left(s - \frac{1}{4}\right) \dot{\Theta}_t \right]^T,
\end{equation}
and that for the head link is of a similar form. For concreteness, we discuss the case when $\theta_r > 0$ and $\dot{\theta}_r < 0$, as depicted in Figure~\ref{fig:2link_sketch}. The analysis can be easily extended to other cases with minor modifications. Substituting~\eqref{eq:2linkxit} into~\eqref{eq:f} and integrating over the whole link, the frictional force exerted on the tail link is
\begin{equation}
	\label{eq:2linktailf}
	\begin{split}
		\int_0^{\frac{1}{2}} \mathbf{f} \text{d}s  = & -\frac{\mu_t
}{\dot{\Theta}_t} \left[ \sqrt{\left(U_t^n + \dot{\Theta}_t/4 \right)^2 +
(U_t^s)^2} - \sqrt{\left(U_t^n -\dot{\Theta}_t/4 \right)^2 + (U_t^s)^2}\right]
\hat{\mathbf{n}}_t\\
		& - \frac{\mu_b U_t^s}{\dot{\Theta}_t}\ln \left(\frac{U_t^n +
\dot{\Theta}_t/4 + \sqrt{\left(U_t^n+ \dot{\Theta}_t/4\right)^2 +
(U_t^s)^2}}{U_t^n - \dot{\Theta}_t/4 + \sqrt{\left(U_t^n -
\dot{\Theta}_t/4\right)^2 + (U_t^s)^2}}\right)\, \hat{\mathbf{s}}_t.
	\end{split}
\end{equation}
Similarly, the frictional force on the head link is
\begin{equation}
	\label{eq:2linkheadf}
	\begin{split}
		\int_{\frac{1}{2}}^1 \mathbf{f} \text{d}s  = & -\frac{\mu_t
}{\dot{\Theta}_h} \left[ \sqrt{\left(U_h^n + \dot{\Theta}_h/4 \right)^2 +
(U_h^s)^2} - \sqrt{\left(U_h^n -\dot{\Theta}_h/4 \right)^2 + (U_h^s)^2}\right]
\hat{\mathbf{n}}_h\\
		& - \frac{U_h^s}{\dot{\Theta}_h}\ln \left(\frac{U_h^n +
\dot{\Theta}_h/4 + \sqrt{\left(U_h^n+ \dot{\Theta}_h/4\right)^2 +
(U_h^s)^2}}{U_h^n - \dot{\Theta}_h/4 + \sqrt{\left(U_h^n -
\dot{\Theta}_h/4\right)^2 + (U_h^s)^2}}\right)\, \hat{\mathbf{s}}_h.
	\end{split}
\end{equation}
The force equation~\eqref{eq:eom} is $\int_0^{1/2} \mathbf{f} \text{d}s + \int_{1/2}^1 \mathbf{f} \text{d}s = \mathbf{0}$. One can readily derive the torque equation, which is omitted here due to complexity. We have given the force and torque equations in terms of $U_t^s, U_t^n, U_h^s$ and $U_h^n$. The center of mass velocity $\boldsymbol{\xi}_c = [U_c , V_c , \Omega_c]^T$ can be easily obtained by inverting~\eqref{eq:2linkut} and~\eqref{eq:2linkuh}. In general, $\boldsymbol{\xi}_c$ does not have a closed-form solution for large amplitudes of $\theta_r$. 

For small amplitude kinematics $\sup_t \|\theta_r\| = \epsilon \ll 1$, one can make the approximation
\begin{equation}
	\cos(\theta_r/2) \approx 1,\quad \sin(\theta_r/2) \approx \theta_r/2.
\end{equation}
Hence, $\hat{\mathbf{s}}_t$ and $\hat{\mathbf{s}}_h$ are almost parallel to $\mathbf{b}_x$ while $\hat{\mathbf{n}}_t$ and $\hat{\mathbf{n}}_h$ are almost parallel to $\mathbf{b}_y$. One can intuitively see that $U_c,\, V_c,\, \Omega_c \ll O(\epsilon)$ due to the symmetry in kinematics for the two-link model, and this can also be verified later. The dominant term in velocity is due to the rotation about each link center. As an example, in the tail link~\eqref{eq:2linkxit}, the dominant term is $(s- 1/4)\dot{\Theta}_t \hat{\mathbf{n}}_t$, whose magnitude is of the order $O(\epsilon)$ and direction is almost parallel to $\mathbf{b}_y$. All other terms have order of magnitude $O(\epsilon^2)$, and they are comparable to the dominant term only when $s \approx 1/4$. Therefore, the components of the unit velocity vector in~\eqref{eq:2linktailveldir} are: almost sign function in $\mathbf{b}_y$, and mostly zero in $\mathbf{b}_x$ except around $s \approx 1/4$. A similar result holds for the head link.

\begin{figure}[!tb]  
	\centering 
	\includegraphics[width=0.35\textwidth]{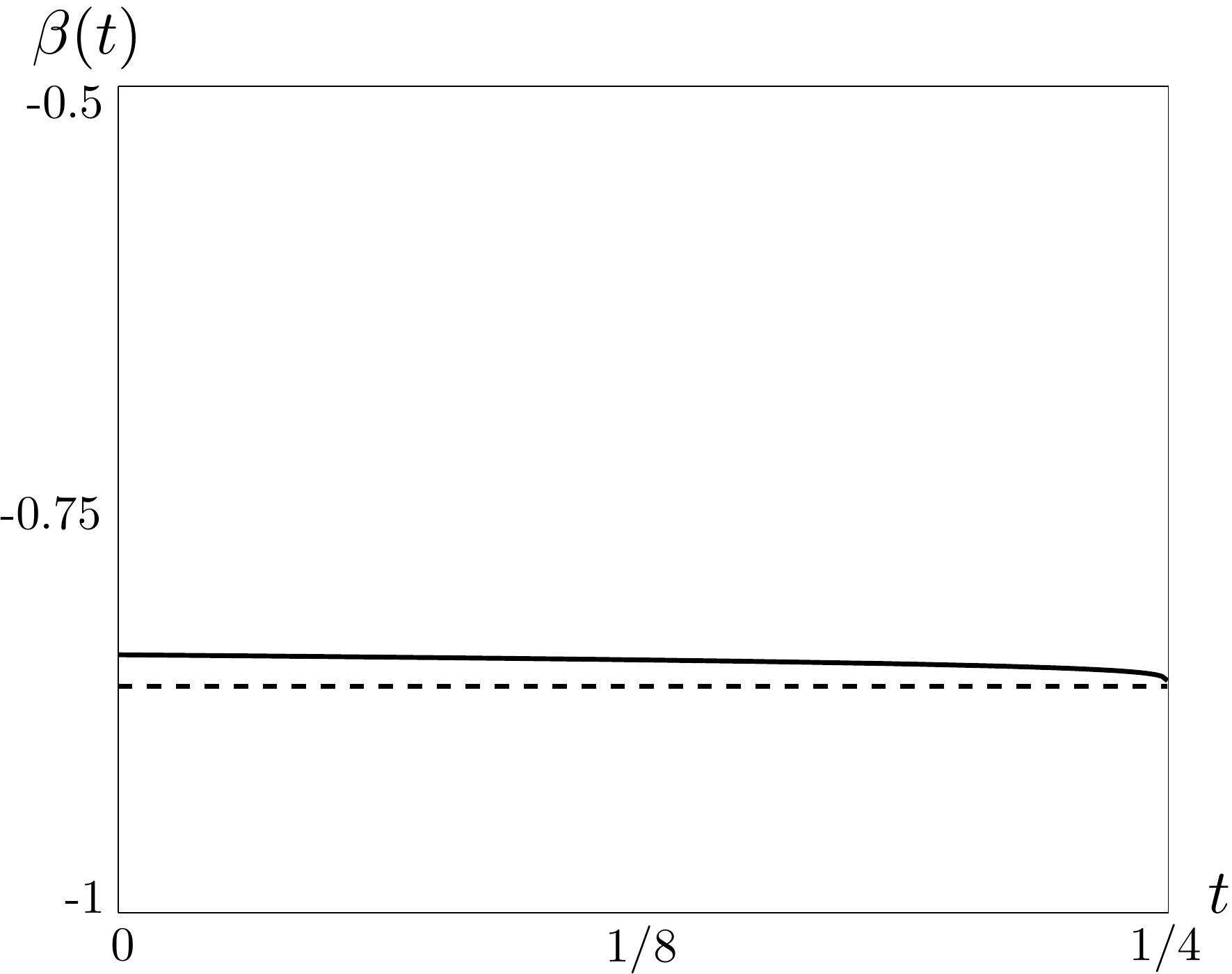}
	\caption{\footnotesize $\beta$ (solid) as a function of $t$ for $\mu_b = 1.3$
and $\epsilon$ = 0.1. It is close to a constant $-2/(\mu_b + 1) = -0.8696$ (dashed).}
\label{fig:2linkbeta}
\end{figure}

We now analyze the orders of magnitudes in the equations of motion. For the $\mathbf{b}_x$ component of the force equation, the term generated from the transverse direction, $\mu_t \theta_r^2/2 \sim O(\epsilon^2)$, is negligible compared to the tangential direction terms $\sim O(\epsilon)$, as long as $\mu_t/\mu_f$ and $\mu_b/\mu_f$ are $O(1)$ in the small-$\epsilon$ limit. Hence, after multiplying by the common denominator $\dot{\theta}_r/2$ and neglecting higher order terms, the equation is reduced to
\begin{equation}
	\label{eq:2linklineqbx}
			\mu_b \left(U_c +
	\dfrac{1}{16}\dot{\theta}_r\theta_r\right)\ln\left|\dfrac{-\dot{\theta}_r}{4U_c+\dot{
	\theta}_r\theta_r/4}\right|+\left(U_c -
	\dfrac{1}{16}\dot{\theta}_r\theta_r\right)\ln\left|\dfrac{-\dot{\theta}_r}{4U_c-\dot{
	\theta}_r\theta_r/4}\right| \approx 0.
\end{equation}
For the $\mathbf{b}_y$ component of the force equation, although the absolute values of the transverse terms are $O(1)$, they almost cancel out due to the symmetry in shape. As a result, contributions from both tangential and transverse directions are comparable in order of magnitude and $\sim O(\epsilon^2)$. Neglecting higher order terms, the equation becomes
\begin{equation}
			-\mu_t \left(2V_c + \dfrac{1}{16}\dot{\theta}_r\theta^2_r\right)
	- \mu_b \left(U_c\theta_r +
	\dfrac{1}{16}\dot{\theta}_r\theta^2_r\right)\ln\left|\dfrac{-\dot{\theta}_r}{4U_c+\dot{
	\theta}_r\theta_r/4}\right| \approx 0,
\end{equation}
in which~\eqref{eq:2linklineqbx} was used. For the torque equation, after multiplying by the common denominator $\dot{\theta}_r/2$, contributions from the tail and head links are comparable and $\sim O(\epsilon^2)$. After manipulation, the linearized torque equation is reduced to
\begin{equation}
\label{eq:2linklineqbx1}
			-\dfrac{3}{4} \mu_t\left(2U_c\theta_r - \Omega_c\right) +
	\left[4\mu_t(1+\mu_b) \dfrac{U_c}{\dot{\theta}_r} +
	\dfrac{\mu_t}{4}(1-\mu_b)\theta_r\right]\left(U_c +
	\dfrac{1}{16}\dot{\theta}_r\theta_r\right)\ln\left|\dfrac{-\dot{\theta}_r}{4U_c+\dot{
	\theta}_r\theta_r/4}\right| \approx 0,
\end{equation}
in which~\eqref{eq:2linklineqbx} was used again. To solve the 3 equations \eqref{eq:2linklineqbx}--\eqref{eq:2linklineqbx1}, notice in~\eqref{eq:2linklineqbx} that $U_c$ appears with $\dot{\theta}_r\theta_r/16$, and hence we seek $U_c$ in the form
\begin{equation}
	\label{eq:Uc}
	U_c \approx -\frac{1}{16}[1 + \beta(t)]\dot{\theta}_r\theta_r,
\end{equation}
where $\beta(t)$ is a function of time. The governing equation for $\beta(t)$ can be obtained by substituting~\eqref{eq:Uc} into~\eqref{eq:2linklineqbx}, obtaining
\begin{equation}
	\mu_b \left[-\frac{\beta(t)}{16}\dot{\theta}_r\theta_r\right] \ln \left|\frac{4}{\beta(t) \theta_r}\right| - \frac{1}{16}\left[2 + \beta(t)\right] \dot{\theta}_r\theta_r \ln \left|\frac{4}{\left[2 + \beta(t)\right]\theta_r}\right| \approx 0,
\end{equation}
which can be further simplified to
\begin{equation}
	\label{eq:betafull}
		 \left[(\mu_b + 1)\beta(t) + 2\right]  \left[2\ln2 - \ln|\theta_r(t)|
	\right] - \mu_b \beta(t) \ln |\beta(t)| - \left[\beta(t) + 2\right]
	\ln|\beta(t)+2| \approx 0.
\end{equation}
Note that $\beta$ depends on $\mu_b$ but not on $\mu_t$. For $\theta_r \sim O(\epsilon) \ll 1$, the dominant term in~\eqref{eq:betafull} is $\ln|\theta_r(t)|$. Hence, for the left side of~\eqref{eq:betafull} to be bounded in the small-$\epsilon$ limit, the coefficient of the dominant term has to tend to zero, i.e.
\begin{equation}
	\beta(t) \approx -\frac{2}{1 + \mu_b}.
\end{equation}
For the parameter $\mu_b = 1.3$, and $\theta_r(t)$ given by the triangular wave in~\eqref{eq:thetar} for $\epsilon = 0.1$, when $\theta_r > 0$ and $\dot{\theta}_r < 0$ (first quarter period), $\beta(t)$ is calculated from~\eqref{eq:betafull} and plotted in Figure~\ref{fig:2linkbeta}. One can see that $-2/(\mu_b + 1) = -0.8696$ is a good approximation of $\beta(t)$ for most times.
One can obtain solutions of $V_c$ and $\Omega_c$ by substituting~\eqref{eq:Uc} into the remaining two equations of motion:
\begin{equation}
	\label{eq:VcOmc}
			V_c \approx \dfrac{1}{32}\left(-1 +
	\dfrac{\mu_b}{\mu_t}\beta\ln\left|\dfrac{4}{\beta\theta_r}\right|\right)\dot{\theta}
	_r\theta^2_r,\quad
			\Omega_c \approx \left\{-\dfrac{1}{8}(1+\beta) -
	\dfrac{1}{48}\beta\left[\beta (1+\mu_b) +
	2\mu_b\right]\ln\left|\dfrac{4}{\beta\theta_r}\right|\right\}\dot{\theta}_r\theta^2_r,
\end{equation}

As stated before, these solutions for $U_c,\, V_c$ and $\Omega_c$ are for the case when $\theta_r > 0$ and $\dot{\theta}_r < 0$. One can easily obtain solutions for other cases based on symmetry. The reader is reminded that one can approximate inertial frame velocity with body frame velocity since $\theta_c \ll 1$ for all time, i.e. $\dot{\mathbf{g}}_c \approx \boldsymbol{\xi}_c$. As an example, for $\mu_b = 1.3, \mu_t = 1.7$ and $\epsilon = 0.1$, the velocities are given in Figure~\ref{fig:2linkvel0.1}. Since the velocity in the $\mathbf{e}_y$-direction $\dot{y}_c$ is almost anti-symmetric about $t = 1/2$, one has $\int_0^1 \dot{y}_c \text{d}t \approx 0$. Hence, the distance is given by
\begin{equation}
\begin{split}
	d & \approx \int_0^1 \dot{x}_c\, \text{d}t \approx \int_0^1 U_c\, \text{d}t \approx 4 \int_0^{1/4} U_c\, \text{d}t \approx 4 \int_0^{1/4} -\frac{1}{16}(1 + \beta)\dot{\theta}_r\theta_r\, \text{d}t\\
	&  = 4 \int_0^{1/4} -\frac{1}{16}(1 + \beta)(-4\epsilon)\epsilon(1-4t)\, \text{d}t = \frac{1}{8}(1 + \beta)\epsilon^2 \approx \frac{\mu_b - 1}{8(\mu_b + 1)}\epsilon^2.
\end{split}	
\end{equation}
The work is given by
\begin{equation}
	\begin{split}
		W & \approx \int_0^1\int_0^1 -\mathbf{f}\cdot\boldsymbol{\xi}_{lin}\,\text{d}s\,\text{d}t \approx - 4\int_0^{1/4}\left(\int_0^{1/2} + \int_{1/2}^1\right) \mathbf{f}\cdot\boldsymbol{\xi}_{lin}\,\text{d}s\,\text{d}t \\
		& \approx \frac{1}{4}\mu_t\epsilon + \gamma \epsilon^3,
	\end{split}	
\end{equation}
where
\begin{equation}
	\gamma \approx \frac{1}{16(\mu_b + 1)^2}\left[\frac{2\mu_t + \mu_b + \mu_b^2 + \mu_t\mu_b^2}{3} + \frac{\mu_t + \mu_b + \mu_b^2}{2}\ln \frac{(\mu_b + 1)^2}{2\epsilon^2} -  \frac{2\mu_b^2 + \mu_t}{2}\ln \mu_b\right].
\end{equation}
Therefore, the efficiency is given by
\begin{equation}
	e = \frac{d}{W} \approx \frac{\mu_b - 1}{2(\mu_b + 1)(\mu_t + 4\gamma \epsilon^2)} \epsilon.
\end{equation}
One can see that when $\mu_t \gg 4\gamma \epsilon^2$,
\begin{equation}
	e \approx \frac{\mu_b - 1}{2\mu_t (\mu_b + 1)} \epsilon.
\end{equation}



\newpage


\begin{thebibliography}{99}
	
\bibitem{AvRa2008}
Avron JE, Raz O (2008) A geometric theory of swimming: Purcell's swimmer and its
symmetrized cousin. {\em New J. Phys.} {\bf 10}:063016.
	
\bibitem{BeKoSt2003}
Becker L, Koehler S, Stone H (2003) On self-propulsion of micro-machines at low
Reynolds number: Purcell's three-link swimmer. {\em J. Fluid Mech.} {\bf
490}:15--35.

\bibitem{BuRaCh1994}
Burdick JW, Radford J, Chirikjian GS (1994) A `sidewinding' locomotion gait for
hyper-redundant robots. {\em Adv. Robotics} 9({\bf 3}):195--216.

\bibitem{Ch2003}
Chernousko FL (2003) Snake-like locomotions of multilink mechanisms. {\em J.
Vib. Control}  9({\bf 1-2}):235--256.

\bibitem{Ch1981}
Childress S (1981) {\em Mechanics of swimming and flying.} Cambridge University
Press.

\bibitem{Ch2005}
Choset HM (2005) {\em Principles of robot motion: theory, algorithms and
implementation.} MIT Press, Cambridge.

\bibitem{Ga1962}
Gans C (1962) Terrestrial locomotion without limbs. {\em Amer. Zool.}
2:167–-182.

\bibitem{Gr1946}
Gray J (1946) The mechanism of locomotion in snakes. {\em J. Exp. Biol.}
23:101–-120.

\bibitem{GuMa2008}
Guo ZV, Mahadevan L (2008) Limbless undulatory locomotion on land. {\em Proc.
Natl.
Acad. Sci. USA} 105:3179–3184

\bibitem{HaCh2011}
Hatton RL, Choset H (2011) Geometric motion planning: The local connection,
Stokes' theorem, and the importance of coordinate choice. {\em Int. J. Robot.
Res.} 30({\bf 8}):988--1014.

\bibitem{Hi1993}
Hirose S (1993) {\em Biologically inspired robots: snake-like locomotors and
manipulators.} Oxford University Press, Oxford.

\bibitem{HuNiScSh2009}
Hu D, Nirody J, Scott T, Shelley M (2009) The mechanics of slithering
locomotion. {\em Proc. Natl. Acad. Sci. USA} 106({\bf 25}):10081.

\bibitem{HuSh2012}
Hu DL, Shelley M (2012) Slithering locomotion, in Natural locomotion in fluids
and on surfaces (Childress S. {\em et al.} eds.) {\em IMA Vol. Math. Appl.}
155({\bf 1}):117--135.

\bibitem{JiKa2011}
Jing F, Kanso E (2011) Effects of body elasticity on stability of underwater
locomotion. {\em J. Fluid Mech.} 690:461--473.

\bibitem{KaMaRoMe2005} 
Kanso E, Marsden JE, Rowley CW, Melli-Huber JB, (2005) Locomotion of articulated
bodies in a perfect fluid. {\em J. Nonlinear Sci.} 15:255--289.

\bibitem{MaDiLiGo2009}
Maladen R, Ding Y, Li C, Goldman D (2009) Undulatory swimming in sand:
subsurface locomotion of the sandﬁsh lizard. {\em Science} 325:314.

\bibitem{MaHu2012}
Marvi H, Hu DL (2012) Friction enhancement in concertina locomotion of snakes.
{\em J. R. Soc. Interface.} 9({\bf 76}):3067--3080.

\bibitem{MeRoRu2006}
Melli JB, Rowley CW, Rufat DS (2006) Motion planning for an articulated body in
a perfect planar fluid. {\em SIAM J. Appl. Dyn. Sys.} 5({\bf 4}):650--669.

\bibitem{Mi2002}
Miller G (2002) Snake robots for search and rescue. {\em Neurotechnology for
biomimetic robots.} (Ayers J.D.J., Rudolph A.) Bradford/MIT Press, Cambridge,
269–-284.

\bibitem{Mo1932}
Mosauer W (1932) On the locomotion of snakes. {\em Science} 76:583–-585.

\bibitem{OsBu1996}
Ostrowksi J, Burdick J (1996) Gait kinematics for a serpentine robot. {\em Proc.
IEEE international conference on robotics and automation}, Minneapolis, MN,
1294--1299.

\bibitem{Pu1977}
Purcell E (1977) Life at low reynolds number. {\em Am. J. Phys.} 45({\bf
1}):3--11.

\bibitem{SeJaBe1992}
Secor SM, Jayne BC, Bennett AC (1992) Locomotor performance and energetic cost
of sidewinding by the snake crotalus cerastes. {\em J. Exp. Biol.} 163:1-–14.

\bibitem{Se1987}
Segel LA (1987) \textit{Mathematics Applied to Continuum
Mechanics} (1st~edn), Dover Publications, New York.

\bibitem{TaHo2007}
Tam D, Hosoi AE (2007) Optimal stroke patterns for Purcell's three-link swimmer.
{\em Phys. Rev. Lett.} {\bf 98}:068105.

\bibitem{VeYe2008}
Verriest EI, Yeung D (2008) Locomotion based on differential friction. In: {\em
Amer. Control Conf.}, Seattle, WA.

\bibitem{WaJaBe1990}
Walton M, Jayne BC, Bennett AF (1990) The energetic cost of limbless locomotion.
{\em Science} 249:524–-527.

\end{thebibliography}
\end{document}